\documentclass[12pt, journal,onecolumn]{IEEEtran}
\usepackage{cite}
\usepackage{amsmath,amssymb,amsfonts}
\usepackage[noend]{algorithmic}
\usepackage{graphicx}
\usepackage{textcomp}
\usepackage{xcolor}
\usepackage{amssymb}
\usepackage{indentfirst}
\usepackage{amsmath}
\usepackage{bm}
\usepackage{booktabs}
\usepackage{subfigure}
\usepackage{epsfig}
\usepackage{algorithm}
\usepackage{changepage}
\usepackage{amsthm} 
\usepackage{soul,color}
\usepackage{hyperref}
\usepackage{setspace}

\usepackage{tikz}
\usepackage{pgfplots}
\usetikzlibrary{plotmarks}
\usetikzlibrary{matrix,positioning,shapes,arrows,decorations,calc,fit}
\usetikzlibrary{decorations.pathreplacing}
\usetikzlibrary{spy,backgrounds}
\usetikzlibrary{external}
\usetikzlibrary{patterns}
\usetikzlibrary{shapes,arrows}
\usetikzlibrary{shadows.blur}
\usetikzlibrary{shapes.symbols}
\usetikzlibrary{
	pgfplots.groupplots,
	matrix
}

\newtheorem{proposition}{Proposition}

\theoremstyle{remark}
\newtheorem*{remark}{Remark}

\ifCLASSINFOpdf
\else
\fi
\IEEEaftertitletext{\vspace{-2\baselineskip}}

\begin{document}

\setlength{\abovedisplayskip}{2.5pt}
\setlength{\belowdisplayskip}{2.5pt}

\title{Linear-Equation Ordered-Statistics Decoding}

\author{Chentao~Yue,~\IEEEmembership{Member,~IEEE,}
        Mahyar~Shirvanimoghaddam,~\IEEEmembership{Senior Member,~IEEE,}\\
        Giyoon~Park, Ok-Sun~Park, 
        Branka~Vucetic,~\IEEEmembership{Life Fellow,~IEEE,}
        and~Yonghui~Li,~\IEEEmembership{Fellow,~IEEE}
\thanks{Chentao Yue, Mahyar Shirvanimoghaddam, Branka Vucetic, and Yonghui Li are with the School of Electrical and Information Engineering, the University of Sydney, NSW 2006, Australia (email:\{chentao.yue, mahyar.shm, branka.vucetic, yonghui.li\}@sydney.edu.au). Giyoon Park and Ok-sun Park are with the Electronics and Telecommunications Research Institute, Daejeon, South Korea (email: gypark@etri.re.kr; ospark@etri.re.kr).}
}

\markboth{Journal of \LaTeX\ Class Files,~Vol.~14, No.~8, August~2015}%
{Shell \MakeLowercase{\textit{et al.}}: Bare Demo of IEEEtran.cls for IEEE Journals}
%



\maketitle

\begin{abstract}
 In this paper, we propose a new linear-equation ordered-statistics decoding (LE-OSD). Unlike the OSD, LE-OSD uses high reliable parity bits rather than information bits to recover the codeword estimates, which is equivalent to solving a system of linear equations (SLE). Only test error patterns (TEPs) that create feasible SLEs, referred to as the valid TEPs, are used to obtain different codeword estimates. We introduce several constraints on the Hamming weight of TEPs to limit the overall decoding complexity. Furthermore, we analyze the block error rate (BLER) and the computational complexity of the proposed approach. It is shown that LE-OSD has a similar performance as OSD in terms of BLER, which can asymptotically approach Maximum-likelihood (ML) performance with proper parameter selections. Simulation results demonstrate that the LE-OSD has a significantly reduced complexity compared to OSD, especially for low-rate codes, that usually require high decoding order in OSD. Nevertheless, the complexity reduction can also be observed for high-rate codes. In addition, we further improve LE-OSD by applying the decoding stopping condition and the TEP discarding condition. As shown by simulations, the improved LE-OSD has a considerably reduced complexity while maintaining the BLER performance, compared to the latest OSD approach from literature.
\end{abstract}

\begin{IEEEkeywords}
Short block codes, URLLC, Ordered-statistics decoding, Soft decoding
\end{IEEEkeywords}

%
\IEEEpeerreviewmaketitle

\vspace{-0.5em}
\section{Introduction}
\vspace{-0.5em}
Since 1948, when Shannon introduced the notion of channel capacity \cite{Shannon}, researchers have been exploring the powerful channel codes that can approach this limit. As remarkable research milestones, low density parity check (LDPC) codes and Turbo codes have been shown to perform very close to Shannon's limit at large block lengths and have been widely applied in the 3rd generation (3G) and 4th generation (4G) of mobile standards \cite{lin2004ECC}. The Polar code proposed by Arikan in 2009 \cite{polar2009} has attracted much attention in the last decade and has been chosen as the standard coding scheme for the 5th generation (5G)-enhanced mobile broadband (eMBB) control channels and the physical broadcast channel \cite{3GPP-1611081}. 

Recently, short code design and related decoding algorithms have attracted a great deal of interest \cite{liva2016codeSurvey,2016Performancecomparison,Mahyar2019ShortCode}, triggered by the stringent requirements of the new ultra-reliable and low-latency communications (URLLC) service. The decisive latency requirements of URLLC mandate the use of finite block-length codes. Under the short block length regime, Polyanskiy \emph{et. al.} derived new bounds for the achievable rate, referred to as the normal approximation (NA) \cite{PPV2010l,erseghe2016coding}. Following this theoretical milestone, researchers have been seeking practical coding schemes achieving NA to fulfill the requirements of URLLC. 

Polar codes, LDPC codes, and Bose-Chaudhuri-Hocquenghem (BCH) codes are conceivable coding scheme candidates for URLLC \cite{liva2016codeSurvey,2016Performancecomparison,Mahyar2019ShortCode}. It has been shown that the cyclic redundancy check (CRC)-aided Polar codes offer a desirable trade-off between decoding complexity and the gap to NA \cite{liva2016codeSurvey}. Nevertheless, the successive cancellation list (SCL) decoding with large list size still has a relatively high decoding complexity \cite{bioglio2020design}. LDPC codes have a low-complexity iterative belief propagation (BP) decoding, but their block error rate (BLER) performance must be improved for the short block length region in URLLC applications. The improvement can be achieved by employing near maximum-likelihood (ML) decoders. As reported by \cite{liva2016codeSurvey}, under ML decoding, the gap from LDPC codes to the NA can be shortened by 0.5 dB compared to that with BP. However, the ML decoding of LDPC codes has a high complexity \cite{liva2016codeSurvey}. A recently proposed primitive rateless (PR) code \cite{Mahyar2021primitive} with its inherent rateless property is also shown to perform very close to the finite length bounds; however, its decoding complexity is the major drawback. Furthermore, short BCH codes have also gained interest from the research community recently \cite{NewOSD-5GNR, liva2016codeSurvey, Mahyar2019ShortCode, yue2021probability}, as it outperforms other existing short channel codes in terms of error correction capability. BCH codes usually have large minimum distances \cite{lin2004ECC}, but its ML decoding is also highly complex.

To address the aforementioned issues, code candidates for URLLC require practical decoders with excellent performance. The ordered-statistics decoding (OSD), as a universal decoder for block codes, rekindled the interests recently \cite{dhakal2016error,NewOSD-5GNR, cavarec2020learning, celebi2021latency,choi2021fast, yue2021revisit, yue2021probability}. For a linear block code $\mathcal{C}(n,k)$ with minimum distance $d_{\mathrm{H}}$, an OSD with the order of $m = \lceil d_{\mathrm{H}}/4-1\rceil$ is asymptotically optimum approaching the same performance as the ML decoding \cite{Fossorier1995OSD}. OSD only requires inputs of the code generator matrix and the transmitted signal, and does not depend on any unique code structure. Apart from OSD, guessing random additive noise decoding (GRAND) \cite{duffy2019capacity, solomon2020soft} is another potential universal decoding for URLLC. GRAND guess noise patterns rather than generating codeword estimates, and the average number of noise guesses to approach ML decoding is $2^{n\min(1-\frac{k}{n},H_{\frac{1}{2}})}$ \cite{duffy2019capacity}, where $H_{\frac{1}{2}}$ is the entropy rate at parameter $\frac{1}{2}$. Compared to the brutal ML decoding that generates $2^{k}$ codeword estimates, GRAND is significantly efficient for high-rate code, i.e., large $\frac{k}{n}$, whereas for low-to-moderate-rate codes, OSD will be more efficient.

In OSD, the bit-wise log-likelihood ratios (LLRs) of the received symbols are sorted in descending order and the columns of the generator matrix are permuted accordingly. Systematic form of the permuted generator matrix is then obtained by performing Gaussian elimination (GE). Then, the first $k$ positions, referred to as the most reliable basis (MRB), are XORed with a set of the test error patterns (TEP) and re-encoded to generate codeword estimates. The re-encoding will continue until all TEPs are processed. The maximum allowed Hamming weight of TEPs is known as the decoding order. It is shown that the decoding complexity of an order-$m$ OSD is as high as $\mathcal{O}(k^m)$ because of the large list size of TEPs \cite{Fossorier1995OSD}.

Many previous works have focused on improving OSD and some significant progress has been achieved\cite{wu2007preprocessing_and_diversification,Wu2007OSDMRB,jin2006probabilisticConditions,FossorierBoxandMatch,Fossorier2002IISR,Fossorier2007OSDbias,choi2021fast,cavarec2020learning}. Some notable accomplishments include the skipping and stopping rules introduced in \cite{wu2007preprocessing_and_diversification,Wu2007OSDMRB,jin2006probabilisticConditions} to prevent processing unpromising TEPs which are unlikely to result in the correct output. Most recently, deep learning is applied recently to help OSD adaptively identify the required decoding order to achieve the optimal performance \cite{cavarec2020learning}. Johannes \emph{et. al.} \cite{NewOSD-5GNR} discussed the potential application of OSD for 5G new radio, and proposed a fast variant reducing the complexity of OSD from $\mathcal{O}(k^m)$ to $\mathcal{O}(k^{m-2})$ at high signal-to-noise ratios (SNRs). Latency in URLLC scenarios where OSD algorithms are used as the decoder was analyzed and optimized in \cite{celebi2021latency}, where the latency is represented as a function of the decoding complexity of OSD. Some attempts were also made to analyze the distance properties and error rate performance of the OSD and its variants \cite{Fossorier1995OSD,Fossorier2002ErrorAnalysis,fossorier1996first,dhakal2016error,yue2021revisit}. The evolution of the Hamming distances and weighted Hamming distances (WHD) from generated codeword estimates to the received signal in OSD are comprehensively studied in \cite{yue2021revisit}, which for the first time reveals how the decoding techniques (e.g., stopping and skipping rules) could improve OSD. Based on the theoretical results provided in \cite{yue2021revisit}, a novel probability-based OSD (PB-OSD) was proposed in \cite{yue2021probability}, which has a significantly reduced complexity compared with existing OSD approaches. 

Although researchers have made remarkable progress on devising low-complexity OSD algorithms, almost all existing approaches can effectively reduce the complexity only when the SNR is relatively high. This is mainly because the bit reliabilities, i.e., LLRs, are relatively accurate at moderate-to-high SNRs. In contrast, when the SNR is low, the decoding techniques based on reliabilities usually fail, as the LLRs are not reliable and the MRB itself might be misleading. This shortcoming can be even observed in recent approaches, e.g., PB-OSD \cite{yue2021probability}. As a result, current OSD decoders still exhibit relatively high complexity at low-to-moderate SNRs. Accordingly, when the channel conditions fluctuate, the decoder will potentially violate the system latency requirement. Furthermore, OSD usually needs a high order in decoding low-rate codes as they commonly have large minimum distances. In this case, the complexity of existing OSD approaches is still forbiddingly high, hindering their potential applications.

In this paper, we propose a linear-equation OSD (LE-OSD) approach, aiming at having a general complexity reduction of OSD, especially at low-to-moderate SNRs. Unlike OSD \cite{Fossorier1995OSD}, LE-OSD recovers the information bits based on the most reliable $n-k$ parity check bits, and TEPs are added to the parity check bits to generate a number of codeword estimates. This process is equivalent to solving a system of linear equations (SLE), and we refer to a TEP that results in a feasible SLE as a \textit{valid TEP}. A feasible SLE is defined as the SLE that has at least one solution. Given a Hamming weight constrain for the TEPs, only valid TEPs are processed to retrieve the codeword estimations, and thus the number of TEPs as well as the number of generated codeword estimates can be significantly reduced.

The BLER performance and the asymptotic BLER performance (in terms of the SNR) of LE-OSD are analyzed. We show that with a proper parameter selection, LE-OSD has a similar error rate performance to the original OSD \cite{Fossorier1995OSD}, and can asymptotically approach the ML performance. Moreover, the number of required TEPs and the computational complexity of LE-OSD are characterized. It is shown that LE-OSD has a significantly reduced number of TEPs compared to OSD in decoding codes with rates lower than 0.5, while needing to perform three GEs in the preprocessing of decoding. Therefore, the proposed LE-OSD is especially efficient when the order of OSD is high, whose complexity is dominated by the size of the TEP list and the overhead of GE is relatively negligible. As advised by simulations, the LE-OSD can be two to four times faster than the original OSD in high-order decoding, achieving the similar error rate performance. For example, regarding the decoding time of the order-5 decoding of rate-$1/4$ BCH codes, LE-OSD requires 5 ms to decode a codeword compared to 20 ms of the original OSD.

Furthermore, we design the decoding stopping condition (SC) and the TEP discarding condition (DC) for the LE-OSD by leveraging the idea in \cite{yue2021revisit} and \cite{yue2021probability}, which further reduces the complexity of LE-OSD. Simulation results show that LE-OSD can reduce the complexity by more than half at low to medium SNRs in terms of the decoding time consumption, compared to the recent approach, PB-OSD.

The rest of this paper is organized as follows. Section \ref{Sec::Preliminaries} describes the preliminaries and reviews the OSD. In Section \ref{sec::LE-OSD}, the details of the LE-OSD algorithm are described. The error rate performance, the computational complexity, and the parameter selection of LE-OSD are discussed in Section \ref{sec::ANA}. Simulation results and the comparison between LE-OSD and the original OSD algorithm are provided in Section \ref{sec::Simulation}. Section \ref{sec::Improve} proposes the improved LE-OSD. Finally, Section \ref{Sec::Conclusion} concludes the paper.

\emph{Notation}: In this paper, we use $\mathrm{Pr}(\cdot)$ to denote the probability of an event. A bold letter, e.g., $\mathbf A$, represents a matrix, and a lowercase bold letter, e.g., $\mathbf{a}$, denotes a row vector. We also use $[a]_u^v$ to denote a row vector containing element $a_{\ell}$ for $u\le \ell\le v$, i.e., $[a]_u^v = [a_u,\ldots,a_v]$. If $\mathbf{a}$ is a binary vector, we use $w(\mathbf{a})$ to denote its Hamming weight, i.e., $\ell_1$-norm. Furthermore, We use $\mathbf{A}^\top$ and $\mathbf{a}^\top$ to denote the transposition of a matrix $\mathbf{A}$ or vector $\mathbf{a}$, respectively, and $r_{\mathbf{A}}$ denotes the rank of $\mathbf{A}$. We use $\mathcal{O}(\cdot)$ to denote the Big-O notation, i.e., $\mathcal{O}(x)$ is bounded above by $x$ with up to a constant factor, asymptotically.
\vspace{-0.5em}
\section{Preliminaries} \label{Sec::Preliminaries}
\vspace{-0.5em}
    \subsection{Preliminaries of System}

    We consider a binary linear block code ${\mathcal C}(n,k)$ with binary phase shift keying (BPSK) modulation over an additive white Gaussian Noise (AWGN) channel, where $k$ and $n$ denote the information block and codeword length, respectively. Let $ \mathbf{b} = [b]_{1}^k$ and $\mathbf{c} = [c]_{1}^n$ denote the information sequence and codeword, respectively. Given the generator matrix $\mathbf{G}$ of ${\mathcal C}(n,k)$, the encoding operation can be described as $\mathbf{c} = \mathbf{b}\mathbf{G}$. At the channel output, the received signal is given by $\bm{\gamma} = \mathbf{s} + \mathbf{w}$, where $\mathbf{s} = [s]_{1}^n$ denotes the sequence of modulated symbols with $s_{i} = (-1)^{c_{i}}\in \{\pm 1\}$, where $\mathbf{w}= [w]_{1}^n$ is the AWGN vector with zero mean and variance $N_0/2$, for $N_0$ being the single side-band noise power spectrum density. The SNR is then given by $\mathrm{SNR} = 2/N_0$.

    At the receiver, the bitwise hard-decision estimate $\mathbf{y}= [y]_{1}^n$ of $\mathbf{c}$ can be obtained according to the following rule: $y_{i} = 1 $ for $\gamma_{i}<0$ and $y_{i} = 0$ for $\gamma_{i}\geq 0$. We use subscripts $\mathrm{B}$ and $\mathrm{P}$ to denote the first $k$ positions and the remaining $n-k$ positions of a length-$n$ vector, respectively, e.g., $\mathbf{y} = [\mathbf{y}_{\mathrm{B}} \ \ \mathbf{y}_{\mathrm{P}}]$. In general, if codewords in $\mathcal{C}(n,k)$ have equal transmission probability, LLR of the $i$-th received symbol is defined as ${\ell}_{i} \triangleq \ln \frac{\mathrm{Pr}(c_{i}=1|\gamma_{i})}{\mathrm{Pr}(c_{i}=0|\gamma_{i})}$, which is further simplified to ${\ell}_{i} = \frac{4\gamma_{i}}{N_{0}}$ if employing BPSK. Thus, we define $\alpha_{i} \triangleq |\gamma_{i}|$ (the scaled magnitude of LLR) as the reliability of $y_i$, where $|\cdot|$ is the absolute operation.
	
     For an arbitrary binary matrix $\mathbf{A}$, let $\mathbf{A}'$ denote its row echelon form (REF) obtained by performing GE. Without loss of generality, performing GE over an $u \times v$ matrix $\mathbf{A}$ is represented by two steps, i.e., 1) permuting columns of $\mathbf{A}$ to obtain $\mathbf{A}_1$ whose first $r_{\mathbf{A}}$ columns are linearly independent, and 2) performing row operations over $\mathbf{A}_1$ to obtain the REF $\mathbf{A}'$. Let a $v\times v$ matrix $\bm{\Pi}_{\mathbf{A}}$ denote the columns permutation in step 1, and let an $u\times u$ matrix $\mathbf{E}_{\mathbf{A}}$ represent the row operations performed in step 2. Then, $\mathbf{A}'$ is represented as $\mathbf{E}_{\mathbf{A}}\mathbf{A}\bm{\Pi}_{\mathbf{A}}$. Particularly, when the first $k$ column of $\mathbf{A}$ are linearly independent, $\bm{\Pi}_{\mathbf{A}}$ is a $v$-dimensional identity matrix. Note that $\bm{\Pi}_{\mathbf{A}}^{-1}=\bm{\Pi}_{\mathbf{A}}^{\top}$ because $\bm{\Pi}_{\mathbf{A}}$ is an orthogonal matrix. For the generator matrix $\mathbf{G}$ of $\mathcal{C}(n,k)$, $\mathbf{G}' = \mathbf{E}_{\mathbf{G}}\mathbf{G}\bm{\Pi}_{\mathbf{G}}$ is also known as the systematical form of $\mathbf{G}$. 
     
    \vspace{-0.5em}
    \subsection{Ordered Statistics Decoding} \label{sec::Pri::OSD}
    \vspace{-0.5em}
        Utilizing the bit reliability, the soft-decision decoding can be effectively conducted by OSD \cite{Fossorier1995OSD}. In OSD, a permutation $\bm{\Pi}_{d}$ is performed first to sort the reliabilities $\bm{\alpha} = [\alpha]_1^n$ in descending order, and the ordered reliabilities is obtained as $\bm{\alpha}\bm{\Pi}_d$. Then, the received signal $\bm{\gamma}$, the hard-decision vector $\mathbf{y}$, and the generator matrix are accordingly permuted to $\bm{\gamma}\bm{\Pi}_{d}$, $\mathbf{y}\bm{\Pi}_{d}$, and $\bar{\mathbf{G}} = \mathbf{G}\bm{\Pi}_{d}$, respectively. Next, OSD obtains the systematic matrix $\bar{\mathbf{G}}' =  [\mathbf{I}_k \  \bar{\mathbf{P}}]$ by performing GE over $\bar{\mathbf{G}}$, i.e., $\bar{\mathbf{G}}' = \mathbf{E}_{\bar{\mathbf{G}}}\bar{\mathbf{G}}\bm{\Pi}_{\bar{\mathbf{G}}}$, where $\mathbf{I}_k$ is a $k$-dimensional identity matrix and $\bar{\mathbf{P}}$ is the parity sub-matrix. Accordingly, $\bm{\gamma}\bm{\Pi}_{d}$, $\mathbf{y}\bm{\Pi}_{d}$, and $\bm{\alpha}\bm{\Pi}_d$ are further permuted to $\bar{\bm{\gamma}} = \bm{\gamma}\bm{\Pi}_{d}\bm{\Pi}_{\bar{\mathbf{G}}}$, $\bar{\mathbf{y}} = \mathbf{y}\bm{\Pi}_{d}\bm{\Pi}_{\bar{\mathbf{G}}}$, and $\bar{\bm{\alpha}} = \bm{\alpha}\bm{\Pi}_{d}\bm{\Pi}_{\bar{\mathbf{G}}}$, respectively.
    
        After the permutations and GE, the first $k$ positions of $\bar{\mathbf{y}}$ are associated with the MRB \cite{Fossorier1995OSD}, denoted by $\bar{\mathbf{y}}_{\mathrm{B}} =[\bar{y}]_1^k$. MRB can be regarded as the set of the most reliable $k$ hard-decision bits in $\mathbf{y}$, because the permutation $\bm{\Pi}_{\bar{\mathbf{G}}} $ only marginally disrupts the descending order of $\bm{\alpha}\bm{\Pi}_d$\cite[Eq. (59)]{Fossorier1995OSD}, i.e., $\bm{\Pi}_{\bar{\mathbf{G}}}$ is close to a identity matrix. With the aim to eliminate the errors in the MRB, a length-$k$ test error pattern $\bar{\mathbf{e}}$ is added to $\bar{\mathbf{y}}_{\mathrm{B}}$ to obtain a codeword estimate by re-encoding according to $\bar{\mathbf{c}}_{\mathbf{e}} = \left(\bar{\mathbf{y}}_{\mathrm{B}}\oplus \bar{\mathbf{e}}\right)\bar{\mathbf{G}}' = \left[\bar{\mathbf{y}}_{\mathrm{B}}\oplus \bar{\mathbf{e}} \  ~\left(\bar{\mathbf{y}}_{\mathrm{B}}\oplus \bar{\mathbf{e}}\right)\mathbf{\bar{P}}\right]$, where $\bar{\mathbf{c}}_{\mathbf{e}}$ is the ordered codeword estimate with respect to TEP $\bar{\mathbf{e}}$. 
    
    	In OSD, a list of TEPs are re-encoded to generate codeword candidates. The maximum Hamming weight of TEPs is limited by a parameter referred to as the decoding order. For BPSK modulation, finding the best ordered codeword estimate $\bar{\mathbf{c}}_{\mathrm{best}}$ is equivalent to minimizing the WHD between $\bar{\mathbf{c}}_{\mathbf{e}}$ and $\bar{\mathbf{y}}$, which is defined as \cite{valembois2002comparison}
    	\begin{equation} \small \label{equ::Prelim::WHD_define}
    		 \mathcal{D}(\bar{\mathbf{c}}_{\mathbf{e}},\bar{\mathbf{y}}) \triangleq \sum_{\substack{0<i \leq n \\ \bar{c}_{\mathbf{e},i}\neq \bar{y}_{i}}} \bar{\alpha}_{i}.
    	\end{equation}
        
        Finally, the best codeword estimate $\mathbf{c}_{\mathrm{best}}$, corresponding to the initial received sequence $\bm{\gamma}$, is obtained by performing inverse permutations over $\bar{\mathbf{c}}_{\mathrm{best}}$, i.e.
    	$\hat{\mathbf{c}}_{\mathrm{best}} = \bar{\mathbf{c}}_{\mathrm{best}}\bm{\Pi}_{\bar{\mathbf{G}}}^{\top}\bm{\Pi}_{d}^{\top}$. 
    	
        Since the MRB bits have higher reliabilities than other bits, OSD avoids massive bit flipping to search the best codeword estimate. For a code with the minimum Hamming weight $d_{\mathrm{H}}$, OSD with the order $\lceil d_{\mathrm{H}}/4-1\rceil$ is asymptotically approaching the ML decoding \cite{Fossorier1995OSD}. In other words,  the optimal codeword estimate can be found by an OSD with a list of maximum $\sum_{i=0}^{\lceil d_{\mathrm{H}}/4-1\rceil}\binom{k}{i}$ TEPs.

    \section{Linear-Equation OSD} \label{sec::LE-OSD}
    \vspace{-0.5em}
    \subsection{Decoding using High Reliable Parity Bits} \label{sec::Pri::MRP}
    \vspace{-0.5em}
        We consider a variant of the original OSD algorithm, which decodes using high reliable parity bits. We first order $\bm{\alpha}$ in the ascending order rather than descending order. Let $\bm{\Pi}_{a}$ denote the ordering permutation, and $\mathbf{y}$ and $\bm{\alpha}$ are ordered into $\mathbf{y}\bm{\Pi}_{a}$ and $\bm{\alpha}\bm{\Pi}_{a}$, respectively. Then, the columns of $\mathbf{G}$ are permuted accordingly to obtain $\widetilde{\mathbf{G}} = \mathbf{G}\bm{\Pi}_{a}$, and GE is performed over $\widetilde{\mathbf{G}}$ to obtain its systematic form $\widetilde{\mathbf{G}}' = \mathbf{E}_{\widetilde{\mathbf{G}}}\widetilde{\mathbf{G}}\bm{\Pi}_{\widetilde{\mathbf{G}}} = [\mathbf{I}_k \ \ \widetilde{\mathbf{P}}]$, where $\widetilde{\mathbf{P}}$ is the $k\times (n-k)$ parity matrix. 
        After GE, the received signal sequence, the hard-decision vector, and the reliability vector are further permuted into $\widetilde{\bm{\gamma}} =\bm{\gamma} \bm{\Pi}_{a}\bm{\Pi}_{\widetilde{\mathbf{G}}}$, $\widetilde{\mathbf{y}} = \mathbf{y}\bm{\Pi}_{a}\bm{\Pi}_{\widetilde{\mathbf{G}}}$, and $\widetilde{\bm\alpha} = \bm{\alpha}\bm{\Pi}_{a}\bm{\Pi}_{\widetilde{\mathbf{G}}}$, respectively. If the disruption introduced by permutation $\bm{\Pi}_{\widetilde{\mathbf{G}}}$ is omitted, it can be seen that the last $n-k$ bits of $\widetilde{\mathbf{y}}$ (i.e., $\widetilde{\mathbf{y}}_{\mathrm{P}}$), referred to as the most reliable parities (MRP), have higher reliabilities than the first $k$ bits.
        
        Next, we consider using bits in MRP to generate a codeword estimate. First, let us recover a codeword estimate $\widetilde{\mathbf{c}}_{0}$ directly from $\widetilde{\mathbf{y}}_{\mathrm{P}}$, i.e., assuming $\widetilde{\mathbf{c}}_{0,\mathrm{P}} = \widetilde{\mathbf{y}}_{\mathrm{P}}$. By considering $\widetilde{\mathbf{c}}_{0,\mathrm{P}} = \widetilde{\mathbf{c}}_{0,\mathrm{B}}\widetilde{\mathbf{P}}$, recovering $\widetilde{\mathbf{c}}_{0}$ is equivalent to solving a SLE described by
        \begin{equation} \small \label{equ::Dec::MRP::LinearEq}
            \widetilde{\mathbf{P}}^{\top}\mathbf{x}^{\top} = \widetilde{\mathbf{y}}_{\mathrm{P}}^{\top},
        \end{equation} 
        where $\mathbf{x} = [x]_1^k$ is a $k$-dimension unknown vector. If the SLE (\ref{equ::Dec::MRP::LinearEq}) is feasible and the $j$-th solution is given by $\mathbf{x}^{(j)}$, a codeword estimate $\widetilde{\mathbf{c}}_{0}^{(j)}$ can be recovered as $\widetilde{\mathbf{c}}_{0}^{(j)} = [\mathbf{x}^{(j)} \ \ \widetilde{\mathbf{y}}_{\mathrm{P}}]$. We note that (\ref{equ::Dec::MRP::LinearEq}) is possible to have no solutions or to have multiple solutions resulting in multiple codeword estimates.
        
        Similar to the re-encoding with TEPs in OSD, TEPs can also be added to $\widetilde{\mathbf{y}}_{\mathrm{P}}$ to generate more codeword estimates. Specifically, let $\mathbf{e}$ denote a length-$(n\!-\!k)$ binary TEP vector and let $\widetilde{\mathbf{c}}_{\mathbf{e}}$ denote the codeword estimate with respect to $\mathbf{e}$. Then, recovering $\widetilde{\mathbf{c}}_{\mathbf{e}}$ is equivalent to solving the following SLE 
        \begin{equation} \small \label{equ::Dec::MRP::LinearEq::TEP}
            \widetilde{\mathbf{P}}^{\top} \mathbf{x}^{\top} = \left(\widetilde{\mathbf{y}}_{\mathrm{P}}\oplus\mathbf{e}\right)^{\top}.
        \end{equation} 
         If (\ref{equ::Dec::MRP::LinearEq::TEP}) has solutions and the $j$-th solutions is given by $\mathbf{x}^{(j)}$, the corresponding codeword estimate is be recovered as $\widetilde{\mathbf{c}}_{\mathbf{e}}^{(j)} = [\mathbf{x}^{(j)} \ \ \widetilde{\mathbf{y}}_{\mathrm{P}}\oplus \mathbf{e}]$. We refer to those TEPs $\mathbf{e}$ that will make (\ref{equ::Dec::MRP::LinearEq::TEP}) feasible, as \textit{valid TEPs}. We can then design a decoder that generates codeword estimates according to (\ref{equ::Dec::MRP::LinearEq::TEP}) with only valid TEPs, by which the number of required codeword candidates can be reduced. This approach is referred to as LE-OSD.
    
    \vspace{-0.5em}
    \subsection{Set of Valid TEPs}  \label{sec::LE-OSD::validTEP}
    \vspace{-0.5em}
    
    As described in Section \ref{sec::Pri::MRP}, if the decoder attempts to generate a codeword estimate according to (\ref{equ::Dec::MRP::LinearEq::TEP}), the valid TEPs must be identified. Let $\widetilde{\mathbf{A}}_{\mathbf{e}}$ denote the augmented matrix associated with $\widetilde{\mathbf{P}}^{\top}$ and $\left(\widetilde{\mathbf{y}}_{\mathrm{P}}\!\oplus\!\mathbf{e}\right)^{\top}$, i.e., $\widetilde{\mathbf{A}}_{\mathbf{e}} = [\widetilde{\mathbf{P}}^{\top} \ \ \left(\widetilde{\mathbf{y}}_{\mathrm{P}}\!\oplus\!\mathbf{e}\right)^{\top}]$. Thus, $\widetilde{\mathbf{A}}_{\mathbf{e}}$ has the dimension $(n\!-\!k)\times (k\!+\!1)$. It is known that the rank of the augmented matrix characterizes the number of solutions of a SLE. Specifically, let $r_{\mathbf{P}}$ and $r_{\mathbf{A}}$ denote the ranks of $\widetilde{\mathbf{P}}^{\top}$ and $\widetilde{\mathbf{A}}_{\mathbf{e}}$, respectively.  Then, it is known that 1) when $r_{\mathbf{P}} < r_{\mathbf{A}} $, Eq. (\ref{equ::Dec::MRP::LinearEq::TEP}) has no solutions, 2) when $r_{\mathbf{P}} = r_{\mathbf{A}}= k$, Eq. (\ref{equ::Dec::MRP::LinearEq::TEP}) has an unique solution (i.e., the estimate $\widetilde{\mathbf{c}}_{\mathbf{e}}$ can be determined uniquely), and 3) when $r_{\mathbf{P}} = r_{\mathbf{A}} < k$, Eq. (\ref{equ::Dec::MRP::LinearEq::TEP}) has $2^{k -r_{\mathbf{P}}}$ different solutions.
    
    Let $\widetilde{\mathbf{P}}^{'\top} = \mathbf{E}_{\mathbf{P}}\widetilde{\mathbf{P}}^{\top}\bm{\Pi}_{\mathbf{P}}$ be the REF of $\widetilde{\mathbf{P}}^{\top}$, and let $\bm{\epsilon}_{i}$, $1\!\leq\!i\!\leq\!n\!-\!k$, denote the $i$-th row of $\mathbf{E}_{\mathbf{P}}$. When $r_{\mathbf{P}} < n\!-\!k$, let us define a matrix $\mathbf{Q}$ which is given by the last $n-k-r_{\mathbf{P}}$ rows of $\mathbf{E}_{\mathbf{P}}$, i.e., $\mathbf{Q} \triangleq [\bm{\epsilon}_{r_{\mathbf{P}}+1}^{\top},\cdots,\bm{\epsilon}_{n\!-\!k}^{\top}]^{\top}$. Moreover, let us define $\mathbf{z}_{\mathbf{e}}$ as a length-$(n\!-\!k)$ vector given by $\mathbf{z}_{\mathbf{e}}\triangleq (\mathbf{e}\oplus \widetilde{\mathbf{y}}_{\mathrm{P}})\mathbf{E}_{\mathbf{P}}^{\top} $. Then, concerning the solution of (\ref{equ::Dec::MRP::LinearEq::TEP}), we have the following Propositions.
    \begin{proposition} \label{pro::LE-OSD::validTEP::rp=n-k}
        If $r_{\mathbf{P}} = n\!-\!k$, Eq. (\ref{equ::Dec::MRP::LinearEq::TEP}) has solutions for an arbitrary TEP $\mathbf{e}$.
    \end{proposition}
    \begin{IEEEproof}
        If $r_{\mathbf{P}} = n\!-\!k$, it is implied that $n-k<k$. Thus, it can be seen that $r_{\mathbf{A}} = n\!-\!k$ for an arbitrary TEP $\mathbf{e}$, because $r_{\mathbf{P}}  \leq r_{\mathbf{A}} \leq n-k$. Proposition \ref{pro::LE-OSD::validTEP::rp=n-k} is proved.
    \end{IEEEproof}
    
    \begin{proposition} \label{pro::LE-OSD::validTEP::rp<n-k}
        If $r_{\mathbf{P}} < n\!-\!k$, Eq. (\ref{equ::Dec::MRP::LinearEq::TEP}) has solutions when the TEP $\mathbf{e}$ satisfies $\mathbf{e}\mathbf{Q}^{\top} = \widetilde{\mathbf{y}}_{\mathrm{P}}\mathbf{Q}^{\top}$
    \end{proposition}
    \begin{IEEEproof}
        We first observe that $\widetilde{\mathbf{A}}_{\mathbf{e}}$ is of the same rank as
        \begin{equation} \small 
             \mathbf{E}_{\mathbf{P}}\left( \widetilde{\mathbf{P}}^{\top}\bm{\Pi}_{\mathbf{P}} +\left(\widetilde{\mathbf{y}}_{\mathrm{P}}\!\oplus\!\mathbf{e}\right)^{\top}\right) = [\widetilde{\mathbf{P}}^{'\top} \ \  \mathbf{z}_{\mathbf{e}}^{\top}].
        \end{equation}
        Thus, $\mathbf{z}_{\mathbf{e}}$ must satisfy $[z_{\mathbf{e}}]_{r_{\mathbf{P}}+1}^{n-k} = \mathbf{0}$ to guarantee $r_{\mathbf{P}} = r_{\mathbf{A}}$. In addition, since $[z_{\mathbf{e}}]_{r_{\mathbf{P}}+1}^{n-k}$ is given by $(\mathbf{e}\oplus\widetilde{\mathbf{y}}_{\mathrm{P}}) \mathbf{Q}^{\top}= \mathbf{e}\mathbf{Q}^{\top} \oplus \widetilde{\mathbf{y}}_{\mathrm{P}}\mathbf{Q}^{\top}$, $\mathbf{e}\mathbf{Q}^{\top} = \widetilde{\mathbf{y}}_{\mathrm{P}}\mathbf{Q}^{\top}$ ensures $r_{\mathbf{P}} = r_{\mathbf{A}}$ and Proposition \ref{pro::LE-OSD::validTEP::rp<n-k} is proved.
    \end{IEEEproof}
    
    Note that $\mathbf{E}_{\mathbf{P}}$ is a full-rank matrix because $\mathbf{E}_{\mathbf{P}}$ represents a row operation and its inverse matrix  always exists. Thus, $\mathbf{Q}$ is also a full-rank matrix and so that $r_{\mathbf{P}} + r_{\mathbf{Q}} = n-k$. Based on Proposition \ref{pro::LE-OSD::validTEP::rp<n-k}, we can characterize the set of valid TEPs when $\widetilde{\mathbf{P}}$ is given. Assuming that $r_{\mathbf{P}} < n\!-\!k$, let $\mathbf{Q}' = \mathbf{E}_{\mathbf{Q}}\mathbf{Q}\bm{\Pi}_{\mathbf{Q}}$ denote the REF of $\mathbf{Q}$, and let ${\mathbf{q}'}_{i}^{\top}$ denote the $i$-th ($1\!\leq\! i\! \leq n\!-\!k$) column of $\mathbf{Q}'$, i.e., $\mathbf{Q}' = [{\mathbf{q}'}_{1}^{\top},\cdots,{\mathbf{q}'}_{n\!-\!k}^{\top}]$. Then, let us define a $r_{\mathbf{Q}} \times r_{\mathbf{P}}$ matrix $\mathbf{Q}_{\mathrm{r}}$ as
    \begin{equation} \small  \label{Mat::LE-OSD::validTEP::Qr}
           \mathbf{Q}_{\mathrm{r}} \triangleq [{\mathbf{q}'}_{r_{\mathbf{Q}+1}}^{\top}, {\mathbf{q}'}_{r_{\mathbf{Q}+2}}^{\top},  \ldots , {\mathbf{q}'}_{n-k}^{\top}] 
    \end{equation}
    and a $(n\!-\!k)\times\!r_{\mathbf{P}}$ matrix $\mathbf{Q}_{\mathrm{t}}$ as
    \begin{equation} \small \label{Mat::LE-OSD::validTEP::TEPspace}
        \mathbf{Q}_{\mathrm{t}} \triangleq
            \begin{bmatrix}
         \mathbf{Q}_{\mathrm{r}}
        \\ 
        \mathbf{I}_{r_{\mathbf{P}}}
        \end{bmatrix}
        =
        \begin{bmatrix}
            {\mathbf{q}'}_{r_{\mathbf{Q}+1}}^{\top}& {\mathbf{q}'}_{r_{\mathbf{Q}+2}}^{\top}  & \ldots & {\mathbf{q}'}_{n-k}^{\top} \\ 
            1 & 0 & \cdots & 0 \\ 
            \vdots & \ddots  &\ddots  & \vdots\\ 
            0 & 0  & \cdots &  1
        \end{bmatrix}.
    \end{equation}
    In particular, when $r_{\mathbf{P}} = n\!-\!k$ (i.e., $r_{\mathbf{Q}} = 0$), we take $\mathbf{Q}_{\mathrm{t}} = \mathbf{I}_{n-k}$ and $\bm{\Pi}_{\mathbf{Q}} = \mathbf{I}_{n-k}$, 
    Then, we summarize the set of valid TEPs in the following proposition.
    \begin{proposition}    \label{pro::LE-OSD::validTEP::TEPset}
        Given $\mathbf{Q}_{\mathrm{t}}$ and its column space $\mathcal{R}_{\mathbf{Q}_{\mathrm{t}}}$, an arbitrary vector $\mathbf{a}_{i} \in \mathcal{R}_{\mathbf{Q}_{\mathrm{t}}}$, $1\!\leq\! i \!\leq\! 2^{r_{\mathbf{P}}}$, determines a valid TEP
        \begin{equation} \small   
            \mathbf{e}_{i} = (\mathbf{a}_{i} \oplus \mathbf{e}_0 )\bm{\Pi}_{\mathbf{Q}}^{\top},
        \end{equation}
        where $\mathbf{e}_0$ is given by
    \begin{equation} \small \label{equ::LE-OSD::validTEP::TEPbasis}
           \mathbf{e}_0 \triangleq 
           \begin{cases}
                [\widetilde{\mathbf{y}}_{\mathrm{P}}\mathbf{Q}^{\top} \mathbf{E}_{\mathbf{Q}}^{\top}  \ \ \underbrace{0,\ldots,0}_{r_{\mathbf{P}}}] \ \ &\textup{for } r_{\mathbf{P}}<n-k,\\ 
                [\underbrace{0,\ldots,0}_{r_{\mathbf{P}}}] \ \ &\textup{for } r_{\mathbf{P}}=n-k,
           \end{cases}
    \end{equation}
    referred to as the TEP basis vector.
    \end{proposition}
    \begin{IEEEproof}
        When $r_{\mathbf{P}} = n\!-\!k$, according to Proposition \ref{pro::LE-OSD::validTEP::rp=n-k}, an arbitrary TEP is a valid TEP. In other words, an arbitrary vector from the column space of $\mathbf{Q}_{\mathrm{t}} = \mathbf{I}_{n-k}$ is a valid TEP, then Proposition \ref{pro::LE-OSD::validTEP::TEPset} naturally holds. When $r_{\mathbf{P}} < n\!-\!k$, according to Proposition \ref{pro::LE-OSD::validTEP::rp<n-k}, a valid TEP $\mathbf{e}$ satisfies $\mathbf{Q}\mathbf{e}^{\top} = \mathbf{Q}\widetilde{\mathbf{y}}_{\mathrm{P}}^{\top}$. On the other hand, according to the theory of linear homogeneous systems \cite{mirsky2012introduction}, $\mathbf{e}_0$ given by (\ref{equ::LE-OSD::validTEP::TEPbasis}) is a particular solution of $\mathbf{Q}\bm{\Pi}_{\mathbf{Q}}\mathbf{e}^{\top} = \mathbf{Q}\widetilde{\mathbf{y}}_{\mathrm{P}}^{\top}$ with respect to $\mathbf{e}$. Then the general solution of $\mathbf{Q}\bm{\Pi}_{\mathbf{Q}}\mathbf{e}^{\top} = \mathbf{Q}\widetilde{\mathbf{y}}_{\mathrm{P}}^{\top}$ is given by $\mathbf{e}_{0} \oplus \mathbf{a}_i$, where $\mathbf{a}_{i} \in \mathcal{R}_{\mathbf{Q}_{\mathrm{t}}}$, $1\leq i \leq 2^{r_{\mathbf{P}}}$. Therefore, by applying the inverse permutation $\bm{\Pi}_{\mathbf{Q}}^{\top}$, $\mathbf{e}_{i} = (\mathbf{a}_{i} \oplus \mathbf{e}_0 )\bm{\Pi}_{\mathbf{Q}}^{\top}$ is a solution of $\mathbf{e}\mathbf{Q}^{\top} = \widetilde{\mathbf{y}}_{\mathrm{P}}\mathbf{Q}^{\top}$, which completes the proof.
        
    \end{IEEEproof}
    
    From Proposition \ref{pro::LE-OSD::validTEP::TEPset}, we can observe that for an arbitrary valid TEP $\mathbf{e} = [e]_1^{n-k}$, it is uniquely determined by its last $r_{\mathbf{P}}$ bits. In other words, given an arbitrary length-$r_{\mathbf{P}}$ binary vector $\mathbf{e}^{\mathrm{pri}}$, $\mathbf{e}$ is uniquely determined as
    \begin{equation} \small \label{equ::LE-OSD::validTEP::generation}
        \mathbf{e} = (\mathbf{e}^{\mathrm{pri}} \mathbf{Q}_{\mathrm{t}}^{\top} \oplus \mathbf{e}_0)\bm{\Pi}_{\mathbf{Q}}^{\top}.
    \end{equation}
    We refer to $\mathbf{e}^{\mathrm{pri}}$as the \textit{primary TEP} of the TEP $\mathbf{e}$.

      In the OSD algorithm introduced in Section \ref{sec::Pri::OSD} \cite{Fossorier1995OSD}, there are in total $2^k$ TEPs that can be applied by flipping maximum $k$ MRB bits. Then, the number of TEPs is constrained by limiting their Hamming weights to the decoding order. According to Proposition \ref{pro::LE-OSD::validTEP::TEPset}, there are maximum $2^{r_{\mathbf{P}}}$ valid TEPs in the proposed LE-OSD. We will further discuss the constrain of TEP Hamming weights in Section \ref{sec::LE-OSD::Reprocessing}.
    
    \vspace{-0.5em}
    \subsection{Recovering Codeword Estimates from Valid TEPs}   \label{sec::LE-OSD::Recover}
    \vspace{-0.5em}
    
    In this section, we will explain how to recover codeword estimates from valid TEPs. Given a valid TEP $\mathbf{e}$, its corresponding codeword estimates can be easily recovered by solving (\ref{equ::Dec::MRP::LinearEq::TEP}) through the row reduction method, i.e., performing GE. Recall that $\mathbf{z}_{\mathbf{e}}=  (\mathbf{e}\oplus \widetilde{\mathbf{y}}_{\mathrm{P}})\mathbf{E}_{\mathbf{P}}^{\top} $ defined in Section \ref{sec::LE-OSD::validTEP} is in fact the last column of $\widetilde{\mathbf{A}}_{\mathbf{e}}$ after performing GE over $\widetilde{\mathbf{P}}^{\top}$. Thus, when $r_{\mathbf{P}} = k$, a unique codeword estimate is directly recovered as 
    \begin{equation} \small \label{equ::LE-OSD::Recover::oneCode}
       \widetilde{\mathbf{c}}_{\mathbf{e}} = [\mathbf{x}_{\mathbf{e}} \bm{\Pi}_{\mathbf{P}}^{\top} \ \ \widetilde{\mathbf{y}}_{\mathrm{P}}\oplus \mathbf{e}] ,     
    \end{equation}
    where $\mathbf{x}_{\mathbf{e}}$ is given by
    \begin{equation} \small \label{equ::LE-OSD::Recover::Codebase}
        \mathbf{x}_{\mathbf{e}} = 
        \begin{cases}
        [z_{\mathbf{e},1}, z_{\mathbf{e},2}, \ldots, z_{\mathbf{e},k}], \ \ &\text{if} \ k\leq n-k, \\
        [z_{\mathbf{e},1}, z_{\mathbf{e},2}, \ldots, z_{\mathbf{e},n-k}, \underbrace{0,\ldots,0}_{2k-n} ], \ \ &\text{if} \ k > n-k. \\
        \end{cases}
    \end{equation}
    When $r_{\mathbf{P}} < k$, maximum $2^{k - r_{\mathbf{P}}}$ codeword estimates can be recovered based on a valid TEP $\mathbf{e}$. Let $\mathbf{p}_{i}'$, $1\!\leq\! i \!\leq\! k $, denote the $i$-th row of $\widetilde{\mathbf{P}}'$, i.e., $\widetilde{\mathbf{P}}'^{\top} = [{\mathbf{p}_{1}'}^{\top}, \ldots, {\mathbf{p}_{k}'}^{\top}] $. Also, let $\mathbf{p}_{i}^{\dag} $ denote the first $r_{\mathbf{P}}$ elements of $\mathbf{p}_{i}'$, i.e., $\mathbf{p}_{i}^{\dag} = [p_{i}']_1^{r_{\mathbf{P}}}$. Then, we define a $k\times (k-r_{\mathbf{P}})$ matrix $\mathbf{P}_{\mathrm{r}}$ as
    \begin{equation} \small  \label{Mat::LE-OSD::Recover::Pr}
        \mathbf{P}_{\mathrm{r}} \triangleq [\mathbf{p}_{r_{\mathbf{P}}+1}^{\dag\top}, \mathbf{p}_{r_{\mathbf{P}}+2}^{\dag\top} , \ldots , \mathbf{p}_{k}^{\dag\top}],
    \end{equation}
    and also define a $r_{\mathbf{P}}\times (k-r_{\mathbf{P}})$ matrix $\mathbf{P}_{\mathrm{t}}$ as
    \begin{equation} \small \label{Mat::LE-OSD::Recover::codeset}
        \mathbf{P}_{\mathrm{t}} \triangleq
        \begin{bmatrix}
         \mathbf{P}_{\mathrm{r}}
        \\ 
        \mathbf{I}_{k - r_{\mathbf{P}}}
        \end{bmatrix}
        =
        \begin{bmatrix}
            \mathbf{p}_{r_{\mathbf{P}}+1}^{\dag\top}& \mathbf{p}_{r_{\mathbf{P}}+2}^{\dag\top} & \ldots & \mathbf{p}_{k}^{\dag\top} \\ 
            1 & 0 & \cdots & 0 \\ 
            \vdots &\ddots  &\ddots  & \vdots\\ 
            0 & 0  & \cdots &  1
        \end{bmatrix}.
    \end{equation}
    Let $\mathcal{R}_{\mathbf{P}_{\mathrm{t}}}$ denote the column space of $\mathbf{P}_{\mathrm{t}}$ and there are $2^{k - r_{\mathbf{P}}}$ vectors in $\mathcal{R}_{\mathbf{P}_{\mathrm{t}}}$. For an arbitrary vector $\mathbf{x}_{j} \in \mathcal{R}_{\mathbf{P}_{\mathrm{t}}}$, according to the theory of linear homogeneous systems \cite{mirsky2012introduction}, $( \mathbf{x}_{\mathbf{e}}\oplus \mathbf{x}_{j})\bm{\Pi}_{\mathbf{P}}^{\top}$ is a solution of (\ref{equ::Dec::MRP::LinearEq::TEP}). Then, the $j$-th codeword estimates associated with the TEP $\mathbf{e}$ is given by
    \begin{equation} \small \label{equ::LE-OSD::Recover::setCode}
         \widetilde{\mathbf{c}}_{\mathbf{e}}^{(j)} = [( \mathbf{x}_{\mathbf{e}}\oplus \mathbf{x}_{j})\bm{\Pi}_{\mathbf{P}}^{\top} \ \ \widetilde{\mathbf{y}}_{\mathrm{P}}\oplus \mathbf{e}].   
    \end{equation}
    Introducing a length-$(k-r_{\mathbf{P}})$ binary vector, the vector $\mathbf{e}_j^{\mathrm{ext}}$, $\mathbf{x}_{j}$ can be represented as 
    \begin{equation} \small  \label{equ::LE-OSD::Recover::Codeaddon}
        \mathbf{x}_{j} = ([\widetilde{y}']_{r_{\mathbf{P}}+1}^{k}\oplus \mathbf{e}_j^{\mathrm{ext}})\mathbf{P}_{\mathrm{t}}^{\top}.
    \end{equation}
    where $[\widetilde{y}']_{r_{\mathbf{P}}+1}^{k}$ is the last $k-r_{\mathbf{P}}$ bits of $[\widetilde{y}]_{1}^{k}\bm{\Pi}_{\mathbf{P}}$, and $\mathbf{e}_j^{\mathrm{ext}}$ is referred to as the \textit{extended} TEP. 
    
    \vspace{-0.5em}
    \subsection{Re-processing}  \label{sec::LE-OSD::Reprocessing}
    \vspace{-0.5em}
    
    Drawing on similar terminologies used in OSD\cite{Fossorier1995OSD}, we refer to recovering a group of codeword estimates by solving (\ref{equ::Dec::MRP::LinearEq::TEP}) with a list of TEPs as the \textit{re-processing} of LE-OSD. Let $\mathcal{E}$ denote the TEP list and $q_{t}$ denote its size, i.e., $\mathcal{E} = \{\mathbf{e}_1,\mathbf{e}_2,\ldots,\mathbf{e}_{q_{t}}\}$. In the list $\mathcal{E}$, each TEP $\mathbf{e}_i$, $1\!\leq \! i \! \leq \! q_{t}$ is a valid TEP as described in Proposition \ref{pro::LE-OSD::validTEP::TEPset}. Depending on the rank $r_{\mathbf{P}}$, each valid TEP $\mathbf{e}_i$ will be used to recover a single codeword estimate according to (\ref{equ::LE-OSD::Recover::oneCode}) or a set of codeword estimates according to (\ref{equ::LE-OSD::Recover::setCode}).
    
    Let $q_{c}$ denote the number of codeword estimates recovered by processing $\mathcal{E}$, and let $\mathcal{C}_{\mathrm{est}}$ denote the set of these $q_{c}$ codeword estimates. For each recovered codeword estimate $\widetilde{\mathbf{c}}_{\mathbf{e}}^{(j)} \in \mathcal{C}_{\mathrm{est}}$, its WHD $\mathcal{D}(\widetilde{\mathbf{c}}_{\mathbf{e}_i}^{(j)},\widetilde{\mathbf{y}})$ to $\widetilde{\mathbf{y}}$
    is computed to evaluate its likelihood. For the sake of brevity, we denote $\mathcal{D}(\widetilde{\mathbf{c}}_{\mathbf{e}_i}^{(j)},\widetilde{\mathbf{y}})$ as $\mathcal{D}_{\mathbf{e}_i}^{(j)}$
    
    Among all estimates in $\mathcal{C}_{\mathrm{est}}$, let $\widetilde{\mathbf{c}}_{\mathrm{best}}$ denote the most likely estimate that has the minimum WHD $\mathcal{D}_{\min} = \mathcal{D}(\widetilde{\mathbf{c}}_{\mathrm{best}},\widetilde{\mathbf{y}})$. At the end of the decoding, $\hat{\mathbf{c}}_{\mathrm{best}} = \widetilde{\mathbf{c}}_{\mathrm{best}}\bm{\Pi}_{\widetilde{\mathbf{G}}}^{\top}\bm{\Pi}_{a}^{\top}$ is output as the decoding result. 
    
    \subsection{Constrains on the Hamming weights of TEPs}
    
    In the subsequent analysis, we omit the permutations introduced by $\bm{\Pi}_{\mathbf{Q}}$ and $\bm{\Pi}_{\mathbf{P}}$ for the simplicity.  In section \ref{sec::ANA}, we will show that $\widetilde{\mathbf{P}}$ tends to be full-rank when $\mathcal{C}(n,k)$ has a random generator matrix, and $\mathbf{Q}$ is a full-rank matrix as being a sub-matrix of $\mathbf{E}_{\mathbf{P}}$. Thus, one can proof that permutations introduced by $\bm{\Pi}_{\mathbf{Q}}$ and $\bm{\Pi}_{\mathbf{P}}$ are minor, similar to
    $\bm{\Pi}_{\widetilde{\mathbf{G}}}$ analyzed by \cite[Eq. (59)]{Fossorier1995OSD}.
    
    We refer to the Hamming weight of a TEP $\mathbf{e}$, i.e., $w(\mathbf{e})$, as the \textit{actual Hamming weight} of $\mathbf{e}$, and define that $w^{\mathrm{pri}}(\mathbf{e})\triangleq w(\mathbf{e}^{\mathrm{pri}})$, namely the \textit{primary Hamming weight} of $\mathbf{e}$. It is worth noting that when $r_{\mathbf{P}}=n-k$, there exists $w^{\mathrm{pri}}(\mathbf{e}) = w(\mathbf{e}) $. Furthermore, let us define $w^{\mathrm{ext}}(\mathbf{e}^{(j)}) \triangleq w(\mathbf{e}_j^{\mathrm{ext}}) + w(\mathbf{e})$ as the \textit{extended Hamming weight} of $\mathbf{e}$ with respect to an extended TEP $\mathbf{e}_j^{\mathrm{ext}}$. Particularly, when $r_{\mathbf{P}} = k$, there exist $w(\mathbf{e}_j^{\mathrm{ext}}) = 0$ and $w^{\mathrm{ext}}(\mathbf{e}^{(j)}) = w(\mathbf{e})$.
    
    The number of TEPs in the LE-OSD is limited by considering the number of errors in high-reliable parity bits. Let $\widetilde{\mathbf{d}}_{\mathbf{e}}^{(j)} = \widetilde{\mathbf{c}}_{\mathbf{e}}^{(j)}\oplus \widetilde{\mathbf{y}} = [\widetilde{d}_{\mathbf{e}}^{(j)}]_1^n$ denote the difference pattern between the estimate $\widetilde{\mathbf{c}}_{\mathbf{e}}^{(j)}$ and the hard-decision vector $\widetilde{\mathbf{y}}$. Then, by omitting the permutations introduced by $\bm{\Pi}_{\widetilde{\mathrm{P}}}$ and $\bm{\Pi}_{\mathrm{Q}}$, there exist $w^{\mathrm{pri}}(\mathbf{e}) = w([\widetilde{d}_{\mathbf{e}}^{(j)}]_{n-r_{\mathbf{P}}+1}^n)$, $w(\mathbf{e}) = w([\widetilde{d}_{\mathbf{e}}^{(j)}]_{k+1}^n)$, and $w^{\mathrm{ext}}(\mathbf{e}^{(j)}) = w([\widetilde{d}_{\mathbf{e}}^{(j)}]_{r_{\mathbf{P}}+1}^n)$. In other words, $w^{\mathrm{pri}}(\mathbf{e})$, $w(\mathbf{e})$, and  $w^{\mathrm{ext}}(\mathbf{e}^{(j)})$ represent the number of bits flipped by $\widetilde{\mathbf{c}}_{\mathbf{e}}^{(j)}$ over the $r_{\mathbf{P}}$, $n-k$, and $n-r_{\mathbf{P}}$ most reliable bits of $\widetilde{\mathbf{y}}$, respectively. If $\hat{\mathbf{c}}_{\mathbf{e}^{(j)}} = \widetilde{\mathbf{c}}_{\mathbf{e}}^{(j)}\bm{\Pi}_{\widetilde{\mathbf{G}}}^{\top}\bm{\Pi}_{a}^{\top}$ is the correct decoding result, then $\widetilde{\mathbf{d}}_{\mathbf{e}}^{(j)}$ in fact represents the ordered hard-decision errors in $\widetilde{\mathbf{y}}$. Therefore, it is reasonable to limit the Hamming weights $w(\mathbf{e})$, $w^{\mathrm{pri}}(\mathbf{e})$, and $w^{\mathrm{ext}}(\mathbf{e}^{(j)})$ at the same time to constrain the number of required TEPs and recovered estimates.
    
    We introduce three non-negative integer parameters $\rho$, $\tau$, and $\xi$, as the constrains of $w^{\mathrm{pri}}(\mathbf{e})$, $w(\mathbf{e})$, and $w^{\mathrm{ext}}(\mathbf{e}^{(j)})$, respectively, satisfying $0\leq \rho\leq r_{\mathbf{P}}$, $0\leq \tau -\rho \leq \ n-k- r_{\mathbf{P}}$ and $0\leq \xi - \tau \leq k-r_{\mathbf{P}}$. Specifically, in the reprocessing of LE-OSD, the following conditions are satisfied: $\{w^{\mathrm{pri}}(\mathbf{e}) \leq \rho\}$, $\{(\mathbf{e})\leq \tau\}$, and $\{w^{\mathrm{ext}}(\mathbf{e}^{(j)}) \leq \xi\}$. Conditions $\{w^{\mathrm{pri}}(\mathbf{e})\leq \rho\}$ and $\{w(\mathbf{e}) \leq \tau\}$ constrain the number of applied TEPs to reduce the complexity. On the other hand, the condition $\{w^{\mathrm{ext}}(\mathbf{e}^{(j)}) \leq \xi\}$ constrains the number of generated codeword estimates with respect to a specific $\mathbf{e}$, when $r_{\mathbf{P}}<k$. Let $q_{\mathbf{e}}$ denote the number of estimates generated concerning $\mathbf{e}$, then we will have
    \begin{equation} \small  \label{equ::LE-OSD::Reprocessing:qe}
        q_{\mathbf{e}} = \sum_{\ell = 0}^{\xi-w(\mathbf{e})} \binom{k-r_{\mathbf{P}}}{\ell}.
    \end{equation}
    Thus, the number of codeword estimates, i.e., $q_{c}$, recovered from $q_{t}$ TEPs can be represented as $q_{c} = \sum_{i = 1}^{q_{t}}q_{\mathbf{e}_{i}}$
    
    It is worth noting that the number of TEPs may not equal the number of codeword estimates in LE-OSD, i.e., $q_{t} \neq q_{c}$, which is different from OSD. In particular, only when $r_{\mathbf{P}} = k$ and $k = n-k$, we have $w^{\mathrm{ext}}(\mathbf{e}^{(j)}) = w^{\mathrm{pri}}(\mathbf{e}) = w(\mathbf{e})$ and $q_{t} = q_{c}$, which implies that the LE-OSD with parameters $\rho=\tau=\xi$ is equivalent to an order-$\rho$ OSD; that is, recovering codeword estimates by flipping at most $\rho$ bits of the $k$ most reliable bits.
    
    In Section \ref{sec::ANA}, we will further analyze the error rate performance and the computational complexity of LE-OSD. It will be shown that the LE-OSD can achieve a similar error-rate performance to the OSD with a reduced complexity. 
    
    \subsection{LE-OSD Algorithm} 
    
    We summarize the LE-OSD algorithm in Algorithm \ref{ago::LE-OSD}. We note that since (\ref{equ::LE-OSD::Recover::oneCode}) is equivalent to (\ref{Mat::LE-OSD::Recover::codeset}) when $r_{\mathbf{P}} = k$, the cases of $\{r_{\mathbf{P}} = k\}$ and $\{r_{\mathbf{P}} < k\}$ are both implicit in steps 16-17. 
    
    \begin{spacing}{1.2}
        \begin{algorithm}
            \small
        	\caption{LE-OSD}
        	\label{ago::LE-OSD}
        	\begin{algorithmic} [1]
        		\REQUIRE Received signal $\bm{\gamma}$, Parameters $\rho$, $\tau$ and $\xi$
        		\ENSURE ~Optimal codeword estimate $\hat{\mathbf{c}}_{\mathrm{best}}$
        		
        		$//$ \textbf{Prepossessing}
        		
        		\STATE Obtain the hard-decision vector $\mathbf{y}$ and reliabilities $\bm{\alpha}$
        		\STATE Obtain $\bm{\Pi}_{a}$ by sorting $\bm{\alpha}$ in ascending order and obtain $\widetilde{\mathbf{G}} = \mathbf{G}\bm{\Pi}_{a}$ and $\mathbf{y}\bm{\Pi}_{a}$
        		\STATE Perform GE over $\widetilde{\mathbf{G}}$ and obtain $\widetilde{\mathbf{G}}' =\mathbf{E}_{\widetilde{\mathbf{G}}}\widetilde{\mathbf{G}}\bm{\Pi}_{\widetilde{\mathbf{G}}}=[\mathbf{I}_k \ \ \widetilde{\mathbf{P}}]$, $\widetilde{\mathbf{y}} = \mathbf{y}\bm{\Pi}_{a}\bm{\Pi}_{\widetilde{\mathbf{G}}}$, and $\widetilde{\bm{\alpha}} = \bm{\alpha}\bm{\Pi}_{a}\bm{\Pi}_{\widetilde{\mathbf{G}}}$
        		\STATE Perform GE over $\widetilde{\mathbf{P}}^{\top}$ and obtain $\widetilde{\mathbf{P}}_{\mathrm{s}}^{\top} = \mathbf{E}_{\mathbf{P}} \widetilde{\mathbf{P}}^{\top} \pi_{\widetilde{\mathbf{P}}}$ and rank $r_{\mathbf{P}}$
        		\STATE    Form $\mathbf{Q}$ by the last $n-k-r_{\mathbf{P}}$ rows of $\mathbf{E}_{\mathrm{P}}$
        		\STATE    Perform GE over $\mathbf{Q}$ and obtain $\mathbf{Q}' = \mathbf{E}_{\mathbf{Q}}\mathbf{Q}\bm{\Pi}_{\mathbf{Q}}$
        		\STATE Construct matrices $\mathbf{Q}_{\mathrm{t}}$ and $\mathbf{P}_{\mathrm{t}}$ according to (\ref{Mat::LE-OSD::validTEP::TEPspace}) and (\ref{Mat::LE-OSD::Recover::codeset}), respectively
        		\STATE Calculate the TEP basis vector according to (\ref{equ::LE-OSD::validTEP::TEPbasis})
        		
        		$//$ \textbf{Re-processing}
        		
        		\FOR{$i=1:\sum\limits_{\ell = 0}^{\rho}\binom{r_{\mathbf{P}}}{\ell}$}
                		\STATE Select an unprocessed primary TEP $\mathbf{e}_i^{\mathrm{pri}}$ with $w(\mathbf{e}_i^{\mathrm{pri}}) \leq \rho$
                		\STATE Generate a valid TEP $\mathbf{e}_i$ according to (\ref{equ::LE-OSD::validTEP::generation})
                		\IF{$w(\mathbf{e}_i)>\tau$}
                		\STATE \textbf{Continue}
                		\ENDIF
                		\STATE Calculate $\mathbf{z}_{\mathbf{e}_i}=  (\mathbf{e}_i\oplus \widetilde{\mathbf{y}}_{\mathrm{P}})\mathbf{E}_{\mathbf{P}}^{\top}$ and obtain $\mathbf{x}_{\mathbf{e}_i}$ according to (\ref{equ::LE-OSD::Recover::Codebase})
		        		\FOR{$j=1:q_{\mathbf{e}_{i}}$}
                    		\STATE Obtain $\mathbf{x}_{j}$ according to (\ref{equ::LE-OSD::Recover::Codeaddon}) with an extended TEP $\mathbf{e}_{j}^{\mathrm{ext}}$ satisfies $w(\mathbf{e}_{j}^{\mathrm{ext}})\leq \xi - w(\mathbf{e})$
                    		\STATE Recover a codeword estimate $\widetilde{\mathbf{c}}_{\mathbf{e}_i}^{(j)}$ according to (\ref{equ::LE-OSD::Recover::setCode})
		        		\ENDFOR 
                		\STATE Calculate the WHD from recovered codeword estimates to $\widetilde{\mathbf{y}}$, update  $\widetilde{\mathbf{c}}_{\mathrm{best}} $ and $\mathcal{D}_{\min}$
        		\ENDFOR 
        		\RETURN $\hat{\mathbf{c}}_{\mathrm{best}} = \widetilde{\mathbf{c}}_{\mathrm{best}}\bm{\Pi}_{\widetilde{\mathbf{G}}}^{-1}\bm{\Pi}_{a}^{-1}$
        	\end{algorithmic}
        \end{algorithm} 
    \end{spacing}
    
    \vspace{-0.5em}
 \section{Error Rate and Complexity Analyses}    \label{sec::ANA}
    \vspace{-0.5em}
    
    \subsection{Ordered Statistics}
    \vspace{-0.5em}
    
        For the simplicity of analysis and without loss of generality, we assume an all-zero codeword transmission. Thus, the $i$-th symbol of $\bm{\gamma}$ is given by $\gamma_i = 1 + w_i$, $1\leq i\leq n$. Let us consider the $i$-th reliability as a random variable denoted by $A_i$, then the sequence of random variables representing reliabilities is denoted by $[A]_1^n$. Note that $[A]_1^n$ is a sequences of independent and identically distributed (i.i.d.) random variables. Accordingly, after ordering the reliability in ascending order, the random variables of ordered reliabilities $\widetilde{\bm{\alpha}} = [\widetilde\alpha]_1^n$ are denoted by $[\widetilde{A}]_1^n$. Thus, the $\mathrm{pdf}$ of $A_i$, $1 \leq i \leq {n}$, is given by \cite{yue2021revisit}
        \begin{equation} \small 
            f_{A}(\alpha)=
            \begin{cases}
                0,                           & \text{if} \ \alpha<0,\\
                \frac{e^{-\frac{(\alpha+1)^2}{N_0}}}{\sqrt{\pi N_0}} + \frac{e^{-\frac{(\alpha-1)^2}{N_0}}}{\sqrt{\pi N_0}},        & \text{if} \ \alpha\geq 0.
            \end{cases}
        \end{equation}
        Given the $Q$-function defined as $Q(x) \triangleq \frac{1}{\sqrt{2\pi }} \int _{x}^{\infty} \exp(-\frac{u^2}{2})du$, the $\mathrm{cdf}$ of $A_i$ is derived as \cite{yue2021revisit}
        \begin{equation} \small 
        	F_{A}(\alpha)=
        	\begin{cases}
        		0,                           & \text{if} \ \alpha<0,\\
        		1 -  Q(\frac{\alpha+1}{\sqrt{N_0/2}}) -  Q(\frac{\alpha-1}{\sqrt{N_0/2}}),       & \text{if} \ \alpha\geq 0.
        	\end{cases}
        \end{equation}
        By omitting the permutation introduced by $\bm{\Pi}_{\widetilde{\mathbf{G}}}$, the $\mathrm{pdf}$ of $\widetilde{A}_i$ can be derived as \cite{papoulis2002probability}
        \begin{equation} \small  \label{equ::Ana::OrderStat::pdfAu}
        	f_{\widetilde{A}_i}(\widetilde{\alpha}_i) = \frac{n!}{(i-1)!(n-i)!} F_{A}(\widetilde{\alpha}_i)^{i-1}  (1-F_{A}(\widetilde{\alpha}_i))^{n-i}  f_{A}(\widetilde{\alpha}_i) .
        \end{equation}
        
        When the random variables of ordered reliabilities are given by $[\widetilde{A}]_1^n = [\widetilde{\alpha}]_1^n$ (i.e., the channel output random variable is sampled as received sequence in each receiving), the error probability of the $i$-th ordered hard-decision bit $\widetilde{y}_i$ is derived as 
        \begin{equation} \small \label{equ::Ana::OrderStat::Pbit}
            \mathrm{Pe}(i|\widetilde{\alpha}_i) =\frac{1}{1+\exp\left(4\widetilde{\alpha}_i/N_0\right)}.
        \end{equation}
        For the sake of simplicity, we denote $\mathrm{Pe}(i|\widetilde{\alpha}_i)$ as $\mathrm{Pe}(\widetilde{i})$.
        
    \vspace{-0.5em}
    \subsection{Block Error Rate} \label{sec::Ana::BLER}
    \vspace{-0.5em}
    
    Let $\mathrm{P}_{\mathrm{e}}$ denote the BLER of the LE-OSD algorithm. Then, we can upper bound $\mathrm{P}_{\mathrm{e}}$ by 
    \begin{equation} \small \label{equ::Ana::Pe::Pe}
        \mathrm{P}_{\mathrm{e}} \leq \mathrm{P}_{\mathrm{est}} + \mathrm{P}_{\mathrm{ML}},
    \end{equation}
    where $\mathrm{P}_{\mathrm{est}}$ is the probability that the correct codeword is not in the set $\mathcal{C}_{\mathrm{est}}$ of $q_{c}$ recovered estimates, and $\mathrm{P}_{\mathrm{ML}}$ denotes the ML error rate of $\mathcal{C}(n,k)$, representing the probability that a decoding error occurs even if the correct codeword is contained by $\mathcal{C}_{\mathrm{est}}$. $\mathrm{P}_{\mathrm{ML}}$ is determined by the structure of $\mathcal{C}(n,k)$ and its minimum distance $d_{\mathrm{H}}$. Generally, $\mathrm{P}_{\mathrm{ML}}$ can be obtained by computer search and available theoretical bounds of specific codes, e.g., the tangential sphere bound (TSB)  \cite{poltyrev1994bounds}.
    
    To characterize $\mathrm{P}_{\mathrm{est}}$, let us define $\widetilde{\mathbf{e}} = [\widetilde{e}]_1^n$ as the hard-decision error pattern over $\widetilde{\mathbf{y}}$. Thus, by assuming that the ordered transmitted codeword is $\widetilde{\mathbf{c}} = \mathbf{c} \bm{\Pi}_{a} \bm{\Pi}_{\widetilde{\mathbf{G}}} $, we can obtain that $\widetilde{\mathbf{c}} = \widetilde{\mathbf{y}} \oplus \widetilde{\mathbf{e}}$. Note that $\widetilde{\mathbf{e}}$ and $\widetilde{\mathbf{c}}$ are unknown to the decoder, but defined for simplicity of analysis. Then, $\mathrm{P}_{\mathrm{est}}$ can be represented as $\mathrm{P}_{\mathrm{est}} = \mathrm{Pr}(\widetilde{\mathbf{c}} \notin \mathcal{C}_{\mathrm{est}})$. Furthermore, one can prove that if the errors over the $n-r_{\mathbf{P}}$ most reliable bits of $\widetilde{\mathbf{y}}$ is eliminated a TEP $\mathbf{e}\in\mathcal{E}$, there must exist $\widetilde{\mathbf{c}} \in \mathcal{C}_{\mathrm{est}}$, which is summarized in the following proposition.
    \begin{proposition} \label{pro::Ana::Pe::Equiv}
        If errors over the $n-r_{\mathbf{P}}$ most reliable bits of $\widetilde{\mathbf{y}}$ are eliminated by some TEP in the LE-OSD, $ \mathcal{C}_{\mathrm{est}}$ includes the correct codeword.
    \end{proposition}
    \begin{IEEEproof}
        As discussed in Section \ref{sec::LE-OSD}, there are maximum $2^{r_{\mathbf{P}}}$ valid TEPs, and for each TEP, maximum $2^{k-r_{\mathbf{P}}}$ codeword estimates can be recovered. Therefore, there are maximum $2^{r_{\mathbf{P}}}\times 2^{k-r_{\mathbf{P}}}=2^k$ codeword estimate can be recovered, which covers all codewords in $\mathcal{C}(n,k)$. Then, it can be obtained that if an estimate cannot eliminate the errors over the $n-r_{\mathbf{P}}$ most reliable bits of $\widetilde{\mathbf{y}}$, it is not the correct estimate. Moreover, because LE-OSD can retrieve the whole codebook of $\mathcal{C}(n,k)$, if a TEP can eliminate the errors over the $n-r_{\mathbf{P}}$ most reliable bits, the LE-OSD can generate an estimate list $ \mathcal{C}_{\mathrm{est}}$, where the correct codeword must be included.
    \end{IEEEproof}
    
    For a LE-OSD with given parameters $\rho$, $\tau$, and $\xi$, the probability that it can eliminate the errors over the $n-r_{\mathbf{P}}$ most reliable bits is given by $\mathrm{Pr}(w([\widetilde{e}]_{n-r_{\mathbf{P}}+1}^{n})\geq \rho,w([\widetilde{e}]_{k+1}^{n})\geq \tau,w([\widetilde{e}]_{r_{\mathbf{P}}+1}^{n})\geq \xi)$. Let us define a random variable $E_a^b$ representing the number of errors over position $a$ to $b$ of $\widetilde{\mathbf{y}}$, i.e., $[\widetilde{y}]_a^b$, $1\geq a <b\geq n$. Therefore, according to Proposition \ref{pro::Ana::Pe::Equiv}, we obtain
    \begin{equation} \small \label{equ::Ana::Pe::Pest}
    \begin{split}
           \mathrm{P}_{\mathrm{est}} =& \mathrm{Pr}\left(E_{n\!-\!r_{\mathbf{P}}\!+\!1}^{n}\geq \rho,E_{k\!+\!1}^{n}\geq \tau,E_{r_{\mathbf{P}}\!+\!1}^{n}\geq \xi\right) \leq  \max  \left\{ \mathrm{Pr}\!\left(E_{n\!-\!r_{\mathbf{P}}\!+\!1}^{n}\geq \rho\right),\mathrm{Pr}\!\left(E_{k+\!1}^{n}\geq \tau\right), \mathrm{Pr}\left(E_{r_{\mathbf{P}}+\!1}^{n}\geq \xi\right) \right\}.
    \end{split}
    \end{equation}
    
    According to \cite[Lemma 1]{yue2021revisit}, the probability mass function ($\mathrm{pmf}$) of $E_a^b$, for $1\leq a<n$ and a fixed $b = n$, can be derived as 
    \begin{equation} \small  \label{equ::Ana::Pe::Ean}
        	p_{E_a^n}(j) = 
    	            \displaystyle\int_{0}^{\infty}\binom{n - a +1}{j} p(x)^j (1 - p(x))^{n - a +1-j} f_{\widetilde{A}_{a - 1} }(x)dx ,
	\end{equation}	
     where $p(x)$ is given by
     \begin{equation} \small  \label{equ::Ana::Pe::px}
         p(x) = \frac{1 - Q(\frac{-2x-2}{\sqrt{2N_0}}) }{1 - Q(\frac{-2x-2}{\sqrt{2N_0}}) +  Q(\frac{2x-2}{\sqrt{2N_0}}) } .
     \end{equation}
     We omit the proof of (\ref{equ::Ana::Pe::Ean}) as it can be directly obtained from \cite[Lemma 1]{yue2021revisit}. By using the $\mathrm{pdf}$ given by (\ref{equ::Ana::Pe::Ean}), the probability that the LE-OSD can eliminate all errors over the $n-r_{\mathbf{P}}$ most reliable bits can be characterized, i.e.
     \begin{equation} \small \label{equ::Ana::Pe::Pest2}
     \begin{split}
           \mathrm{P}_{\mathrm{est}} \leq & \max \left\{\!1\!-\! \sum_{\ell=0}^{\rho}p_{E_{n\!-\!r_{\mathbf{P}}\!+\!1}^{n}}(\ell),1 \!-\! \sum_{\ell=0}^{\tau}p_{E_{k\!+\!1}^{n}}(\ell), 1 \!-\! \sum_{\ell=0}^{\xi}p_{E_{r_{\mathbf{P}}\!+\!1}^{n}}(\ell) \!\right\}  = 1 \!-\! \min \! \left\{\sum_{\ell=0}^{\rho}p_{E_{n\!-\!r_{\mathbf{P}}\!+\!1}^{n}}(\ell),\sum_{\ell=0}^{\tau}p_{E_{k\!+\!1}^{n}}(\ell),\sum_{\ell=0}^{\xi}p_{E_{r_{\mathbf{P}}\!+\!1}^{n}}(\ell) \!\right\} \!. 
     \end{split}
     \end{equation}
     
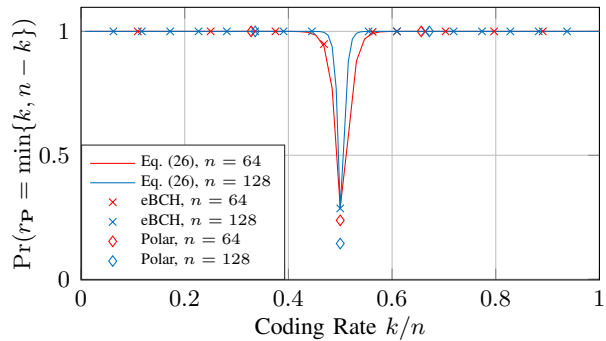
\begin{figure} 
     \centering
    \definecolor{mycolor1}{rgb}{0.00000,0.44706,0.74118}%
    \definecolor{mycolor2}{rgb}{0.00000,0.44700,0.74100}%
    \begin{tikzpicture}
    
    \begin{axis}[%
    width=0.38\textwidth,
    height=0.2\textwidth,
    at={(0.822in,0.529in)},
    scale only axis,
    xmin=0,
    xmax=1,
    xlabel style={at={(0.5,1ex)},font=\color{white!15!black},font = \footnotesize},
    xlabel={Coding Rate $k/n$},
    ymin=-0.0029341157642038,
    ymax=1.0970658842358,
    ylabel style={at={(2.5ex,0.5)},font=\color{white!15!black},font = \footnotesize},
    ylabel={$\mathrm{Pr}(r_{\mathbf{P}} = \min\{k,n-k\})$},
    axis background/.style={fill=white},
    tick label style={font=\footnotesize},
    xmajorgrids,
    ymajorgrids,
    legend style={at={(0,0)}, anchor=south west, legend cell align=left, align=left, draw=white!15!black,font = \tiny,row sep=-2.5pt}
    ]
    \addplot [color=red]
      table[row sep=crcr]{%
    0.015625	1\\
    0.03125	1\\
    0.046875	1\\
    0.0625	1\\
    0.078125	1\\
    0.09375	1\\
    0.109375	0.999999999999999\\
    0.125	0.999999999999997\\
    0.140625	0.999999999999986\\
    0.15625	0.999999999999943\\
    0.171875	0.999999999999773\\
    0.1875	0.999999999999091\\
    0.203125	0.999999999996362\\
    0.21875	0.999999999985449\\
    0.234375	0.999999999941794\\
    0.25	0.999999999767173\\
    0.265625	0.999999999068685\\
    0.28125	0.999999996274724\\
    0.296875	0.999999985098867\\
    0.3125	0.999999940395413\\
    0.328125	0.999999761581553\\
    0.34375	0.999999046326214\\
    0.359375	0.99999618530804\\
    0.375	0.999984741289457\\
    0.390625	0.999938966087321\\
    0.40625	0.999755879246159\\
    0.421875	0.999023755354362\\
    0.4375	0.996098833439939\\
    0.453125	0.98445619877386\\
    0.46875	0.938790505986965\\
    0.484375	0.770101586987258\\
    0.5	0.288788095153841\\
    0.515625	0.577576190307682\\
    0.53125	0.880116099414009\\
    0.546875	0.969074070696222\\
    0.5625	0.992207822386252\\
    0.578125	0.998048146225535\\
    0.59375	0.999511798224589\\
    0.609375	0.999877934658105\\
    0.625	0.999969482734133\\
    0.640625	0.999992370625781\\
    0.65625	0.999998092653035\\
    0.671875	0.999999523163145\\
    0.6875	0.999999880790829\\
    0.703125	0.999999970197735\\
    0.71875	0.999999992549448\\
    0.734375	0.999999998137369\\
    0.75	0.999999999534346\\
    0.765625	0.999999999883588\\
    0.78125	0.999999999970898\\
    0.796875	0.999999999992725\\
    0.8125	0.999999999998181\\
    0.828125	0.999999999999545\\
    0.84375	0.999999999999886\\
    0.859375	0.999999999999972\\
    0.875	0.999999999999993\\
    0.890625	0.999999999999998\\
    0.90625	1\\
    0.921875	1\\
    0.9375	1\\
    0.953125	1\\
    0.96875	1\\
    0.984375	1\\
    1	1\\
    };
    \addlegendentry{Eq. (\ref{equ::Ana::Pe::Pfullrank}), $n=64$}

    \addplot [color=mycolor1]
      table[row sep=crcr]{%
    0.0078125	1\\
    0.015625	1\\
    0.0234375	1\\
    0.03125	1\\
    0.0390625	1\\
    0.046875	1\\
    0.0546875	1\\
    0.0625	1\\
    0.0703125	1\\
    0.078125	1\\
    0.0859375	1\\
    0.09375	1\\
    0.1015625	1\\
    0.109375	1\\
    0.1171875	1\\
    0.125	1\\
    0.1328125	1\\
    0.140625	1\\
    0.1484375	1\\
    0.15625	1\\
    0.1640625	1\\
    0.171875	1\\
    0.1796875	1\\
    0.1875	1\\
    0.1953125	1\\
    0.203125	1\\
    0.2109375	1\\
    0.21875	1\\
    0.2265625	1\\
    0.234375	1\\
    0.2421875	1\\
    0.25	1\\
    0.2578125	1\\
    0.265625	1\\
    0.2734375	1\\
    0.28125	1\\
    0.2890625	1\\
    0.296875	1\\
    0.3046875	0.999999999999999\\
    0.3125	0.999999999999997\\
    0.3203125	0.999999999999986\\
    0.328125	0.999999999999943\\
    0.3359375	0.999999999999773\\
    0.34375	0.999999999999091\\
    0.3515625	0.999999999996362\\
    0.359375	0.999999999985448\\
    0.3671875	0.999999999941792\\
    0.375	0.999999999767169\\
    0.3828125	0.999999999068678\\
    0.390625	0.99999999627471\\
    0.3984375	0.999999985098839\\
    0.40625	0.999999940395356\\
    0.4140625	0.99999976158144\\
    0.421875	0.999999046325987\\
    0.4296875	0.999996185307585\\
    0.4375	0.999984741288548\\
    0.4453125	0.999938966085503\\
    0.453125	0.999755879242522\\
    0.4609375	0.999023755347093\\
    0.46875	0.996098833425443\\
    0.4765625	0.984456198745208\\
    0.484375	0.93879050593232\\
    0.4921875	0.770101586897607\\
    0.5	0.288788095086602\\
    0.5078125	0.577576190173205\\
    0.515625	0.88011609931155\\
    0.5234375	0.969074070639815\\
    0.53125	0.992207822357375\\
    0.5390625	0.998048146211012\\
    0.546875	0.999511798217316\\
    0.5546875	0.999877934654467\\
    0.5625	0.999969482732315\\
    0.5703125	0.999992370624871\\
    0.578125	0.99999809265258\\
    0.5859375	0.999999523162918\\
    0.59375	0.999999880790715\\
    0.6015625	0.999999970197678\\
    0.609375	0.99999999254942\\
    0.6171875	0.999999998137355\\
    0.625	0.999999999534339\\
    0.6328125	0.999999999883585\\
    0.640625	0.999999999970896\\
    0.6484375	0.999999999992724\\
    0.65625	0.999999999998181\\
    0.6640625	0.999999999999545\\
    0.671875	0.999999999999886\\
    0.6796875	0.999999999999972\\
    0.6875	0.999999999999993\\
    0.6953125	0.999999999999998\\
    0.703125	1\\
    0.7109375	1\\
    0.71875	1\\
    0.7265625	1\\
    0.734375	1\\
    0.7421875	1\\
    0.75	1\\
    0.7578125	1\\
    0.765625	1\\
    0.7734375	1\\
    0.78125	1\\
    0.7890625	1\\
    0.796875	1\\
    0.8046875	1\\
    0.8125	1\\
    0.8203125	1\\
    0.828125	1\\
    0.8359375	1\\
    0.84375	1\\
    0.8515625	1\\
    0.859375	1\\
    0.8671875	1\\
    0.875	1\\
    0.8828125	1\\
    0.890625	1\\
    0.8984375	1\\
    0.90625	1\\
    0.9140625	1\\
    0.921875	1\\
    0.9296875	1\\
    0.9375	1\\
    0.9453125	1\\
    0.953125	1\\
    0.9609375	1\\
    0.96875	1\\
    0.9765625	1\\
    0.984375	1\\
    0.9921875	1\\
    1	1\\
    };
    \addlegendentry{Eq. (\ref{equ::Ana::Pe::Pfullrank}), $n=128$}
    
    \addplot [color=red, only marks, mark size=2.0pt, mark=x, mark options={solid, red}]
      table[row sep=crcr]{%
    0.109375	1\\
    0.25	1\\
    0.375	1\\
    0.46875	0.94875\\
    0.5625	0.99805\\
    0.609375	0.9999\\
    0.703125	1\\
    0.796875	1\\
    0.890625	1\\
    };
    \addlegendentry{eBCH, $n=64$}

    \addplot [color=mycolor1, only marks, mark size=2.0pt, mark=x, mark options={solid, mycolor1}]
      table[row sep=crcr]{%
    0.0625	1\\
    0.1171875	1\\
    0.171875	1\\
    0.2265625	1\\
    0.28125	1\\
    0.3359375	1\\
    0.390625	1\\
    0.4453125	0.99995\\
    0.5	0.2857\\
    0.5546875	0.9999\\
    0.609375	1\\
    0.6640625	1\\
    0.71875	1\\
    0.7734375	1\\
    0.828125	1\\
    0.8828125	1\\
    0.9375	1\\
    };
    \addlegendentry{eBCH, $n=128$}
    
    \addplot [color=red, only marks,mark size=2.0pt, mark=diamond, mark options={solid, red}]
      table[row sep=crcr]{%
    0.328125	1\\
    0.5	0.23745\\
    0.65625	1\\
    };
    \addlegendentry{Polar, $n=64$}

    \addplot [color=mycolor2, only marks,mark size=2.0pt, mark=diamond, mark options={solid, mycolor2}]
      table[row sep=crcr]{%
    0.3359375	1\\
    0.5	0.1439\\
    0.671875	1\\
    };
    \addlegendentry{Polar, $n=128$}
    
    \end{axis}
    \end{tikzpicture}%
	\vspace{-0.5em}
    \caption{The probability $\mathrm{Pr}(r_{\mathbf{P}} = \min\{k,n-k\})$ of various codes with length $n=64$ and $n=128$. }
	\vspace{-0.5em}
	\label{Fig::Fullrank}
        
	\end{figure}

     Eq. (\ref{equ::Ana::Pe::Pest2}) depends on the rank $r_{\mathbf{P}}$. Let $\mathrm{Pr}(r_{\mathbf{P}} = \min\{k,n-k\})$ denote the probability that $\widetilde{\mathbf{P}}$ is full rank. Then, assuming that $\mathcal{C}(n,k)$ is a random code with a randomly constructed binary generator matrix, we can obtain by induction that
     \begin{equation} \small  \label{equ::Ana::Pe::Pfullrank}
        \mathrm{Pr}(r_{\mathbf{P}} = \min\{k,n-k\}) = 
        \begin{cases}
            \prod_{\ell = 1}^{k} \frac{2^{n-k} - 2^{\ell-1}}{2^{n-k}},  \ \  &\text{for  } k\leq n-k, \\
            \prod_{\ell = 1}^{n-k} \frac{2^{k} - 2^{\ell-1}}{2^{k}},  \ \  &\text{for  } n-k< k.
        \end{cases}
     \end{equation}
     We compare (\ref{equ::Ana::Pe::Pfullrank}) with the simulation results of various codes in Fig. \ref{Fig::Fullrank}. As shown, the probability $\mathrm{Pr}(r_{\mathbf{P}} = \min\{k,n-k\})$ for extended BCH (eBCH) codes with various rates can be precisely approximated by (\ref{equ::Ana::Pe::Pfullrank}). This is because BCH codes have the binomial-like weight spectrum and are close to random codes \cite{dorsch1974decoding}. Note that despite $\mathrm{Pr}(r_{\mathbf{P}} = \min\{k,n-k\})$ tends to be small when $k = n-k$, it has been shown that for a large $k$, the random binary square matrix has the expected rank $\mathbb{E}[r_{\mathbf{P}}] \approx k - 0.85$ \cite{kolchin1999random}, which is still close to the full rank. It is validated from our simulation that $(128,64)$ eBCH has $\mathbb{E}[r_{\mathbf{P}}] \approx 63.150$, and $(128,64)$ Polar code has $\mathbb{E}[r_{\mathbf{P}}] \approx 62.701$.
     
     For the above reasons, we can take $r_{\mathbf{P}} = \min\{k,n-k\}$ for random code with $k\neq n-k$, and then (\ref{equ::Ana::Pe::Pest2}) can be simplified as follows
     
      \begin{equation} \small \label{equ::Ana::Pest::appNotHalf}
        \mathrm{P}_{\mathrm{est}} \leq
      \begin{cases}
            1 - \min \left\{\sum_{\ell=0}^{\rho}p_{E_{n\!-\!k\!+\!1}^{n}}(\ell),\sum_{\ell=0}^{\tau}p_{E_{k\!+\!1}^{n}}(\ell) \right\}, \ \ &\text{for $k<n-k$}\\
            1 - \min \left\{\sum_{\ell=0}^{\tau}p_{E_{k\!+\!1}^{n}}(\ell), \sum_{\ell=0}^{\xi}p_{E_{n\!-\!k\!+\!1}^{n}}(\ell), \right\}, \ \ &\text{for $n-k < k$}.\\
      \end{cases}
     \end{equation}
     Eq. (\ref{equ::Ana::Pest::appNotHalf}) implies that for codes that are not half-rate, the probability $\mathrm{P}_{\mathrm{est}}$ is determined by the number of errors over the $k$ and $n-k$ most reliable bits of $\widetilde{\mathbf{y}}$, respectively. Particularly, for the half-rate codes, i.e., $k=n-k$, we can approximate $r_{\widetilde{\mathrm{P}}} \approx k$ according to the argument from \cite{kolchin1999random}, i.e., $\mathbb{E}[r_{\mathbf{P}}] \approx k - 0.85\approx k$. Then, by taking $\rho = \tau = \xi$, we approximate $\mathrm{P}_{\mathrm{est}}$ as 
     \begin{equation} \small \label{equ::Ana::Pe::appHalf}
     \begin{split}
          \mathrm{P}_{\mathrm{est}} =& \mathrm{Pr}\left(E_{n\!-\!r_{\mathbf{P}}\!+\!1}^{n}\geq \rho,E_{k\!+\!1}^{n}\geq \tau,E_{r_{\mathbf{P}}\!+\!1}^{n}\geq \xi\right) \approx  1 - \sum_{\ell=0}^{\rho}p_{E_{n\!-\!k\!+\!1}^{n}}(\ell).
     \end{split}
     \end{equation}
     The left-hand side of (\ref{equ::Ana::Pe::appHalf}) is the same as the probability that the correct codeword is not in the list of codeword estimates generated by an order-$\rho$ OSD \cite{dhakal2016error}, denoted by $\mathrm{P}_{\mathrm{list}}$.
     
     Overall, substituting $\mathrm{P}_{\mathrm{est}}$ given by (\ref{equ::Ana::Pe::Pest}) into (\ref{equ::Ana::Pe::Pe}), we can obtain the upper-bound of the BLER $\mathrm{P}_{\mathrm{e}}$ of the LE-OSD algorithm. For codes that are not half-rate, $\mathrm{P}_{\mathrm{est}}$ can be replaced by its simplification (\ref{equ::Ana::Pest::appNotHalf}), while for half-rate codes, approximation (\ref{equ::Ana::Pe::appHalf}) can be adopted. 
     
     \vspace{-0.5em}
     \subsection{Asymptotic Error Rate}
     \vspace{-0.5em}
     In this section, we discuss the performance of LE-OSD in the asymptotic scenario, i.e., the noise power density $N_0\to 0$. First, let us investigated the asymptotic behavior of $p(x)$ given by (\ref{equ::Ana::Pe::px}). We can re-write $p(x)$ as 
     \begin{equation} \small  
         p(x) = \frac{Q(\frac{2x+2}{\sqrt{2N_0}}) }{ Q(\frac{2x+2}{\sqrt{2N_0}}) +  Q(\frac{2x-2}{\sqrt{2N_0}}) } ,
     \end{equation}
     which is obtained by taking $Q(x) = 1 - Q(-x)$ in (\ref{equ::Ana::Pe::px}). Then, with the L'Hospital's rule, we obtain that
     \begin{equation} \small  \label{equ::Ana::Asm::pxlim}
     \begin{split}
          \lim_{N_0 \to 0}p(x) = &\lim_{N_0 \to 0}\frac{\exp\left(-\frac{(x+1)^2}{N_0}\right)}{\exp\left(-\frac{(x+1)^2}{N_0}\right)+\exp\left(-\frac{(x-1)^2}{N_0}\right)} = \frac{1}{1 + \lim\limits_{N_0 \to 0}\exp\left(\frac{4x}{N_0}\right)} \overset{(a)}{\approx}  \lim\limits_{N_0\to 0}\exp\left(\frac{-4x}{N_0}\right),
     \end{split}
     \end{equation}
     where the approximation (a) is accurate for a small $N_0$. Eq. (\ref{equ::Ana::Asm::pxlim}) infers that $p(x)$ converges to the error probability of the $i$-th ordered bits by taking $x=\widetilde{\alpha}_i$, i.e., $p(\widetilde{\alpha}_i) \to \mathrm{Pe}(\widetilde{i}) = \frac{1}{1 + \exp\left(\frac{4\widetilde{\alpha}_i}{N_0}\right)}$ when $N_0\to 0$. 
     
     Then, when the noise power is small (i.e., $N_0 \to 0$), for any $i$, $1\leq i\leq n$, we have $\widetilde{\alpha}_i \to 1$ (for all-zero transmission) and $f_{\widetilde{A}_{i}}(x) = \delta(x-1)$, where $\delta(x)$ is the Dirac delta function. Therefore, we can upper-bound $p_{E_a^n}(j)$ in (\ref{equ::Ana::Pe::Ean}) by 
     \begin{equation} \small  \label{equ::Ana::Asm::Eanlim}
     \begin{split}
          p_{E_a^n}(j) &< \binom{n-a+1}{j} \int_{0}^{+\infty}\left(e^{-4x/N_0}\right)^j\delta(x-1) dx = \binom{n-a+1}{j} e^{-4j/N_0}.        
     \end{split}
     \end{equation}
     By noticing that $p_{E_a^n}(0) \gg p_{E_a^n}(1) \gg \ldots \gg p_{E_a^n}(n-a+1)$ when $N_0 \to 0$ and $\sum_{j=0}^{n-a+1}p_{E_a^n}(j)=1$, we further obtain that 
     \begin{equation} \small  \label{equ::Ana::Asm::SumEanlim}
     \begin{split}
         \sum_{\ell = 0}^{j}  p_{E_a^n}(\ell) &= 1 - p_{E_a^n}(j +1 ) - \mathcal{O}\left(\frac{1}{e^{(j+2)/N_0}}\right) = 1 - \binom{n-a+1}{j+1} e^{-\frac{4(j+1)}{N_0}} - \mathcal{O}\left(\frac{1}{e^{(j+2)/N_0}}\right),         
     \end{split}
     \end{equation}
     for any integer $j$, $0 \leq j \leq n-k-2$. Substituting (\ref{equ::Ana::Asm::SumEanlim}) into (\ref{equ::Ana::Pe::Pest2}), we can obtain the asymptotic upper-bound of $\mathrm{P}_{\mathrm{est}}$ for a LE-OSD algorithm with parameters $\rho$, $\tau$, and $\xi$, i.e.,
     \begin{equation} \small \label{equ::Ana::Asm::ErrorRate}
     \begin{split}
          \mathrm{P}_{\mathrm{est}} &\leq \max\left\{\binom{r_{\mathbf{P}}}{\rho+1} e^{-4(\rho+1)/N_0}, \binom{n-k}{\tau+1} e^{-4(\tau +1)/N_0}, \binom{n-r_{\mathbf{P}}}{\xi+1} e^{-4(\xi+1)/N_0} \right\} - \mathcal{O}\left(\frac{1}{e^{(\min\{\rho,\tau,\xi\}+2)/N_0}}\right)\\
          &\leq \max\left\{\binom{r_{\mathbf{P}}}{\rho+1} e^{-4(\rho+1)/N_0}, \binom{n-k}{\tau+1} e^{-4(\tau +1)/N_0}, \binom{n-r_{\mathbf{P}}}{\xi+1} e^{-4(\xi+1)/N_0} \right\} .
     \end{split}
     \end{equation}
     
     \begin{remark} \label{rem::OSDorder}
        As previously introduced by (\ref{equ::Ana::Pe::appHalf}), the BLER of an order-$\rho$ OSD algorithm can be upper-bounded by $\mathrm{P_e}\leq \mathrm{P}_{\mathrm{list}} + \mathrm{P}_{\mathrm{ML}}$ \cite{Fossorier1995OSD}, where $\mathrm{P}_{\mathrm{list}} = 1 - \sum_{\ell=0}^{\rho}p_{E_{n\!-\!k\!+\!1}^{n}}(\ell)$. According to (\ref{equ::Ana::Asm::SumEanlim}), $\mathrm{P}_{\mathrm{list}}$ can be upper-bounded by $\mathrm{P}_{\mathrm{list}} < \binom{k}{\rho +1} e^{-4(\rho+1)/N_0}$ asymptotically. In previous work \cite[Eq. (72)]{Fossorier1995OSD}, the authors obtained that $\mathrm{P}_{\mathrm{list}} \approx e^{-4(\rho+1)/N_0}$ for $N_0\to 0$ using a different approach. Then, by considering that $\mathrm{P}_{\mathrm{ML}} = \exp\left(-\frac{d_{\mathrm{H}}}{N_0}\right)$ for codes with the minimum Hamming distance $d_{\mathrm{H}}$ when $N_0 \to 0$ \cite{forney1965concatenated}, it has been obtained that the OSD with order $\rho \geq \lceil \frac{d_{\mathrm{H}}}{4} -1 \rceil$ is near ML, i.e., guaranteeing $\mathrm{P}_{\mathrm{list}} \leq  \mathrm{P}_{\mathrm{ML}}$ \cite{Fossorier1995OSD}. However, by 
        omitting $\binom{k}{\rho +1}$ as a constant factor, the effect of $k$ over the error rate is overlooked in \cite{Fossorier1995OSD}.
        
        More insights for \cite[Eq. (72)]{Fossorier1995OSD} could be provided by (\ref{equ::Ana::Asm::SumEanlim}). By taking $N_0\to 0$ and noticing $\binom{k}{\rho +1} \leq k^{\rho +1}$, we can obtain from (\ref{equ::Ana::Pe::appHalf}) that
         \begin{equation} \small \label{equ::Ana::Asm::OSDBLER}
         \begin{split}
               \mathrm{P}_{\mathrm{list}} &< \binom{k}{\rho +1} e^{-4(\rho+1)/N_0} \ \leq \ e^{(\rho+1)\log k} \  e^{-4(\rho+1)/N_0} \ = \ e^{(\rho+1)(-\frac{4}{N_0}+\log k)}.       
         \end{split}
         \end{equation}
         To approach the ML error rate performance by an order-$\rho$ OSD algorithm, i.e., $e^{(\rho+1)(-\frac{4}{N_0}+\log k)} \leq e^{\left(-\frac{d_{\mathrm{H}}}{N_0}\right)}$ for a small enough $N_0$, we can conclude that
         \begin{equation} \small  \label{equ::Ana::Asm::OSDOrder}
             \rho >  \frac{d_{\mathrm{H}}}{4 - N_0 \log k} -1 .
         \end{equation}
         It can be seen that if $N_0 = 0$, (\ref{equ::Ana::Asm::OSDOrder}) gives the same results as in \cite{Fossorier1995OSD}, i.e., $\rho \geq \left\lceil \frac{d_{\mathrm{H}}}{4} -1 \right\rceil $ . However, when $N_0$ is small but not negligible, (\ref{equ::Ana::Asm::OSDOrder}) implies that the order of OSD that approaches ML decoding depends on not only $d_{\mathrm{H}}$, but the information block length $k$.
     \end{remark}
     \vspace{-0.5em}
     \subsection{Parameter Selections of the LE-OSD Algorithm} \label{sec::Ana::Para}
     \vspace{-0.5em}
     In this subsection, we discuss the selection of parameters $\rho$, $\tau$, and $\xi$ in the LE-OSD algorithm. As shown in Section \ref{sec::Ana::BLER}, usually $\widetilde{\mathbf{P}}$ tends to be full rank when $k\neq n-k$ and be close to full-rank when $k = n-k$. In this regard, for the sake of simplicity, we assume that $\widetilde{\mathbf{P}}$ is full-rank in the analysis of this subsection, i.e., $r_{\mathbf{P}} = \min\{k,n-k\}$.

     \subsubsection{Low-Rate Codes $(k< n-k )$} Since $0\leq \xi - \tau \leq k-r_{\mathbf{P}}$ and $r_{\mathbf{P}} = k$, we can assume that $\xi = \tau$. Thus, we only discuss the selection of $\rho$ and $\tau$, and (\ref{equ::Ana::Asm::ErrorRate}) can be simplified as 
     \begin{equation} \small  \label{equ::Ana::PestAsm::k<n-k}
         \mathrm{P}_{\mathrm{est}} \leq \max\left\{\binom{k}{\rho+1} e^{-4(\rho+1)/N_0}, \binom{n-k}{\tau+1} e^{-4(\tau +1)/N_0}\right\}.
     \end{equation}
     To approach the near ML error rate performance, i.e., $\mathrm{P}_{\mathrm{est}} \leq \mathrm{P}_{\mathrm{ML}}$, parameters $\rho$ and $\tau$ should satisfy
     \begin{equation} \small \label{equ::Ana::ParaAsm::k<n-k}
         \rho >  \frac{d_{\mathrm{H}}}{4 - N_0 \log k} -1 \ \  \text{and} \ \ \tau  >  \frac{d_{\mathrm{H}}}{4 - N_0 \log (n-k)} -1
     \end{equation}
     From (\ref{equ::Ana::ParaAsm::k<n-k}), we can conclude that if $N_0 \to 0$, the parameter selection of $\tau = \rho =  \left\lceil \frac{d_{\mathrm{H}}}{4} -1 \right\rceil $ makes the LE-OSD algorithm approach the ML decoding asymptotically. Furthermore, comparing (\ref{equ::Ana::PestAsm::k<n-k}) with (\ref{equ::Ana::Asm::OSDBLER}), the LE-OSD with $\tau = \rho $ has the same asymptotic performance with the order-$\rho$ OSD.
     
     On the other hand, when $N_0$ is not negligible, it can be seen from (\ref{equ::Ana::PestAsm::k<n-k}) that $\binom{k}{\rho+1} e^{-4(\rho+1)/N_0} < \binom{n-k}{\tau+1} e^{-4(\tau +1)/N_0}$ for $\tau = \rho$. This implies that for a fixed $\rho$, $\tau $ should be larger than $\rho$ to avoid the performance degradation, i.e. maintaining $\binom{k}{\rho+1} e^{-4(\rho+1)/N_0} \approx \binom{n-k}{\tau+1} e^{-4(\tau +1)/N_0}$. However, for arbitrary values of $N_0$, it is not easy to find a closed-form expression of $\tau$ as a function of $\rho$. This is because in the non-asymptotic scenario, $\mathrm{P}_{\mathrm{est}}$ is given by (\ref{equ::Ana::Pest::appNotHalf}) rather than (\ref{equ::Ana::PestAsm::k<n-k}). That is, $\sum_{\ell=0}^{\rho}p_{E_{n\!-\!k\!+\!1}^{n}}(\ell)\approx \sum_{\ell=0}^{\tau}p_{E_{k\!+\!1}^{n}}(\ell)$ should be maintained.
     
      To estimate the value of $\tau$, we can consider the asymptotic scenario when $N_0 \to +\infty$, where $k$ and $n-k$ will significantly affect $\mathrm{P}_{\mathrm{est}}$ as shown in (\ref{equ::Ana::Pest::appNotHalf}) and (\ref{equ::Ana::PestAsm::k<n-k}). Recalling the $\mathrm{pmf}$ $p_{E_a^n}(j)$ given by (\ref{equ::Ana::Pe::Ean}) and $p(x)$ given by (\ref{equ::Ana::Pe::px}), we can obtain that
      \begin{equation} \small \label{equ::Ana::pEapp1}
          \lim_{N_0 \to +\infty}p(x) =  \lim_{N_0\to +\infty} \frac{1 - Q(\frac{-2x-2}{\sqrt{2N_0}}) }{1 - Q(\frac{-2x-2}{\sqrt{2N_0}}) +  Q(\frac{2x-2}{\sqrt{2N_0}}) }  = \frac{1}{2},
      \end{equation}
      and 
      \begin{equation} \small \label{equ::Ana::pEapp2}
          \lim_{N_0 \to +\infty} p_{E_a^n}(j) = \binom{n-a+1}{j} 2^{a-n-1},
      \end{equation}
     which indicates that $p_{E_a^n}(j)$ tends to be the $\mathrm{pmf}$ of a binomial distribution $\mathcal{B}(n-a+1,\frac{1}{2})$ when $N_0\to +\infty$. Then, for $k$ satisfying $k^3(\frac{1}{2})^3 \gg 1$, $\sum_{\ell=0}^{\rho}p_{E_{n\!-\!k\!+\!1}^{n}}(\ell)$ and $\sum_{\ell=0}^{\tau}p_{E_{k\!+\!1}^{n}}(\ell)$ in (\ref{equ::Ana::Pest::appNotHalf}) can be well approximated to $Q$-functions according to the Demoivre-Laplace theorem \cite[Eq. 3-27]{papoulis2002probability}, i.e., 
     \begin{equation} \small   
         \sum_{\ell=0}^{\rho}p_{E_{n\!-\!k\!+\!1}^{n}}(\ell) \approx 1-Q\left(\frac{2\rho - k}{\sqrt{k}}\right)  \ \ \textup{and} \ \  \sum_{\ell=0}^{\tau}p_{E_{k\!+\!1}^{n}}(\ell) \approx 1-Q\left(\frac{2\tau - (n-k)}{\sqrt{n-k}}\right).
     \end{equation}
     Therefore, when $N_0 \to +\infty$ and for a fixed $\rho$, to satisfy $Q\left(\frac{2\rho - k}{\sqrt{k}}\right) \geq Q\left(\frac{2\tau - (n-k)}{\sqrt{n-k}}\right)$, we have
     \begin{equation} \small  \label{equ::Ana::ParaAsm::rhotau}
         \tau \geq \left\lceil m(\rho) \right\rceil= \left\lceil \rho \sqrt{\frac{n-k}{k}} + \frac{1}{2}\left(n-k-\sqrt{(n-k)k} \right)\right\rceil.
     \end{equation}
     From (\ref{equ::Ana::ParaAsm::rhotau}), we obtain that $\tau \geq \left\lceil m(\rho) \right\rceil $ approximately indicates $\sum_{\ell=0}^{\rho}p_{E_{n\!-\!k\!+\!1}^{n}}(\ell) \leq \sum_{\ell=0}^{\tau}p_{E_{k\!+\!1}^{n}}(\ell)$ for $N_0 \to +\infty$ and a fixed $\rho$. Note that because of approximation (\ref{equ::Ana::pEapp1}) and (\ref{equ::Ana::pEapp2}), $\tau \geq \left\lceil m(\rho) \right\rceil$ is not a rigorous sufficient condition of that no performance degradation is introduced by $\tau$. We can however conjecture that for a code with $k<n-k$, a LE-OSD algorithm with parameters $\rho$ and $\tau = \left\lceil m(\rho) \right\rceil$ has the similar error rate performance to an order-$\rho$ OSD algorithm \cite{Fossorier1995OSD} when $N_0 \to +\infty$, by comparing (\ref{equ::Ana::Pe::Pest2}) and (\ref{equ::Ana::Pe::appHalf}).
     
     Therefore, LE-OSD with $\tau = \rho $ and $\tau = \left\lceil m(\rho) \right\rceil$ has the similar performance with the order-$\rho$ OSD when $N_0\to 0$ and $N_0\to +\infty$, respectively. Thus, for an arbitrary $N_0$ and a fix $\rho$, the parameter $\tau$ can be selected from $\rho\leq \tau\leq \left\lceil m(\rho) \right\rceil$ according to the system requirements. We will further show the performance of different parameters in Section \ref{sec::Simulation}.

     \subsubsection{High-Rate Codes $(k> n-k)$} Since $0\leq \tau -\rho \leq \ n-k- r_{\mathbf{P}}$ and $r_{\mathbf{P}} = n-k$, it can be taken that $\tau = \rho$. Thus, we can only discuss the selections of $\tau$ and $\xi$, and (\ref{equ::Ana::Asm::ErrorRate}) is simplified as $\mathrm{P}_{\mathrm{est}} \leq \min\left\{\binom{n-k}{\tau+1} e^{-4(\tau+1)/N_0}, \binom{k}{\xi+1} e^{-4(\xi +1)/N_0}\right\}$.
     Then, similar to (\ref{equ::Ana::ParaAsm::k<n-k}), it is concluded that the parameter $\tau = \xi = \left\lceil \frac{d_{\mathrm{H}}}{4} -1 \right\rceil $ makes the LE-OSD algorithm approach ML decoding asymptotically. Moreover, following the analysis of (\ref{equ::Ana::ParaAsm::rhotau}), $\tau >= \left\lceil m(\rho) \right\rceil$ ensures that no performance degradation is introduced by $\tau$ for a fixed $\xi$ and $N_0\to +\infty$. Therefore, in the practical implementation, the parameter $\tau$ can be selected from $\left\lceil m(\xi) \right\rceil \leq \tau\leq \xi$ according to the system requirements.

     \subsubsection{Half-Rate Codes $(k = n-k)$} Since $0\leq \tau -\rho \leq \ n-k- r_{\mathbf{P}}$ and $0\leq \xi - \tau \leq k-r_{\mathbf{P}}$, it can be taken that $\rho = \tau = \xi $ if $r_{\mathbf{P}}$ is full-rank. Then, the LE-OSD is equivalent to an order-$\rho$ OSD, and $ \rho  = \left\lceil \frac{d_{\mathrm{H}}}{4} -1 \right\rceil $makes the LE-OSD algorithm approach the near ML decoding asymptotically.
      
     \vspace{-0.5em}
     \subsection{Numbers of TEPs and Codeword Estimations}
     \vspace{-0.5em}
     Recall Algorithm \ref{ago::LE-OSD}, for each TEP in $\mathcal{E}$, the LE-OSD is potentially retrieving multiple codeword estimates. Thus, the overall decoding complexity of the LE-OSD is determined by  both the number of processed TEPs $q_{t}$ and the number of retrieved codeword estimates $q_{c}$. For a LE-OSD with parameters $\rho$, $\tau$, and $\xi$. The average number of processed valid TEPs, i.e., $\mu_{t} = \mathbb{E}[q_{t}]$, can be represented as 
     \begin{equation} \small
     \begin{split}
         \mu_{t} &= 2^{r_{\mathbf{P}}}\cdot\mathrm{Pr}(w^{\mathrm{pri}}(\mathbf{e})\leq\rho,w(\mathbf{e})\leq\tau)     \\
          &=  2^{r_{\mathbf{P}}}\cdot \sum_{\ell=0}^{\rho}  \mathrm{Pr}(w([e]_1^{r_{\mathbf{Q}}})\leq \tau - \ell | w(\mathbf{e}^{\mathrm{pri}})= \ell) \cdot\mathrm{Pr}(w(\mathbf{e}^{\mathrm{pri}})= \ell|w(\mathbf{e}^{\mathrm{pri}})\leq \rho) \cdot\mathrm{Pr}(w(\mathbf{e}^{\mathrm{pri}})\leq \rho),
     \end{split}
     \end{equation}
     where $\mathbf{e} = (\mathbf{e}^{\mathrm{pri}} \mathbf{Q}_{\mathrm{t}}^{\top} \oplus \mathbf{e}_0)$ is a valid TEP as described in Proposition \ref{pro::LE-OSD::validTEP::TEPset}. It can be directly obtained that $\mathrm{Pr}(w(\mathbf{e}^{\mathrm{pri}})= \ell) = \binom{r_{\mathbf{P}}}{\ell}2^{-r_{\mathbf{P}}}$. Furthermore, the Hamming weight $w([e]_1^{r_{\mathbf{Q}}})$ is given by $ w(\mathbf{e}^{\mathrm{pri}} \mathbf{Q}_{\mathrm{r}}^{\top} \oplus [e_0]_1^{r_{\mathbf{Q}}})$ depending on $\mathbf{Q}_{\mathrm{r}}^{\top}$. Let us assume that $\mathcal{C}(n,k)$ has a randomly constructed generator matrix, so that $\mathbf{Q}_{\mathrm{r}}^{\top}$ is a $r_{\mathbf{Q}}\times r_{\mathbf{P}}$ random matrix. This is because $\mathbf{Q}_{\mathrm{r}}^{\top}$ is  a sub-matrix of $\mathbf{E}_{\mathbf{P}}$, and $\mathbf{E}_{\mathbf{P}}$ is a random matrix if $\widetilde{\mathbf{P}}$ is constructed randomly. Then, for any $\ell$, $0\leq \ell \leq \rho$, we can obtain that
     \begin{equation} \small
         \mathrm{Pr}(w([e]_1^{r_{\mathbf{Q}}})\leq \tau - \ell | w(\mathbf{e}^{\mathrm{pri}})= \ell)  = \sum_{j = 0}^{\tau-\ell}\binom{n-k-r_{\mathbf{P}}}{j} 2^{-(n-k-r_{\mathbf{P}})}.
     \end{equation}
     Then, we can also derive that $\mathrm{Pr}(w(\mathbf{e}^{\mathrm{pri}})\leq \rho) = \sum_{i=0}^{\rho}\binom{r_{\mathbf{P}}}{i} 2^{-r_{\mathbf{P}}}$, and $\mathrm{Pr}(w(\mathbf{e}^{\mathrm{pri}})= \ell|w(\mathbf{e}^{\mathrm{pri}})\leq \rho) = \binom{r_{\mathbf{P}}}{\ell}/\sum_{i=0}^{\rho}\binom{r_{\mathbf{P}}}{i} . $
     Therefore, we obtain $\mu_{t}$ as 
     \begin{equation} \small \label{equ::Ana::Na::Na}
         \mu_{t} = \sum_{\ell = 0}^{\rho} \sum_{j=0}^{\tau-\ell} \binom{r_{\mathbf{P}}}{\ell}\binom{n-k-r_{\mathbf{P}}}{j}2^{-(n-k-r_{\mathbf{P}})}.
     \end{equation}
     By taking $\widetilde{\mathbf{P}}$ as full-rank, (\ref{equ::Ana::Na::Na}) is simplified as 
     \begin{equation} \small \label{equ::Ana::Na::Na::Fullrank}
         \mu_{t} \approx
         \begin{cases}
              \displaystyle\sum_{\ell = 0}^{\rho}\binom{n-k}{\ell},  \ \ & \text{for $n-k\leq k$},\\
              \displaystyle\sum_{\ell = 0}^{\rho} \sum_{j=0}^{\tau-\ell} \binom{k}{\ell}\binom{n-2k}{j}2^{-(n-2k)}, \ \ & \text{for $k\leq n-k$,}
         \end{cases}
     \end{equation}
     
     Let $\mu_{t}^{(j)}$ denote the average number of processed TEP with the Hamming weight $j$, $0\leq j \leq \tau$. From (\ref{equ::Ana::Na::Na}), $\mu_{t}^{(j)}$ can be derived as
     \begin{equation} \small
          \mu_{t}^{(j)} = \sum_{\ell = 0}^{\min(\rho,j)}  \binom{r_{\mathbf{P}}}{\ell}\binom{n-k-r_{\mathbf{P}}}{j-\ell}2^{-(n-k-r_{\mathbf{P}})}.
     \end{equation}
     As shown by (\ref{equ::LE-OSD::Reprocessing:qe}), with the condition $\{w^{\mathrm{ext}}(\mathbf{e}^{(j)})\leq \xi\}$ with respect to the estimate $\widetilde{\mathbf{c}}_{\mathbf{e}}^{(j)}$, the number of estimates that can be retrieved from $\mathbf{e}$ is given by $q_{\mathbf{e}} = \sum_{\ell = 0}^{\xi-w(\mathbf{e})} \binom{k-r_{\mathbf{P}}}{\ell}$. Thus, let $\mu_{c}$ denote the average number of recovered codeword estimates in the LE-OSD algorithm, i.e., $\mu_{c} = \mathbb{E}[q_{c}]$. $\mu_{c}$ is derived as 
    \begin{equation} \small  \label{equ::Ana::Na::Ca}
        \mu_{c} = \sum_{j=0}^{\tau}\sum_{u = 0}^{\xi-j} \mu_{t}^{(j)}\binom{k-r_{\mathbf{P}}}{u}.
    \end{equation}
    By taking $\widetilde{\mathbf{P}}$ as full-rank, (\ref{equ::Ana::Na::Ca}) is simplified as 
    \begin{equation} \small  \label{equ::Ana::Na::Ca::Fullrank}
        \mu_{c} \approx
        \begin{cases}
            \displaystyle \sum_{\ell=0}^{\rho}\sum_{j=0}^{\xi-\ell} \binom{n-k}{\ell}\binom{2k-n}{j}  \ \ & \text{for $n-k < k$},\\
            \displaystyle\sum_{\ell = 0}^{\rho} \sum_{j=0}^{\tau-\ell} \binom{k}{\ell}\binom{n-2k}{j}2^{-(n-2k)}, \ \ & \text{for $k < n-k$,}\\
            \displaystyle \sum_{\ell=0}^{\rho}  \binom{k}{\ell} \ \ & \text{for $k = n-k$.}\\
        \end{cases}
    \end{equation}
     \vspace{-0.5em}
     \subsection{ Computational Complexity}
     \vspace{-0.5em}
            In this section, we characterize the computational complexity of the LE-OSD algorithm with the parameters $\rho,\tau$ and $\xi$, by measuring the average complexity of each step of Algorithm \ref{ago::LE-OSD}. Let $C_{\mathrm{pre}}$ represent the complexity of ``Preprocessing'' of Algorithm \ref{ago::LE-OSD}, i.e., step 1 to 8, and let $C_{\mathrm{re}}$ represent the complexity of ``Re-processing'', i,e., step 9 to 24. Then, the overall complexity of LE-OSD, $C_{\mathrm{LEOSD}}$, is represented as 
            \begin{equation} \small  \label{equ::Ana::ComCpx::LE-OSD}
                C_{\mathrm{LEOSD}} = C_{\mathrm{pre}} + C_{\mathrm{re}}.
            \end{equation}
            
            In ``Preprocessing'', step 1 obtains $\mathbf{y}$ with performing $n$ comparisons and obtains $\bm{\alpha}$ with $n$ symbolic operations. We regard one comparison as one FLOP and one symbolic operation as one binary operation (BOP). Step 2 sorts $\bm{\alpha}$ with $n\log n$ FLOPs by ``Quick Sorting'', and obtains $\widetilde{\mathbf{G}} = \mathbf{G}\bm{\Pi}_{a}$ and $\mathbf{y}\bm{\Pi}_{a}$ with $2n$ BOPs\footnote{Although $\widetilde{\mathbf{G}} = \mathbf{G}\bm{\Pi}_{a}$ is presented as a matrix multiplication, $\widetilde{\mathbf{G}}$ is simply obtained with $n$ BOPs because $\bm{\Pi}_{a}$ is an orthogonal matrix and represents a set of column permutations.}. Then, in step 3, $\widetilde{\mathbf{G}}'$ is obtained by performing GE with $C_{\mathrm{GE}}(k,n)$ BOPs, where $C_{\mathrm{GE}}(k,n) = (\min(n,k)-1)kn - \frac{1}{2}(\min(n,k)-1)^2\min(n,k)$. Also, step 3 obtains $\widetilde{\mathbf{y}}$ and $\widetilde{\bm{\alpha}}$ with $2n$ BOPs. Step 4 and step 6 performs the GE over $\widetilde{\mathbf{P}}^{\top}$ and $\mathbf{Q}$, respectively. In step 8, the basic TEP is obtained with $2(n-k-r_{\mathbf{P}})(n-k) + (n-k-r_{\mathbf{P}})^2$ FLOPs. We summarize the computational complexity of each step of ``Preprocessing'' in Table \ref{tab::PreCpx}. Note that we omit the complexity of step 5 and step 7 because they do not involve any computations. Thus, the complexity of ``Preprocessing'' is given by
            \begin{equation} \small
            \begin{split}
                C_{\mathrm{pre}} &= \left[5n+C_{\mathrm{GE}}(k,n)+C_{\mathrm{GE}}(k,n-k)+C_{\mathrm{GE}}(n-k-r_{\mathbf{P}},n-k) +2(n-k-r_{\mathbf{P}})(n-k)+ (n-k-r_{\mathbf{P}})^2 \right]_{(\mathrm{BOP})}   \\
                & + \left[n+n\log n\right]_{(\mathrm{FLOP})}
            \end{split}
            \end{equation}
            
            \begin{table} [t]
            	\centering
    	        \footnotesize	
    	        \tabcolsep=0.11cm
            	\caption{Computational Complexity of ``Preprocessing'' of Algorithm \ref{ago::LE-OSD}}
            	\label{tab::PreCpx}
            	\begin{tabular}{|c|c|c|}
            		\hline
            		Step & BOPs & FLOPs\\
            		\hline
            		\hline
            		1 & $n$ & $n$  \\
                    \hline
            		2 & $2n$ & $n\log n$   \\
                    \hline
            		3 & $C_{\mathrm{GE}}(k,n)+2n$ & -\\
            		\hline
            		4 & $C_{\mathrm{GE}}(k,n-k)$ & - \\
            		\hline
            		6 & $C_{\mathrm{GE}}(n-k-r_{\mathbf{P}},n-k)$ & - \\
            		\hline
            		8 & $(n-k-r_{\mathbf{P}})(3n-3k-r_{\mathbf{P}})$ & - \\
            		\hline
            	\end{tabular}
            \end{table}
            
            In ``Re-processing'', step 10 first selects a primary TEP from the memory. In step 11, a valid TEP is computed with $(n-k)r_{\mathbf{P}} + 2(n-k)$ BOPs. Step 14 computes the vectors $\mathbf{z}_{\mathbf{e}_i}$ and $\mathbf{x}_{\mathbf{e}_i}$ with $(n-k)(n-k+1)$ BOPs. Step 16 and step 17 generate the codeword estimates with the complexity of $(k-r_{\mathbf{P}})k+k$ BOPs. After that, step 18 computes the WHD and updates the best codeword found so far with $n+1$ FLOPs. Among the above steps, step 10 and 11 are repeated $\sum_{\ell=0}^{\rho}\binom{r_{\mathbf{P}}}{\ell}$ times, step 16 and 17 are repeated $\mu_c$ times, and other steps are repeated $\mu_t$ times. We summarize the computational complexity of each step of ``Re-processing'' in Table \ref{tab::ReCpx}. Then, we derive the complexity of ``Re-processing'' as
            \begin{equation} \small
            \begin{split}
                C_{\mathrm{re}} &= \left[2\sum_{i = 0}^{\rho}\binom{r_{\mathbf{P}}}{i}(n-k)(r_{\mathbf{P}}+ 2)+2\mu_t(n-k)(n-k+1)+ 2\mu_c(k-r_{\mathbf{P}})(k+1)\right]_{(\mathrm{BOP})}\\
                 &+ \left[\mu_c(n+1) + \mu_t(n-k+1) \right]_{(\mathrm{FLOP})}        
            \end{split}
            \end{equation}
            
            \begin{table} [t]
            	\centering
    	        \footnotesize	
    	        \tabcolsep=0.11cm
            	\caption{Computational Complexity of ``Reprocessing'' of Algorithm \ref{ago::LE-OSD}}
            	\label{tab::ReCpx}
            	\begin{tabular}{|c|c|c|c|}
            		\hline
            		Step & BOPs & FLOPs & Repetitions\\
            		\hline
            		\hline
            		11 & $2(n-k)r_{\mathbf{P}} + 2(n-k-r_{\mathbf{P}})$ & - & $\sum_{\ell = 0}^{\rho}\binom{r_{\mathbf{P}}}{\ell}$ \\
                    \hline
            		12-13 & - & $n-k+1$ & $\mu_t$  \\
                    \hline
            		14 & $2(n-k)(n-k+1)$ & - & $\mu_t$\\
            		\hline
            		16-17 & $2(k-r_{\mathbf{P}})k+2k$ & - & $\mu_c$\\
            		\hline
            		18 & - & $n+1$ & $\mu_c$\\
            		\hline

            	\end{tabular}
            \end{table}
                        
        As the benchmark of comparison, the computational complexity of an order-$\rho$ OSD can be derived as \cite{yue2021revisit}
        \begin{equation} \small  \label{equ::Ana::ComCpx::OSD}
        \begin{split}
               C_{\mathrm{OSD}} &= \left[5n + C_{\mathrm{GE}}(k,n)+ \sum_{i=0}^{\rho}\binom{k}{i}(2kn+n)\right]_{(\mathrm{BOP})} + \left[n + n\log n +\sum_{i=0}^{\rho}\binom{k}{i}(n+1)\right]_{(\mathrm{FLOP})}.
        \end{split}
        \end{equation}
        Comparing the complexity of LE-OSD (\ref{equ::Ana::ComCpx::LE-OSD}) with the complexity of OSD (\ref{equ::Ana::ComCpx::OSD}), it can be seen that the LE-OSD has a higher overhead in ``Preprocessing''. Precisely, the LE-OSD performs three times of GEs, while the OSD only needs one GE over the generator matrix. However, it will be shown in Section \ref{sec::Simulation} that with proper parameter selection, the LE-OSD can will have a lower overall complexity than OSD with achieving the similar performance.

        \vspace{-0.5em}   
    \section{Simulations and Comparisons} \label{sec::Simulation}
       \vspace{-0.5em}
        In this section, we conduct several simulations to demonstrate the error-rate performance and the complexity of the proposed LE-OSD.
           \vspace{-0.5em}
        \subsection{Low-Rate Codes $(k< n-k )$}
   \vspace{-0.5em}
         Fig. \ref{Fig::64-30-BLER} shows the BLER performance of decoding $(64,30,14)$ eBCH code with various decoders. As discussed in Section \ref{sec::Ana::Para}, we select different settings of $\rho$ and $\tau$ and take $\xi= \tau$ for low rate codes. As shown in Fig. \ref{Fig::64-30-BLER}, LE-OSD ($\tau=3$, $\rho=3$) and LE-OSD ($\tau=3$, $\rho=2$)  exhibit the same BLER performance as order-3 and order-2 OSD, respectively. Although (\ref{equ::Ana::ParaAsm::rhotau}) shows that $\tau\geq m(\rho)$ ensures LE-OSD has the similar performance to an order-$\rho$ OSD when $N_0\to + \infty$, simulations advise that $\tau<m(\rho)$ can also introduce acceptable performance of LE-OSD in a practical SNR range \footnote{$m(\rho) = 4.2250$ for $\rho = 3$ and $(64,30)$ codes.}. On the other hand, LE-OSD ($\tau=2$, $\rho=2$) shows a performance degradation compared to the order-2 OSD. The analytical BLER given by (\ref{equ::Ana::Pest::appNotHalf}) is also depicted for LE-OSD ($\tau=2$, $\rho=2$), which tightly upper bounds the simulation results especially for high SNRs. For clear figures, we omit the bounds for other simulations.
         
         We record the average number of TEPs, the average number of generated codeword estimates, the average decoding time of various decoders, and the number of operations in Table \ref{tab::64-30}. Specifically, the decoding time is measured by performing the algorithm to decode a single received block on a 2.9 GHz CPU in MATLAB 2020a. The number of FLOPs and BOPs are estimated by (\ref{equ::Ana::ComCpx::LE-OSD}) and (\ref{equ::Ana::ComCpx::OSD}) for LE-OSD and OSD, respectively \footnote{In the simulations, there is an inconsistent relation between the number of operations and the decoding times. The reasons are 1) the number of operations can only be estimated as it highly depends on the implementation and 2) BOPs are treated as FLOPs in simulations with high-level compilers, although BOPs are generally much faster than FLOPs 
        in chip-based computing.}. As shown in Table \ref{tab::64-30}, LE-OSD($\tau=3$, $\rho=3$) requires much fewer number of TEPs and generated codewords compared to the order-3 OSD. Moreover, it also presents a shorter decoding time and a fewer number of operations.  We can also observe that LE-OSD ($\tau=3$, $\rho=2$) only generates about 150 TEPs, compared to 466 codeword estimates of the order-2 OSD. However, it still exhibits a similar decoding time to the order-2 OSD. This is because the LE-OSD performs three GEs in the ``Preprocessing'', which brings a marginal effect to the complexity reduction as decreasing the number of TEPs. This marginal effect can be observed in the LE-OSD ($\tau=2$, $\rho=2$) when comparing to the order-2 OSD. Furthermore, this is worth noting that there is a discrepancy between (\ref{equ::Ana::Na::Na::Fullrank}) and the simulated number of TEPs, because (\ref{equ::Ana::Na::Na::Fullrank}) takes the assumption that $\widetilde{\mathbf{P}}$ is full-rank.

     \begin{figure} 
     \centering

	\end{figure}
	
    \begin{table*} [t]
        \footnotesize	
    	\tabcolsep=0.11cm
    	\centering
    	\caption{Complexity comparison of decoding $(64,30,14)$ eBCH code with various decoders}
    	\label{tab::64-30}
        \begin{tabular}{|c|c|c|c|c|c|}
        \hline
        Decoder                     & \multicolumn{2}{c|}{OSD} & \multicolumn{3}{c|}{LE-OSD} \\ \hline\hline
        Parameters           & order 3 & order 2  & \begin{tabular}[c]{@{}c@{}}$\tau = 3$\\ $\rho=3$\end{tabular} & \begin{tabular}[c]{@{}c@{}}$\tau = 3$\\ $\rho=2$\end{tabular} & \begin{tabular}[c]{@{}c@{}}$\tau = 2$\\ $\rho=2$\end{tabular} \\ \hline\hline
        Number of TEPs       &  4526  &  466   & \begin{tabular}[c]{@{}c@{}}Simulation: 400\\ Eq. (\ref{equ::Ana::Na::Na::Fullrank}): 411\end{tabular}  & \begin{tabular}[c]{@{}c@{}}Simulation: 154\\ Eq. (\ref{equ::Ana::Na::Na::Fullrank}): 158\end{tabular} & \begin{tabular}[c]{@{}c@{}}Simulation: 36\\ Eq. (\ref{equ::Ana::Na::Na::Fullrank}): 37\end{tabular} \\ \hline
        Number of codewords  &  4526  &  466   & \begin{tabular}[c]{@{}c@{}}Simulation: 411\\ Eq. (\ref{equ::Ana::Na::Ca::Fullrank}): 411\end{tabular} & \begin{tabular}[c]{@{}c@{}}Simulation: 158\\  Eq. (\ref{equ::Ana::Na::Ca::Fullrank}): 158\end{tabular} & \begin{tabular}[c]{@{}c@{}}Simulation: 37\\  Eq. (\ref{equ::Ana::Na::Ca::Fullrank}): 37\end{tabular}  \\ \hline\hline
        Decoding time (ms)  &  11.78  &  2.23  & 4.98 & 2.173 & 1.577\\ \hline
        Number of FLOPs &  $2.94\times 10^5$  & $3.05\times 10^4$ & $4.14\times 10^4$ & $1.60\times 10^4$ & $4.05\times 10^3$ \\ \hline
        Number of BOPs &  $1.77\times 10^7$ & $1.85\times 10^6$ & $1.08\times 10^7$ & $1.44\times 10^6$ & $1.16\times 10^6$\\ \hline
       
        \end{tabular}
    \end{table*}

    We further simulate the decoding of $(64,16,24)$ eBCH codes and the BLER performance is illustrated in Fig. \ref{Fig::64-16-BLER}. As shown, the LE-OSD with $\tau = 12$ and $\rho = 5,4$ and $2$ show the similar BLER to the OSD with order $5,4$ and $2$, respectively. Note that although order-6 OSD decoding achieves the near-optimum according to \cite{Fossorier1995OSD}, order-4 and order-5 decoding are also close enough to the near-optimal performance, and overlapped in Fig. \ref{Fig::64-16-BLER}. The complexities of different decoders are summarized in Table \ref{tab::64-16}. It is seen that the LE-OSD ($\tau = 12$, $\rho = 5$) has less than the quarter of decoding time of the order-5 OSD and only generates 20 codeword estimates. This improvements of decoding speed can be also found on LE-OSD ($\tau = 12$, $\rho = 4$) compared to the order-4 OSD. We note that LE-OSD ($\tau = 12$, $\rho = 2$) requires a longer decoding time compared to the order-2 OSD, because of the overhead of ``Preprocessing''.
  	    \begin{figure}[t]
            \centering
            \subfigure[$(64,16,24)$ eBCH codes]
            {
                \definecolor{mycolor1}{rgb}{0.14902,0.14902,0.14902}%
                \definecolor{mycolor2}{rgb}{0.00000,0.44700,0.74100}%
                \begin{tikzpicture}
                
                \begin{axis}[%
                width=0.38\textwidth,
                height=0.2\textwidth,
                at={(0.772in,0.55in)},
                scale only axis,
                xmin=0.000,
                xmax=3.500,
                xlabel style={at={(0.5,1ex)},font=\color{white!15!black}, font = \footnotesize},
                xlabel={SNR (dB)},
                ymode=log,
                ymin=0.0001,
                ymax=1,
                yminorticks=true,
                ylabel style={at={(2ex,0.5)},font=\color{white!15!black}, font = \footnotesize},
                ylabel={BLER},
                axis background/.style={fill=white},
                tick label style={font=\footnotesize},
                xmajorgrids,
                ymajorgrids,
                yminorgrids,
                legend style={at={(0,0)}, anchor=south west, legend cell align=left, align=left, draw=white!15!black,font = \tiny,,row sep=-2.5pt}
                ]
        
                \addplot [color=mycolor1, mark=square, mark options={solid, mycolor1}]
                  table[row sep=crcr]{%
                0	0.231481481481481\\
                0.5	0.157977883096367\\
                1	0.078125\\
                1.5	0.0328083989501312\\
                2	0.0113122171945701\\
                2.5	0.00360010080282248\\
                3	0.000690288281863862\\
                3.5	0.00013\\
                };
                \addlegendentry{OSD, order-3}
                
                \addplot [color=mycolor1, mark=triangle, mark options={solid, mycolor1}]
                  table[row sep=crcr]{%
                0	0.270758122743682\\
                0.5	0.175438596491228\\
                1	0.0827586206896552\\
                1.5	0.034944670937682\\
                2	0.014636288237303\\
                2.5	0.00472091522809889\\
                3	0.001215\\
                3.5	0.00023\\
                };
                \addlegendentry{OSD, order-2}

                \addplot [color=red, mark=square, mark options={solid, red}]
                  table[row sep=crcr]{%
                0	0.268817204301075\\
                0.5	0.164473684210526\\
                1	0.0728862973760933\\
                1.5	0.0358551452133381\\
                2	0.012639029322548\\
                2.5	0.00384985563041386\\
                3	0.00079\\
                3.5	0.00011\\
                };
                \addlegendentry{LE-OSD, $\tau = 3$, $\rho = 3$}

                \addplot [color=red, mark=triangle, mark options={solid, red}]
                  table[row sep=crcr]{%
                0	0.291545189504373\\
                0.5	0.16366612111293\\
                1	0.0856898029134533\\
                1.5	0.0393313667649951\\
                2	0.0151366078861727\\
                2.5	0.00484484387490613\\
                3	0.00114721656580721\\
                3.5	0.000215\\
                };
                \addlegendentry{LE-OSD, $\tau = 3$, $\rho = 2$}

                \addplot [color=red, mark=diamond, mark options={solid, red}]
                  table[row sep=crcr]{%
                0	0.329489291598023\\
                0.5	0.189035916824197\\
                1	0.105876124933827\\
                1.5	0.051374261494991\\
                2	0.0228284442415249\\
                2.5	0.00729820464165815\\
                3	0.00214185336670533\\
                3.5	0.000465\\
                };
                \addlegendentry{LE-OSD, $\tau = 2$, $\rho = 2$}
        
                \addplot [color=mycolor1, dashed]
                  table[row sep=crcr]{%
                0	0.476673303375596\\
                0.25	0.375835556990118\\
                0.5	0.287575880602106\\
                0.75	0.212975911507371\\
                1	0.152234697848934\\
                1.25	0.104720587396754\\
                1.5	0.0691181168733928\\
                1.75	0.0436442002469769\\
                2	0.0262955779907616\\
                2.25	0.0150852786225422\\
                2.5	0.00823119294771893\\
                2.75	0.00427353355386204\\
                3	0.00211610741585098\\
                3.25	0.0010035860148743\\
                3.5	0.000458254299598241\\
                };
                \addlegendentry{Eq (\ref{equ::Ana::Pest::appNotHalf}), $\tau = 2$, $\rho = 2$};

                \end{axis}
                \end{tikzpicture}%
        	\label{Fig::64-30-BLER}
            }
            \hspace{0em}
            \subfigure[$(64,30,14)$ eBCH codes]
            {
                 \definecolor{mycolor1}{rgb}{0.14902,0.14902,0.14902}%
                \begin{tikzpicture}
                
                \begin{axis}[%
                width=0.40\textwidth,
                height=0.2\textwidth,
                at={(0.887in,0.561in)},
                scale only axis,
                xmin=-2.5,
                xmax=1,
                xlabel style={at={(0.5,1ex)},font=\color{white!15!black},font=\footnotesize},
                xlabel={SNR (dB)},
                ymode=log,
                ymin=9e-05,
                ymax=0.145243282498184,
                yminorticks=true,
                ylabel style={at={(1.5ex,0.5)},font=\color{white!15!black},font=\footnotesize},
                ylabel={BLER},
                axis background/.style={fill=white},
                tick label style={font=\footnotesize},
                xmajorgrids,
                ymajorgrids,
                yminorgrids,
                legend style={at={(0,0)}, anchor=south west, legend cell align=left, align=left, draw=white!15!black, font = \tiny,row sep=-2.5pt}
                ]
                
                \addplot [color=mycolor1, mark=diamond, mark options={solid, mycolor1}]
                  table[row sep=crcr]{%
                -3	0.184162062615101\\
                -2.5	0.121212121212121\\
                -2	0.0588235294117647\\
                -1.5	0.031640563202025\\
                -1	0.0131527028804419\\
                -0.5	0.00533390228291018\\
                0	0.00191868608376983\\
                0.5	0.000404080403918772\\
                1	9e-05\\
                };
                \addlegendentry{OSD, order 5}
                
                \addplot [color=mycolor1, mark=square, mark options={solid, mycolor1}]
                  table[row sep=crcr]{%
                -2.5	0.112951807228916\\
                -2	0.062879899392161\\
                -1.5	0.0280085893007189\\
                -1	0.0129293625824247\\
                -0.5	0.00473761508456643\\
                0	0.00170374198531374\\
                0.5	0.0005\\
                1	0.0001\\
                };
                \addlegendentry{OSD, order 4}
                
                \addplot [color=mycolor1, mark=triangle, mark options={solid, mycolor1}]
                  table[row sep=crcr]{%
                -3	0.186393289841566\\
                -2.5	0.126502213788741\\
                -2	0.0845665961945032\\
                -1.5	0.0421052631578947\\
                -1	0.0207835394367661\\
                -0.5	0.00908100254268071\\
                0	0.00336072322763859\\
                0.5	0.00124528348878622\\
                1	0.000455\\
                };
                \addlegendentry{OSD, order 2}
                
                \addplot [color=red, mark=diamond, mark options={solid, red}]
                  table[row sep=crcr]{%
                -3	0.175438596491228\\
                -2.5	0.128865979381443\\
                -2	0.0715819613457409\\
                -1.5	0.0320872773945131\\
                -1	0.0153869826127096\\
                -0.5	0.0053714347102111\\
                0	0.00164341238146888\\
                0.5	0.000506666666666667\\
                1	9.33333333333333e-05\\
                };
                \addlegendentry{LE-OSD, $\tau =12$, $\rho = 5$}
                
                \addplot [color=red, mark=square, mark options={solid, red}]
                  table[row sep=crcr]{%
                -2.5	0.11142061281337\\
                -2	0.0678656260604004\\
                -1.5	0.0325256139209628\\
                -1	0.014519056261343\\
                -0.5	0.00533632167347048\\
                0	0.00172060771864623\\
                0.5	0.000485\\
                1	9.5e-05\\
                };
                \addlegendentry{LE-OSD, $\tau =12$, $\rho = 4$}

                \addplot [color=red, mark=triangle, mark options={solid, red}]
                  table[row sep=crcr]{%
                -3	0.221729490022173\\
                -2.5	0.145243282498184\\
                -2	0.0805477245267821\\
                -1.5	0.0448430493273543\\
                -1	0.0226474917902842\\
                -0.5	0.00914745700695207\\
                0	0.00360581256986262\\
                0.5	0.00124589632896647\\
                1	0.00036\\
                };
                \addlegendentry{LE-OSD, $\tau =12$, $\rho = 2$}

                \end{axis}
                \end{tikzpicture}%
                \label{Fig::64-16-BLER}
            }
        	\vspace{-0.3em}
            \caption{BLER of decoding low-rate eBCH codes with various decoders.}
            \label{Fig::lowrate}
        	\vspace{-0.3em}
        \end{figure}
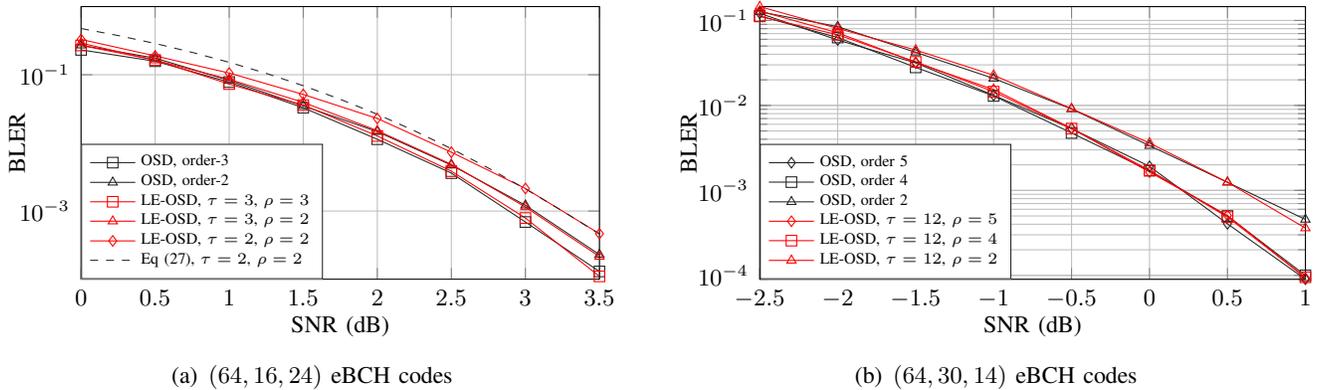

    \begin{table*} [t]
        \footnotesize	
    	\tabcolsep=0.11cm
    	\centering
    	\caption{Complexity comparison of decoding $(64,16,24)$ eBCH code with various decoders}
    	\label{tab::64-16}
        \begin{tabular}{|c|c|c|c|c|c|c|}
        \hline
        Decoder                     & \multicolumn{3}{c|}{OSD} & \multicolumn{3}{c|}{LE-OSD} \\ \hline\hline
        Parameters           & order 5 & order 4 & order 2  & \begin{tabular}[c]{@{}c@{}}$\tau = 12$\\ $\rho=5$\end{tabular} & \begin{tabular}[c]{@{}c@{}}$\tau = 12$\\ $\rho=4$\end{tabular} & \begin{tabular}[c]{@{}c@{}}$\tau = 12$\\ $\rho=2$\end{tabular} \\ \hline\hline
        Number of TEPs       &  6885  &  2517 &  137  & \begin{tabular}[c]{@{}c@{}}Simulation: 20\\ Eq. (\ref{equ::Ana::Na::Na::Fullrank}): 20\end{tabular}  & \begin{tabular}[c]{@{}c@{}}Simulation: 16\\ Eq. (\ref{equ::Ana::Na::Na::Fullrank}): 16\end{tabular} & \begin{tabular}[c]{@{}c@{}}Simulation: 4\\ Eq. (\ref{equ::Ana::Na::Na::Fullrank}): 4\end{tabular} \\ \hline
        Number of codewords  &  6885  &  2517 &  137  & \begin{tabular}[c]{@{}c@{}}Simulation: 20\\ Eq. (\ref{equ::Ana::Na::Ca::Fullrank}): 20\end{tabular} & \begin{tabular}[c]{@{}c@{}}Simulation: 16\\  Eq. (\ref{equ::Ana::Na::Ca::Fullrank}): 16\end{tabular} & \begin{tabular}[c]{@{}c@{}}Simulation: 4\\  Eq. (\ref{equ::Ana::Na::Ca::Fullrank}): 4\end{tabular}  \\ \hline\hline
        Decoding time (ms)  &  20.24  &  8.09 &  0.88 & 5.50 & 2.85 & 1.36\\ \hline
        Number of FLOPs &  $4.47\times 10^5$  & $1.63\times 10^5$ &  $9.17\times 10^3$ & $2.67\times 10^3$ & $2.15\times 10^3$ & $8.75\times 10^2$ \\ \hline
        Number of BOPs &  $1.45\times 10^7$ & $5.32\times 10^6$ &  $3.01\times 10^5$ & $1.20\times 10^7$ & $4.48\times 10^6$ & $3.15\times 10^5$\\ \hline
       
        \end{tabular}
    \end{table*}	
    
       \vspace{-0.5em}
    \subsection{High-Rate Codes $(n-k<k)$}
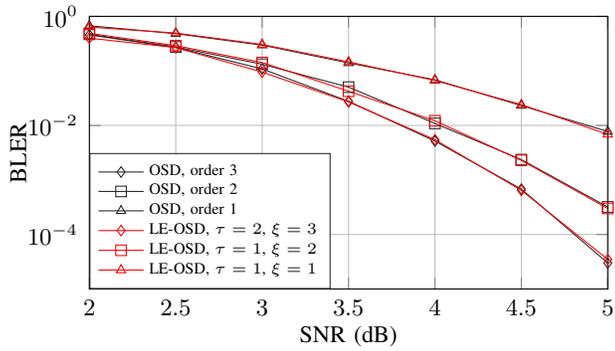
\begin{figure} 
         \centering
        \definecolor{mycolor1}{rgb}{0.00000,0.44700,0.74100}%
        \definecolor{mycolor2}{rgb}{0.14902,0.14902,0.14902}%
        \begin{tikzpicture}
        
        \begin{axis}[%
        width=0.38\textwidth,,
        height=0.2\textwidth,,
        at={(0.739in,0.506in)},
        scale only axis,
        xmin=2,
        xmax=5,
        xlabel style={at={(0.5,1ex)},font=\color{white!15!black},font=\footnotesize},
        xlabel={SNR (dB)},
        ymode=log,
        ymin=1e-05,
        ymax=1,
        yminorticks=true,
        ylabel style={at={(1.5ex,0.5)},font=\color{white!15!black},font=\footnotesize},
        ylabel={BLER},
        axis background/.style={fill=white},
        tick label style={font=\footnotesize},
        xmajorgrids,
        ymajorgrids,
        yminorgrids,
        legend style={at={(0,0)}, anchor=south west, legend cell align=left, align=left, draw=white!15!black, font = \tiny, row sep=-2.5pt}
        ]
        \addplot [color=mycolor2, mark=diamond, mark options={solid, mycolor2}]
          table[row sep=crcr]{%
        2	0.45766590389016\\
        2.5	0.269179004037685\\
        3	0.108695652173913\\
        3.5	0.0277046682365979\\
        4	0.00515929317683477\\
        4.5	0.00068\\
        5	3e-05\\
        };
        \addlegendentry{OSD, order 3}

        \addplot [color=mycolor2, mark=square, mark options={solid, mycolor2}]
          table[row sep=crcr]{%
        2	0.46875\\
        2.5	0.275229357798165\\
        3	0.13262599469496\\
        3.5	0.0504201680672269\\
        4	0.0108861310690181\\
        4.5	0.002361208172935\\
        5	0.00032\\
        };
        \addlegendentry{OSD, order 2}
        
        \addplot [color=mycolor2, mark=triangle, mark options={solid, mycolor2}]
          table[row sep=crcr]{%
        2	0.674157303370786\\
        2.5	0.481540930979133\\
        3	0.29673590504451\\
        3.5	0.139925373134328\\
        4	0.0686184812442818\\
        4.5	0.0231107002542177\\
        5	0.00774893452150329\\
        };
        \addlegendentry{OSD, order 1} 
        
        \addplot [color=red, mark=diamond, mark options={solid, red}]
          table[row sep=crcr]{%
        2	0.398936170212766\\
        2.5	0.267379679144385\\
        3	0.0951776649746193\\
        3.5	0.0269978401727862\\
        4	0.00547026001969294\\
        4.5	0.000645\\
        5	3.5e-05\\
        };
        \addlegendentry{LE-OSD, $\tau = 2$, $\xi = 3$}
        
        \addplot [color=red, mark=square, mark options={solid, red}]
          table[row sep=crcr]{%
        2	0.488599348534202\\
        2.5	0.290697674418605\\
        3	0.142585551330798\\
        3.5	0.0424328147100424\\
        4	0.0121654501216545\\
        4.5	0.00228609747919651\\
        5	0.0003\\
        };
        \addlegendentry{LE-OSD, $\tau = 1$, $\xi = 2$}
        
        \addplot [color=red, mark=triangle, mark options={solid, red}]
          table[row sep=crcr]{%
        2	0.635593220338983\\
        2.5	0.495049504950495\\
        3	0.306122448979592\\
        3.5	0.146627565982405\\
        4	0.0667408231368187\\
        4.5	0.0240847784200385\\
        5	0.00686341798215511\\
        };
        \addlegendentry{LE-OSD, $\tau = 1$, $\xi = 1$}
        
        \end{axis}
        \end{tikzpicture}%
    	\vspace{-0.3em}
        \caption{BLER of decoding $(128,85,14)$ eBCH codes with various decoders.}
    	\vspace{-0.3em}
    	\label{Fig::128-85-BLER}
	\end{figure}

       \vspace{-0.5em}
    We illustrate the performance of LE-OSD for high rate codes by taking $(128,85,14)$ eBCH code as an example. We select different values of $\tau$ and $\xi$ and set $\rho = \tau$ in the LE-OSD, and their performance is depicted in Fig. \ref{Fig::128-85-BLER}. As shown, the LE-OSD ($\tau = 2$, $\xi = 3$) shows the same BLER performance as the order-3 OSD. When $\tau = 1$, the LE-OSD with $\xi = 2$ and $\xi = 1$ also exhibit the very similar performance to order-2 and order-1 OSD, respectively. We further summarize the numbers of operations and required decoding time of the compared decoders in Table \ref{tab::128-85}. We highlight that the decoding time of LE-OSD ($\tau = 2$, $\xi = 3$) is four times shorter than the order-3 OSD. From all simulations conducted above, it can be concluded that the LE-OSD is particularly efficient for codes needing a high decoding order.

    \begin{table*} [t]
        \footnotesize	
    	\tabcolsep=0.11cm
    	\centering
    	\caption{Complexity comparison of decoding $(128,85,14)$ eBCH code with various decoders}
    	\label{tab::128-85}
        \begin{tabular}{|c|c|c|c|c|c|c|}
        \hline
        Decoder                     & \multicolumn{3}{c|}{OSD} & \multicolumn{3}{c|}{LE-OSD} \\ \hline\hline
        Parameters           & order 3 & order 2 & order 1  & \begin{tabular}[c]{@{}c@{}}$\tau = 2$\\ $\xi=3$\end{tabular} & \begin{tabular}[c]{@{}c@{}}$\tau = 1$\\ $\xi=2$\end{tabular} & \begin{tabular}[c]{@{}c@{}}$\tau = 1$\\ $\xi=1$\end{tabular} \\ \hline\hline
        Number of TEPs       &  102426  &  3656 &  86  & \begin{tabular}[c]{@{}c@{}}Simulation: 947\\ Eq. (\ref{equ::Ana::Na::Na::Fullrank}): 947\end{tabular}  & \begin{tabular}[c]{@{}c@{}}Simulation: 44\\ Eq. (\ref{equ::Ana::Na::Na::Fullrank}): 44\end{tabular} & \begin{tabular}[c]{@{}c@{}}Simulation: 44\\ Eq. (\ref{equ::Ana::Na::Na::Fullrank}): 44\end{tabular} \\ \hline
        Number of codewords  &  102426  &  3656 &  86  & \begin{tabular}[c]{@{}c@{}}Simulation: 90085\\ Eq. (\ref{equ::Ana::Na::Ca::Fullrank}): 90085\end{tabular} & \begin{tabular}[c]{@{}c@{}}Simulation: 2753\\  Eq. (\ref{equ::Ana::Na::Ca::Fullrank}): 2753\end{tabular} & \begin{tabular}[c]{@{}c@{}}Simulation: 86\\  Eq. (\ref{equ::Ana::Na::Ca::Fullrank}): 86\end{tabular}  \\ \hline\hline
        Decoding time (ms)  &  451.10  &  21.52 &  5.85 & 107.09 & 11.84 & 8.39\\ \hline
        Number of FLOPs &  $1.32\times 10^7$  & $4.72\times 10^5$ &  $1.17\times 10^4$ & $1.16\times 10^7$ & $3.57\times 10^5$ & $1.37\times 10^4$ \\ \hline
        Number of BOPs &  $2.24\times 10^9$ & $8.06\times 10^7$ &  $2.47\times 10^6$ & $6.62\times 10^8$ & $2.11\times 10^7$ & $1.84\times 10^6$\\ \hline
       
        \end{tabular}
    \end{table*}
    
       \vspace{-0.5em}
    \section{Improved LE-OSD with Decoding Conditions}  \label{sec::Improve}
       \vspace{-0.5em}
    The original OSD has been improved by combining with various decoding conditions, including decoding SC \cite{yue2021probability,yue2021revisit,jin2006probabilisticConditions} and the TEP DC \cite{yue2021revisit,Wu2007OSDMRB}. These conditions reduce the overall decoding complexity by either terminating decoding early or discarding unpromising TEPs. In this section, we demonstrate the complexity of LE-OSD improved by decoding conditions, and compare it with the latest OSD approach, PB-OSD \cite{yue2021probability}.
    
       \vspace{-0.5em}
    \subsection{Decoding conditions for LE-OSD} 
   \vspace{-0.5em}
        
        \subsubsection{TEP discarding condition}
        We modify and devise the DC proposed in \cite{yue2021probability} for the LE-OSD. In \cite{yue2021probability}, a promising probability is computed for each TEP in OSD before re-encoding, then the TEP is discarded directly if the promising probability is less than a threshold. The promising probability is defined as the probability that a TEP can result in a codeword estimate having a higher likelihood than the existing ones.
        
        Next, we introduce the promising probability for LE-OSD. Recall (\ref{equ::LE-OSD::Recover::setCode}), a codeword estimates $\widetilde{\mathbf{c}}_{\mathbf{e}}^{(j)}$ in LE-OSD is uniquely determined by TEP $\mathbf{e}$ and its corresponding extended TEP $\mathbf{e}_j^{\mathrm{ext}}$. Moreover, a TEP $\mathbf{e}$ is uniquely determined by a primary TEP $\mathbf{e}^{\mathrm{pri}}$ as shown in (\ref{equ::LE-OSD::validTEP::generation}). Therefore, leveraging  the idea from \cite{yue2021probability}, we can compute a promising probability for a combination of a primary TEP $\mathbf{e}^{\mathrm{pri}}$ and an extended $\mathbf{e}_j^{\mathrm{ext}}$ based on the minimum WHD $\mathcal{D}_{\min}$ recorded so far. Let us first define $p$ as the arithmetic mean of the bit-wise error probability of $[\widetilde{y}_1,\ldots,\widetilde{y}_{r_{\mathbf{P}}},\widetilde{y}_{k+1},\ldots,\widetilde{y}_{n-r_{\mathbf{P}}}]$, i.e.,
        \begin{equation} \small  \label{equ::Condition::avePe}
            p \triangleq \frac{1}{n-k}\left(\sum_{i=1}^{r_{\mathbf{P}}}\mathrm{Pe}(\widetilde{i}) + \sum_{i=k+1}^{n-r_{\mathbf{P}}}\mathrm{Pe}(\widetilde{i}) \right) ,    
        \end{equation}
        where $\mathrm{Pe}(\widetilde{i})$ is given by (\ref{equ::Ana::OrderStat::Pbit}).
        Next, let $\mathrm{P_d}(\mathbf{e}^{\mathrm{pri}},\mathbf{e}_j^{\mathrm{ext}})$ denote the promising probability of $\mathbf{e}^{\mathrm{pri}}$ and $\mathbf{e}_j^{\mathrm{ext}}$, and then it is computed as
        \begin{equation} \small \label{equ::Condition::Ppro}
        \begin{split}
            \mathrm{P_d}(\mathbf{e}^{\mathrm{pri}},\mathbf{e}_j^{\mathrm{ext}}) &=  \lambda\sum_{i=0}^{\beta} \! \binom{n-k}{j} p^i (1-p)^{n\!- \!k \!- \! i}  
             +  \left(1 - \lambda\right)\sum_{i=0}^{\beta} \! \binom{n-k}{i}  2^{k\!- \!n},   
        \end{split}
        \end{equation}
        where $\lambda = \mathrm{P}(\mathbf{e}^{\mathrm{pri}})\cdot\mathrm{P}(\mathbf{e}_j^{\mathrm{ext}})$ and $\beta$ is a function of $\mathcal{D}_{\min}$. Specifically, $\mathrm{P}(\mathbf{e}^{\mathrm{pri}})$ and $\mathrm{P}(\mathbf{e}_j^{\mathrm{ext}})$ are respectively given by
        \begin{equation} \small 
            \mathrm{P}(\mathbf{e}^{\mathrm{pri}}) = \prod_{\substack{ 1 \leq i <  r_{\mathbf{P}} \\ e_i^{\mathrm{pri}} \neq 0}} \mathrm{Pe}(\widetilde{i})\prod_{\substack{1 \leq i < r_{\mathbf{P}}  \\ e_i^{\mathrm{pri}} = 0}} (1- \mathrm{Pe}(\widetilde{i})),
        \end{equation}
        and
        \begin{equation} \small 
            \mathrm{P}(\mathbf{e}_j^{\mathrm{ext}}) = \prod_{\substack{1 \leq i < k - r_{\mathbf{P}}\\ e_{j,i}^{\mathrm{ext}} \neq 0}} \mathrm{Pe}(\widetilde{i})\prod_{\substack{1 \leq i < k - r_{\mathbf{P}} \\ e_{j,i}^{\mathrm{ext}}  = 0}} (1- \mathrm{Pe}(\widetilde{i})).
        \end{equation} 
        The detailed derivation of (\ref{equ::Condition::Ppro}) and expresion of $\beta$ are introdcued in Appendix \ref{Appdix::Ppro}.
        
        An implementation concern of the promising probability is that if computing (\ref{equ::Condition::Ppro}) for a TEP is more efficient than directly recovering a estimates by a TEP. We next show that by using a monotonicity trick, the DC can be implemented very efficiently. According to \cite[Proposition 1]{yue2021probability}, we can conclude that for a non-increasing $\mathcal{D}_{\min}$ (which holds naturally in a decoding searching for the minimum WHD), $\mathrm{P_d}(\mathbf{e}^{\mathrm{pri}},\mathbf{e}_j^{\mathrm{ext}})$ is monotonically increasing function of $\lambda = \mathrm{P}(\mathbf{e}^{\mathrm{pri}})\cdot\mathrm{P}(\mathbf{e}_j^{\mathrm{ext}})$. Utilizing the monotonicity of $\mathrm{P_d}(\mathbf{e}^{\mathrm{pri}},\mathbf{e}_j^{\mathrm{ext}})$, the DC is implemented in the following manner. 
        Assuming that there are a predetermined threshold $\mathrm{P_d'}$ and $\lambda_{\max} = 0$, we have
        \begin{itemize}
            \item \textit{Case (a)}: For a given $\mathbf{e}^{\mathrm{pri}}$, if $\mathrm{P}(\mathbf{e}^{\mathrm{pri}}) \leq \lambda_{\max}$, discard $\mathbf{e}^{\mathrm{pri}}$; otherwise, perform case (b).
            \item \textit{Case (b)}: For given $\mathbf{e}^{\mathrm{pri}}$ and $\mathbf{e}_j^{\mathrm{ext}}$, if $\lambda \leq \lambda_{\max}$, discard $\mathbf{e}^{\mathrm{pri}}$ and $\mathbf{e}_j^{\mathrm{ext}}$; otherwise, perform case (c).
            \item \textit{Case (c)}: Compute $\mathrm{P_d}(\mathbf{e}^{\mathrm{pri}},\mathbf{e}_j^{\mathrm{ext}})$ according to (\ref{equ::Condition::Ppro}). If $\mathrm{P_d}(\mathbf{e}^{\mathrm{pri}},\mathbf{e}_j^{\mathrm{ext}}) \leq \mathrm{P_d'}$, discard $\mathbf{e}^{\mathrm{pri}}$ and $\mathbf{e}_j^{\mathrm{ext}}$, and set $\lambda_{\max} = \lambda$ when $\lambda > \lambda_{\max}$.
       \end{itemize}
       Details of this implementation are presented in Algorithms \ref{ago::impLE-OSD}. With the above implementation, the DC can efficiently determine if TEPs can be discarded, because (\ref{equ::Condition::Ppro}) is only computed in case (c), while in case (a) and case (b), $\mathrm{P}(\mathbf{e}^{\mathrm{pri}})$ and $\mathrm{P}(\mathbf{e}_j^{\mathrm{ext}})$ are obtained with $\mathcal{O}(n)$ FLOPs by storing and reusing $\mathrm{Pe}(\widetilde{i})$.

        \subsubsection{Decoding stopping condition}
        Generally, the decoding SC identifies if the decoder has found the correct decoding results and terminates the decoding early. In \cite{yue2021revisit}, various SCs were proposed for the OSD decoding. These SCs were developed based on the Hamming distance or WHD from each codeword estimate to the received signal. We design the SC for LE-OSD by leveraging the idea of \textit{Soft individual stopping rule} from \cite{yue2021revisit}, where a success probability of each codeword estimate is computed and then the decoding is terminated if a large enough success probability is found. The success probability is defined as the probability that a codeword estimate is the correct estimate conditioning on the difference pattern between the codeword estimate and the received signal.
        
        Given an ordered codeword estimate $\widetilde{\mathbf{c}}_{\mathbf{e}}^{(j)}$ and its difference pattern $\mathbf{d}_{\mathbf{e}}^{(j)} = \widetilde{\mathbf{c}}_{\mathbf{e}}^{(j)} \oplus \widetilde{\mathbf{c}}_{\mathbf{e}}^{(j)}$, the success probability of $\widetilde{\mathbf{c}}_{\mathbf{e}}^{(j)}$, denoted by $\mathrm{P_s}(\widetilde{\mathbf{c}}_{\mathbf{e}}^{(j)})$, is computed as
        \begin{equation} \small \label{equ::Condition::Psuc}
            \mathrm{P_s}(\widetilde{\mathbf{c}}_{\mathbf{e}}^{(j)}) = \left(1 + \frac{\left(1-\lambda\right)\cdot 2^{k-n}}{\mathrm{P}(\mathbf{d}_{\mathbf{e}}^{(j)} )}\right) ^{-1}
        \end{equation}
        where
        \begin{equation} \small
            \mathrm{P}(\mathbf{d}_{\mathbf{e}}^{(j)} ) = \prod_{\substack{1 \leq i \leq n\\ d_{\mathbf{e},i}^{(j)} \neq 0}} \mathrm{Pe}(\widetilde{i})\prod_{\substack{1 \leq  i \leq n \\ d_{\mathbf{e},i}^{(j)} = 0}} (1- \mathrm{Pe}(\widetilde{i})).
        \end{equation} 
        and $\lambda = \mathrm{P}(\mathbf{e}^{\mathrm{pri}})\cdot\mathrm{P}(\mathbf{e}_j^{\mathrm{ext}})$. The derivation of (\ref{equ::Condition::Psuc}) is elaborated in detail in Appendix \ref{Appdix::Psuc}. 
        
        In the LE-OSD, the SC is implemented  as follows. Once a codeword estimate $\widetilde{\mathbf{c}}_{\mathbf{e}}^{(j)}$ is generated, if it results in a lower WHD than the recorded minimum one $\mathcal{D}_{\min}$, its success probability $\mathrm{P_s}(\widetilde{\mathbf{c}}_{\mathbf{e}}^{(j)})$ is computed accordingly. Then, with a predetermined treshold $\mathrm{P_s'}$, the decoding is terminated if 
        \begin{equation} \small  
            \mathrm{P_s}(\widetilde{\mathbf{c}}_{\mathbf{e}}^{(j)})\geq \mathrm{P_s'},
        \end{equation}
         and $\widetilde{\mathbf{c}}_{\mathbf{e}}^{(j)}\bm{\Pi}_{\widetilde{\mathbf{G}}}^{-1}\bm{\Pi}_{a}^{-1}$ is output as the final result.
        
        By storing and reusing $\mathrm{Pe}(\widetilde{i})$ in computing (\ref{equ::Condition::Psuc}), the success probability $\mathrm{P_s}(\widetilde{\mathbf{c}}_{\mathbf{e}}^{(j)}) $ is obtained with $\mathcal{O}(n)$ FLOPs. Furthermore, since the LE-OSD requires much fewer codeword estimates than the OSD as shown in Section \ref{sec::Simulation}, the overhead of performing SC checks will be significantly reduced compared to other SC-aided OSD algorithms.
        
        The LE-OSD algorithm employing both SC and DC is summarized in Algorithm \ref{ago::impLE-OSD}. We note that the ``Preprocessing'' is omitted for the sake of brevity. 
        
        \begin{spacing}{1.2}
            \begin{algorithm}
            \small
        	\caption{Improved LE-OSD}
        	\label{ago::impLE-OSD}
        	\begin{algorithmic} [1]
        		\REQUIRE Received signal $\bm{\gamma}$, Parameters $\rho$, $\tau$ and $\xi$, Thresholds $\mathrm{P_d'}$ and $\mathrm{P_s'}$
        		\ENSURE ~Optimal codeword estimate $\hat{\mathbf{c}}_{\mathrm{best}}$
        		
        		$//$ \textbf{Prepossessing}
        		
        		\STATE Perform processing part of Algorithm \ref{ago::LE-OSD}, and initialize $\lambda_{\max} = 0$, $\mathcal{D}_{\min} = \infty$
        		
        		$//$ \textbf{Re-processing}
        		
        		\FOR{$i=1:\sum\limits_{\ell = 0}^{\rho}\binom{r_{\mathbf{P}}}{\ell}$}
                		\STATE Select an unprocessed primary TEP $\mathbf{e}_i^{\mathrm{pri}}$ with $w(\mathbf{e}_i^{\mathrm{pri}}) \leq \rho$
                        \IF{$\mathrm{P}(\mathbf{e}^{\mathrm{pri}}) \leq \lambda_{\max}$}
                		\STATE \textbf{Continue} \hspace*{\fill}  $//$\textbf{Case (a) of DC}
            		    \ENDIF                		
                		\STATE Generate a valid TEP $\mathbf{e}_i$ according to (\ref{equ::LE-OSD::validTEP::generation})
                		\IF{$w(\mathbf{e}_i)>\tau$}
                		\STATE \textbf{Continue}
                		\ENDIF
                		\STATE Calculate $\mathbf{z}_{\mathbf{e}_i}=  (\mathbf{e}_i\oplus \widetilde{\mathbf{y}}_{\mathrm{P}})\mathbf{E}_{\mathbf{P}}^{\top}$ and obtain $\mathbf{x}_{\mathbf{e}_i}$ according to (\ref{equ::LE-OSD::Recover::Codebase})
		        		\FOR{$j=1:q_{\mathbf{e}_{i}}$}
		        		    \STATE Select an extended TEP $\mathbf{e}_{j}^{\mathrm{ext}}$ satisfies $w(\mathbf{e}_{j}^{\mathrm{ext}})\leq \xi - w(\mathbf{e})$
		        		    \STATE Compute $\lambda = \mathrm{P}(\mathbf{e}^{\mathrm{pri}})\cdot\mathrm{P}(\mathbf{e}_j^{\mathrm{ext}})$
		        		    \IF{$\lambda \leq \lambda_{\max}$}
                		        \STATE \textbf{Continue} \hspace*{\fill}  $//$\textbf{Case (b) of DC}       		     
		        		    \ELSE
		        		        \STATE Compute $\mathrm{P_d}(\mathbf{e}^{\mathrm{pri}},\mathbf{e}_j^{\mathrm{ext}})$ according to (\ref{equ::Condition::Ppro})
		        		        \IF{$\mathrm{P_d}(\mathbf{e}^{\mathrm{pri}},\mathbf{e}_j^{\mathrm{ext}}) \leq \mathrm{P_d'}$}
		        		        \STATE Set $\lambda_{\max} = \max(\lambda_{\max},\lambda)$
                		        \STATE \textbf{Continue} \hspace*{\fill}  $//$\textbf{Case (c) of DC}
		        		        \ENDIF
		        		    \ENDIF
                    		\STATE Obtain $\mathbf{x}_{j}$ according to (\ref{equ::LE-OSD::Recover::Codeaddon})
                    		\STATE Recover a codeword estimate $\widetilde{\mathbf{c}}_{\mathbf{e}_i}^{(j)}$ according to (\ref{equ::LE-OSD::Recover::setCode})                    		\IF{$\mathcal{D}(\widetilde{\mathbf{c}}_{\mathbf{e}_i}^{(j)},\bar{\mathbf{y}}) \leq \mathcal{D}_{\min}$}
                        		\STATE Update $\widetilde{\mathbf{c}}_{\mathrm{best}} = \widetilde{\mathbf{c}}_{\mathbf{e}_i}^{(j)}$ and $\mathcal{D}_{\min} = \mathcal{D}(\widetilde{\mathbf{c}}_{\mathbf{e}_i}^{(j)},\bar{\mathbf{y}})$
                        		\STATE Compute $\mathrm{P_s}(\widetilde{\mathbf{c}}_{\mathbf{e}}^{(j)})$
                    		    \IF{$\mathrm{P_s}(\widetilde{\mathbf{c}}_{\mathbf{e}}^{(j)}) \geq \mathrm{P_s'}$}
                    		    \RETURN $\hat{\mathbf{c}}_{\mathrm{best}} = \widetilde{\mathbf{c}}_{\mathrm{best}}\bm{\Pi}_{\widetilde{\mathbf{G}}}^{-1}\bm{\Pi}_{a}^{-1}$  \hspace*{\fill}  $//$\textbf{SC}
                    		    \ENDIF
                    		\ENDIF
            		    \ENDFOR
        		\ENDFOR 
        		\RETURN $\hat{\mathbf{c}}_{\mathrm{best}} = \widetilde{\mathbf{c}}_{\mathrm{best}}\bm{\Pi}_{\widetilde{\mathbf{G}}}^{-1}\bm{\Pi}_{a}^{-1}$
        	\end{algorithmic}
        \end{algorithm} 
        \end{spacing}
        
           \vspace{-0.5em}
        \subsection{Comparison with PB-OSD}
           \vspace{-0.5em}
           
        In this section, we compare the performance of the improved LE-OSD (ILE-OSD) and the PB-OSD in terms of the decoding complexity, where $(128,50,28)$ eBCH and $(128,78,16)$ eBCH codes are considered. In the implementation of ILE-OSD, the thresholds are set to $\mathrm{P_s'} = 0.99\epsilon(x)$ and $\mathrm{P_d'} = 0.002\sqrt{\frac{1 - \epsilon(x)}{N(x)}}$, where $\epsilon(x) = \sum_{i=0}^{x}\binom{k}{i}p^i(1-p)^{k-i}$, $N(x) = \sum_{i=0}^{x}\binom{k}{i}$, and $p$ is given by (\ref{equ::Condition::avePe}). When $k\leq n-k$, we select $x=\rho$, while when $k > n-k$, we select $x=\xi$.
        
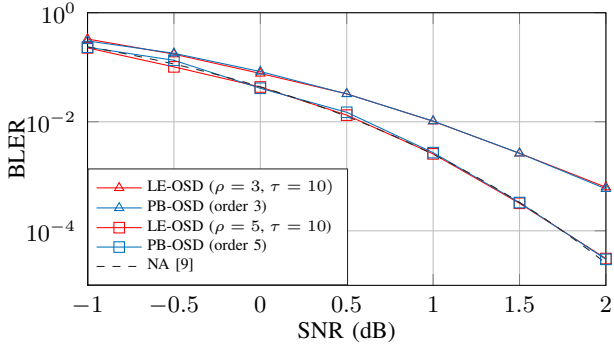
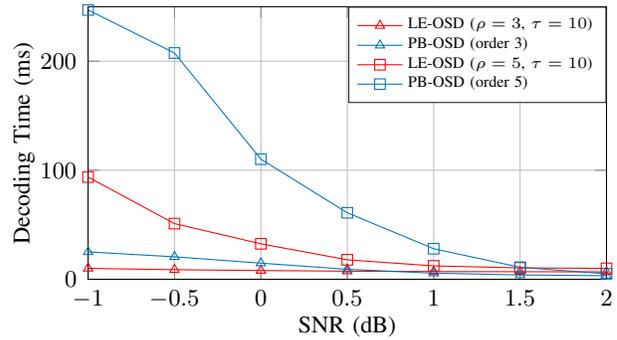
\begin{figure}[t]
            \centering
	    	\vspace{-1.5em}
            \subfigure[Block Error Rate]
            {
                \definecolor{mycolor1}{rgb}{0.00000,0.44700,0.74100}%
                \definecolor{mycolor2}{rgb}{0.14902,0.14902,0.14902}%
                \begin{tikzpicture}
                
                \begin{axis}[%
                width=0.38\textwidth,
                height=0.2\textwidth,
                at={(0.844in,0.559in)},
                scale only axis,
                xmin=-1,
                xmax=2,
                xlabel style={at={(0.5,1ex)},font=\color{white!15!black},font = \footnotesize},
                xlabel={SNR (dB)},
                ymode=log,
                ymin=1e-05,
                ymax=1,
                yminorticks=true,
                ylabel style={at={(1.5ex,0.5)},font=\color{white!15!black},font = \footnotesize},
                ylabel={BLER},
                axis background/.style={fill=white},
                tick label style={font=\footnotesize},
                xmajorgrids,
                ymajorgrids,
                yminorgrids,
                legend style={at={(0,0)}, anchor=south west, legend cell align=left, align=left, draw=white!15!black, font = \tiny, row sep=-2.5pt}
                ]
                
                \addplot [color=red, mark=triangle, mark options={solid, red}]
                  table[row sep=crcr]{%
                -1	0.329489291598023\\
                -0.5	0.173761946133797\\
                0	0.0770119368502118\\
                0.5	0.0325732899022801\\
                1	0.0102139829426485\\
                1.5	0.00263998521608279\\
                2	0.00063\\
                };
                \addlegendentry{LE-OSD ($\rho = 3$, $\tau = 10)$}
                
                \addplot [color=mycolor1, mark=triangle, mark options={solid, mycolor1}]
                  table[row sep=crcr]{%
                -1	0.303951367781155\\
                -0.5	0.180505415162455\\
                0	0.0834724540901502\\
                0.5	0.0323676970383557\\
                1	0.010306622004638\\
                1.5	0.00264274105101812\\
                2	0.00059\\
                };
                \addlegendentry{PB-OSD (order 3)}
                
                \addplot [color=red, mark=square, mark options={solid, red}]
                  table[row sep=crcr]{%
                -1	0.226500566251416\\
                -0.5	0.101626016260163\\
                0	0.043449923962633\\
                0.5	0.013245033112583\\
                1	0.00257921411346\\
                1.5	0.00032\\
                2	3.1e-05\\
                };
                \addlegendentry{LE-OSD ($\rho = 5$, $\tau = 10$)}
                
                \addplot [color=mycolor1, mark=square, mark options={solid, mycolor1}]
                  table[row sep=crcr]{%
                -1	0.233372228704784\\
                -0.5	0.134138162307176\\
                0	0.041135335252982\\
                0.5	0.014926487051272\\
                1	0.002716763790972\\
                1.5	0.00033\\
                2	3e-05\\
                };
                \addlegendentry{PB-OSD (order 5)}
                
                \addplot [color=mycolor2, dashed]
                  table[row sep=crcr]{%
                -1	0.236758297134104\\
                -0.9	0.208260853701435\\
                -0.8	0.181713269767827\\
                -0.7	0.157218413352623\\
                -0.6	0.134838731442401\\
                -0.5	0.114596719660018\\
                -0.4	0.0964766942642718\\
                -0.3	0.0804277480388215\\
                -0.2	0.0663677168257806\\
                -0.1	0.0541879394053937\\
                0	0.0437585628924649\\
                0.1	0.0349341306231101\\
                0.2	0.0275591903267109\\
                0.3	0.0214736767085331\\
                0.4	0.0165178527425736\\
                0.5	0.0125366353120444\\
                0.6	0.00938317988043082\\
                0.7	0.0069216517082532\\
                0.8	0.00502916371775775\\
                0.9	0.0035969096681827\\
                1	0.00253056263573499\\
                1.1	0.00175004053414459\\
                1.2	0.00118876121131961\\
                1.3	0.000792519226844437\\
                1.4	0.000518115470355533\\
                1.5	0.000331860881436437\\
                1.6	0.000208058838640719\\
                1.7	0.000127549775371578\\
                1.8	7.63787531356534e-05\\
                1.9	4.46243373296082e-05\\
                2	2.5406993386249e-05\\
                2.1	1.40786022383114e-05\\
                2.2	7.58223197903738e-06\\
                2.3	3.96308676477951e-06\\
                2.4	2.00722013925533e-06\\
                2.5	9.83472441133788e-07\\
                2.6	4.6533674338648e-07\\
                2.7	2.12221856823876e-07\\
                2.8	9.31007834477803e-08\\
                2.9	3.92029212692985e-08\\
                3	1.58081223418657e-08\\
                };
                \addlegendentry{NA \cite{erseghe2016coding}}
                
                \end{axis}
                \end{tikzpicture}%
                \label{Fig::12850::Fe}
            }
            \hspace{0em}
            \subfigure[Average decoding time]
            {
                \definecolor{mycolor1}{rgb}{0.00000,0.44700,0.74100}%
                \definecolor{mycolor2}{rgb}{0.00000,0.44706,0.74118}%
                \begin{tikzpicture}
                
                \begin{axis}[%
                 width=0.38\textwidth,
                height=0.2\textwidth,
                at={(0.841in,0.578in)},
                scale only axis,
                xmin=-1,
                xmax=2,
                xlabel style={at={(0.5,1ex)},font=\color{white!15!black},font=\footnotesize},
                xlabel={SNR (dB)},
                ymin=0,
                ymax=250,
                ylabel style={at={(2ex,0.5)},font=\color{white!15!black},font=\footnotesize},
                ylabel={Decoding Time (ms)},
                axis background/.style={fill=white},
                tick label style={font=\footnotesize},
                xmajorgrids,
                ymajorgrids,
                legend style={at={(1,1)}, anchor=north east, legend cell align=left, align=left, draw=white!15!black, font = \tiny, row sep=-2.5pt}
                ]
                
                \addplot [color=red, mark=triangle, mark options={solid, red}]
                  table[row sep=crcr]{%
                -1	10.0320202635914\\
                -0.5	8.78778601216334\\
                0	8.03128436657682\\
                0.5	7.46220780130293\\
                1	7.08594207139574\\
                1.5	6.93769811769054\\
                2	6.780094818\\
                };
                \addlegendentry{LE-OSD ($\rho = 3$, $\tau = 10$)}
                
                \addplot [color=mycolor2, mark=triangle, mark options={solid, mycolor2}]
                  table[row sep=crcr]{%
                -1	25.1407211246201\\
                -0.5	20.6216167870036\\
                0	14.8141727045075\\
                0.5	9.16779459459459\\
                1	5.62885643390879\\
                1.5	4.00327424648846\\
                2	3.288712158\\
                };
                \addlegendentry{PB-OSD (order 3)}
                
                \addplot [color=red, mark=square, mark options={solid, red}]
                  table[row sep=crcr]{%
                -1	93.7465723669309\\
                -0.5	51.0486973069106\\
                0	32.5293728220726\\
                0.5	17.875617\\
                1	12.2870597965\\
                1.5	10.465457895\\
                2	9.997700584\\
                };
                \addlegendentry{LE-OSD ($\rho = 5$, $\tau = 10$)}
                
                \addplot [color=mycolor1, mark=square, mark options={solid, mycolor1}]
                  table[row sep=crcr]{%
                -1	246.990540023337\\
                -0.5	207.505642454728\\
                0	110.164034677088\\
                0.5	60.9813623479364\\
                1	28.0070052705218\\
                1.5	11.015938807\\
                2	5.015734129\\
                };
                \addlegendentry{PB-OSD (order 5)}
                
                \end{axis}
                \end{tikzpicture}%
                \label{Fig::12850::Time}
            }
        	\vspace{0em}
        	\vspace{-0.3em}
            \caption{The comparisons of decoding $(128,50,28)$ eBCH code.}
            \label{Fig::12850}
        	\vspace{-0.3em}
        \end{figure}

        In decoding $(128,50,28)$ eBCH code, we set the parameters $\tau = \xi = 10$, and compare LE-OSD ($\rho =5$) and LE-OSD ($\rho = 3$) with the order $5$ and order $3$ PB-OSD, respectively. As shown by Fig. \ref{Fig::12850::Fe}, the ILE-OSD with $\rho = 5$ has almost the same BLER performance as the order-5 PB-OSD, approaching the NA bound \cite{erseghe2016coding}. However, in terms of the complexity, the ILE-OSD is considerably more efficient than the PB-OSD at the most of the SNRs, as depiected in Fig \ref{Fig::12850::Time}. For example, the LED ($\rho=5$) on average takes less than 100 ms to decode a codeword, compared to around 250 ms of order-5 PB-OSD at SNR = 1 dB. 
        
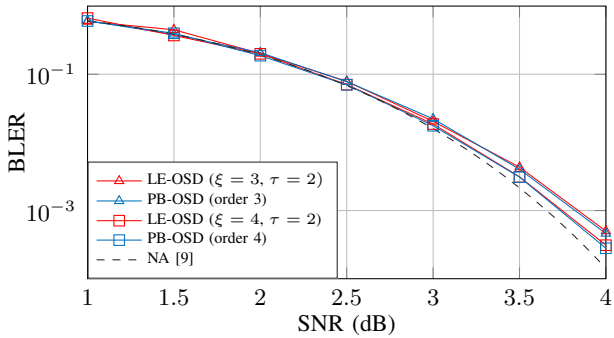
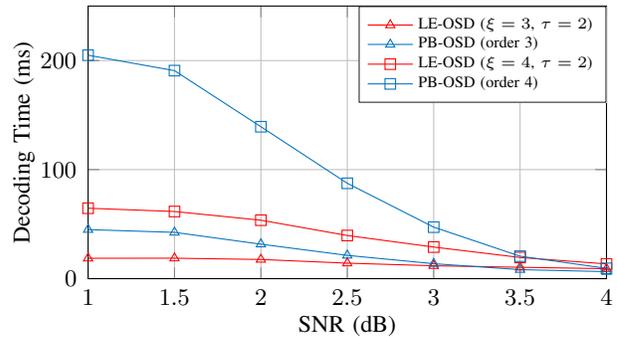
\begin{figure}[t]
            \centering
            \subfigure[Block Error Rate]
            {
                \definecolor{mycolor1}{rgb}{0.00000,0.44700,0.74100}%
                \definecolor{mycolor2}{rgb}{0.14902,0.14902,0.14902}%
                \begin{tikzpicture}
                
                \begin{axis}[%
                width=0.38\textwidth,
                height=0.2\textwidth,
                at={(0.844in,0.559in)},
                scale only axis,
                xmin=1,
                xmax=4,
                xlabel style={at={(0.5,1ex)},font=\color{white!15!black},font = \footnotesize},
                xlabel={SNR (dB)},
                ymode=log,
                ymin=1e-04,
                ymax=1,
                yminorticks=true,
                ylabel style={at={(1.5ex,0.5)},font=\color{white!15!black},font = \footnotesize},
                ylabel={BLER},
                axis background/.style={fill=white},
                tick label style={font=\footnotesize},
                xmajorgrids,
                ymajorgrids,
                yminorgrids,
                legend style={at={(0,0)}, anchor=south west, legend cell align=left, align=left, draw=white!15!black, font = \tiny, row sep=-2.5pt}
                ]
                
                \addplot [color=red, mark=triangle, mark options={solid, red}]
                  table[row sep=crcr]{%
                    1	0.593471810089021\\
                    1.5	0.449438202247191\\
                    2	0.201612903225806\\
                    2.5	0.0788022064617809\\
                    3	0.0203665987780041\\
                    3.5	0.00433228636412867\\
                    4	0.0005\\
                    };
                \addlegendentry{LE-OSD ($\xi = 3$, $\tau = 2)$}
                
                \addplot [color=mycolor1, mark=triangle, mark options={solid, mycolor1}]
                  table[row sep=crcr]{%
                    1	0.609756097560976\\
                    1.5	0.383877159309021\\
                    2	0.20703933747412\\
                    2.5	0.078003120124805\\
                    3	0.0221950948840306\\
                    3.5	0.00404498017959712\\
                    4	0.00046\\
                    };
                \addlegendentry{PB-OSD (order 3)}
                
                \addplot [color=red, mark=square, mark options={solid, red}]
                  table[row sep=crcr]{%
                    1	0.666666666666667\\
                    1.5	0.371057513914657\\
                    2	0.199401794616152\\
                    2.5	0.0704721634954193\\
                    3	0.0186115764005211\\
                    3.5	0.00312763894536015\\
                    4	0.00031\\
                    };
                \addlegendentry{LE-OSD ($\xi = 4$, $\tau = 2$)}
                
                \addplot [color=mycolor1, mark=square, mark options={solid, mycolor1}]
                  table[row sep=crcr]{%
                    1	0.604229607250755\\
                    1.5	0.399201596806387\\
                    2	0.187441424554827\\
                    2.5	0.0699056274030059\\
                    3	0.0175746924428823\\
                    3.5	0.00310207373629271\\
                    4	0.00028\\
                    };
                \addlegendentry{PB-OSD (order 4)}
                
                \addplot [color=mycolor2, dashed]
                  table[row sep=crcr]{%
                    1	0.606562238023019\\
                    1.1	0.56360390775042\\
                    1.2	0.519566937007286\\
                    1.3	0.474957581063885\\
                    1.4	0.430314316960045\\
                    1.5	0.386189747987979\\
                    1.6	0.343130929909008\\
                    1.7	0.301659197337762\\
                    1.8	0.262250675889753\\
                    1.9	0.225318683433818\\
                    2	0.191199142708461\\
                    2.1	0.160139947686713\\
                    2.2	0.132294957936205\\
                    2.3	0.107722959674723\\
                    2.4	0.0863915589264376\\
                    2.5	0.0681855967610463\\
                    2.6	0.0529193367075357\\
                    2.7	0.040351405223975\\
                    2.8	0.0302012956334662\\
                    2.9	0.0221661912612027\\
                    3	0.0159369280936989\\
                    3.1	0.011212090241986\\
                    3.2	0.00770948890492658\\
                    3.3	0.00517458374160003\\
                    3.4	0.00338572601417519\\
                    3.5	0.00215639769416513\\
                    3.6	0.0013348582022704\\
                    3.7	0.000801769277835477\\
                    3.8	0.00046644021324026\\
                    3.9	0.000262324912379742\\
                    4	0.000142324435795215\\
                    };
                \addlegendentry{NA \cite{erseghe2016coding}}
                
                \end{axis}
                \end{tikzpicture}%
                \label{Fig::12878::Fe}
            }
            \hspace{0em}
            \subfigure[Average decoding time]
            {
                \definecolor{mycolor1}{rgb}{0.00000,0.44700,0.74100}%
                \definecolor{mycolor2}{rgb}{0.00000,0.44706,0.74118}%
                \begin{tikzpicture}
                
                \begin{axis}[%
                 width=0.38\textwidth,
                height=0.2\textwidth,
                at={(0.841in,0.578in)},
                scale only axis,
                xmin=1,
                xmax=4,
                xlabel style={at={(0.5,1ex)},font=\color{white!15!black},font=\footnotesize},
                xlabel={SNR (dB)},
                ymin=0,
                ymax=250,
                ylabel style={at={(2ex,0.5)},font=\color{white!15!black},font=\footnotesize},
                ylabel={Decoding Time (ms)},
                axis background/.style={fill=white},
                tick label style={font=\footnotesize},
                xmajorgrids,
                ymajorgrids,
                legend style={at={(1,1)}, anchor=north east, legend cell align=left, align=left, draw=white!15!black, font = \tiny, row sep=-2.5pt}
                ]
                
                \addplot [color=red, mark=triangle, mark options={solid, red}]
                  table[row sep=crcr]{%
                1	18.6938080745342\\
                1.5	18.7538592519685\\
                2	17.5745949570815\\
                2.5	14.1918426895307\\
                3	11.8092593666667\\
                3.5	10.4192428\\
                4	9.12048546666667\\
                };
                \addlegendentry{LE-OSD ($\xi = 3$, $\tau = 2$)}
                
                \addplot [color=mycolor2, mark=triangle, mark options={solid, mycolor2}]
                  table[row sep=crcr]{%
                1	44.9731629969419\\
                1.5	42.4475722440945\\
                2	31.619503286385\\
                2.5	21.3379787189212\\
                3	13.7160291913903\\
                3.5	8.25440556955546\\
                4	6.439039219\\
                };
                \addlegendentry{PB-OSD (order 3)}
                
                \addplot [color=red, mark=square, mark options={solid, red}]
                  table[row sep=crcr]{%
                1	64.5777013333333\\
                1.5	61.587920593692\\
                2	53.5932031904287\\
                2.5	39.5700873854827\\
                3	28.9654508189094\\
                3.5	19.4118664122854\\
                4	13.510835822\\
                };
                \addlegendentry{LE-OSD ($\xi = 4$, $\tau = 2$)}
                
                \addplot [color=mycolor1, mark=square, mark options={solid, mycolor1}]
                  table[row sep=crcr]{%
                1	204.921043202417\\
                1.5	190.765489421158\\
                2	139.324854358013\\
                2.5	87.4689069556099\\
                3	47.2397552196837\\
                3.5	20.5918723713803\\
                4	9.442476441\\
                };
                \addlegendentry{PB-OSD (order 4)}

                \end{axis}
                \end{tikzpicture}%
                \label{Fig::12878::Time}
            }
        	\vspace{0em}
        	\vspace{-0.3em}
            \caption{The comparisons of decoding $(128,78,16)$ eBCH code.}
            \label{Fig::12878}
        	\vspace{-0.3em}
        \end{figure}

        The ILE-OSD and the PB-OSD are further compared in decoding $(128,78,16)$ eBCH code, as depicted in Fig. \ref{Fig::12878}. We set $\tau = \rho = 2$ and select different values of $\xi$ for the LE-OSD. As can be shown, the ILE-OSD with proper parameter selection achieves the same BLER as the PB-OSD, while having a lower complexity at a large range of SNRs. For instance, the ILE-OSD ($\xi =4$) only requires around 60 ms to decode one codeword at SNR = 1.5 dB, while the order-4 PB-OSD spends around 200 ms.  
        
        From Fig. \ref{Fig::12850} and Fig. \ref{Fig::12878}, we can conclude that the ILE-OSD is particularly efficient for high-order decoding at low-to-medium SNRs. This improvements of complexity results from the following reasons: 1) the LE-OSD has a lower computational complexity than the original OSD (as shown in Section \ref{sec::Simulation}), and this advantage is retained in ILE-OSD with the use of decoding conditions, and 2) the overheads of performing DC and SC are reduced in ILE-OSD, as LE-OSD generates fewer codeword estimates than the OSD. However, it is worth noting that ILE-OSD tends to have a slightly higher complexity than the PB-OSD at high SNRs. This is because the SC can effectively terminate the OSD (or its variants) at high SNRs early; thus the number of generated codeword estimates is significantly reduced (e.g., see \cite[Fig. 2]{yue2021probability}). Consequently, the complexity of OSD variants is no longer dominated by the number of generated codeword estimates at high SNRs but is also affected by the consumption of the ``Preprocessing'' part. Since ILE-OSD performs three GEs in preprocessing, its complexity may be unfavorable at high SNRs. Therefore, one can consider adaptively employing different decoding schemes as per the system requirements. In addition, the approach in \cite{choi2021fast} may be considered to reduce the expense of GEs in ILE-OSD.

        \vspace{-0.5em}
 \section{Conclusion} \label{Sec::Conclusion}
    \vspace{-0.5em}
 In this paper, we proposed a new LE-OSD to reduce the decoding complexity of OSD algorithm. Different from the OSD, the LE-OSD uses high reliable parity bits to recover the information bits, which is equivalent to solving a SLE. The TEPs that make the SLE feasible, referred to as the valid TEPs, are added to the parity bits to retrieve a list of code estimates. Three parameters, namely $\rho$, $\tau$ and $\xi$, are introduced to constrain the Hamming weight of the TEPs for limiting the overall complexity. Furthermore, the BLER, the asymptotic error rate, and the complexity are analyzed for the LE-OSD, and it is shown that the LE-OSD has the similar performance as OSD with proper parameter selection and can asymptotically achieve the ML decoding. Simulation results for low-rate and high-rate codes show that the LE-OSD can achieve the similar performance as high-order OSD algorithms with significantly reduced complexity. 
 
 We also improved the LE-OSD by applying the decoding stopping condition and the TEP discarding condition. It is shown that the improved LE-OSD has a superior performance and complexity trade-off than the approach from literature, specially at low-to-medium SNRs, having three to four times of complexity reduction.



   \vspace{-0.5em}
\appendices
   \vspace{-0.5em}

   \vspace{-0.5em}
\section{Derivation of (\ref{equ::Condition::Ppro})} \label{Appdix::Ppro}
   \vspace{-0.5em}
    To derive (\ref{equ::Condition::Ppro}), we first notice that a codeword estimates $\widetilde{\mathbf{c}}_{\mathbf{e}}^{(j)}$ in LE-OSD is uniquely determined by a primary TEP $\mathbf{e}^{\mathrm{pri}}$ and its corresponding extended TEP $\mathbf{e}_j^{\mathrm{ext}}$. Specifically, by omitting the permutation introduced by $\Pi_{\mathbf{P}}$ and $\Pi_{\mathbf{Q}}$, we can rewrite (\ref{equ::LE-OSD::Recover::setCode}) as
    \begin{equation} \small
    \begin{split}
         \widetilde{\mathbf{c}}_{\mathbf{e}}^{(j)} =& \big[\mathbf{x}_{\mathbf{e}}\oplus(([\widetilde{y}]_{r_{\mathbf{P}}+1}^{k}\oplus \mathbf{e}_{j}^{\mathrm{ext}})\mathbf{P}_{\mathrm{t}}^{\top}) \ \ \ \widetilde{\mathbf{y}}_{\mathrm{P}}\oplus (\mathbf{e}^{\mathrm{pri}} \mathbf{Q}_{\mathrm{t}}^{\top} \oplus \mathbf{e}_0)\big]  \\
         = & \big[\mathbf{x}_{\mathbf{e}}\!\oplus\!(([\widetilde{y}]_{r_{\mathbf{P}}+1}^{k}\oplus \mathbf{e}_{j}^{\mathrm{ext}})\mathbf{P}_{\mathrm{r}}^{\top}) \ \ \ [\widetilde{y}]_{r_{\mathbf{P}}+1}^{k}\!\oplus\! \mathbf{e}_{j}^{\mathrm{ext}} \ \ \  \ [\widetilde{y}]_{k+1}^{n-r_{\mathbf{P}}}\!\oplus\!(\mathbf{e}^{\mathrm{pri}} \mathbf{Q}_{\mathrm{r}}^{\top} \oplus \mathbf{e}_0) \ \ \ [\widetilde{y}]_{n-r_{\mathbf{P}}+1}^{n}\!\oplus\! \mathbf{e}^{\mathrm{pri}} \big],
    \end{split}
    \end{equation}
    where $\mathrm{P}_{\mathrm{r}}$ and $\mathrm{Q}_{\mathrm{r}}$ are given by (\ref{Mat::LE-OSD::Recover::Pr}) and (\ref{Mat::LE-OSD::validTEP::Qr}), respectively. Since the $\mathbf{e}^{\mathrm{pri}}$ and $\mathbf{e}_j^{\mathrm{ext}}$ uniquely determine a codeword, we can conclude that  $\widetilde{\mathbf{c}}_{\mathbf{e}}^{(j)}$ is the correct codeword estimate if and only if $\mathbf{e}^{\mathrm{pri}}$ and $\mathbf{e}_j^{\mathrm{ext}}$ eliminate the hard-decision errors over $[\widetilde{y}]_{r_{\mathbf{P}}+1}^{k}$ and $[\widetilde{y}]_{n - r_{\mathbf{P}}+1}^{n}$ respectively. 
    
    Let $\widetilde{\mathbf{e}}$ denote the ordered hard-decision error pattern introduced by the channel noise, and then $\widetilde{\mathbf{y}}$ is rewritten as $\widetilde{\mathbf{y}} = \widetilde{\mathbf{c}} \oplus \widetilde{\mathbf{e}} $, where $\widetilde{\mathbf{c}}$ is the ordered transmitted codeword. Let us represent the difference pattern between $\widetilde{\mathbf{c}}_{\mathbf{e}}^{(j)}$ and $\widetilde{\mathbf{y}}$ as $\widetilde{\mathbf{d}}_{\mathbf{e}}^{(j)} = [\mathbf{h}_1 \ \  \mathbf{e}_{j}^{\mathrm{ext}} \ \ \mathbf{h}_2 \ \ \mathbf{e}^{\mathrm{pri}}]$, where $\mathbf{h}_1$ and $\mathbf{h}_2$ are respectively given by
    \begin{equation} \small
        \mathbf{h}_1 = 
        \begin{cases}
            [\widetilde{e}]_{1}^{r_{\mathbf{P}}} \ \ & \textup{if $\widetilde{\mathbf{c}}_{\mathbf{e}}^{(j)} = \widetilde{\mathbf{c}}$} \\
            [\widetilde{y}]_{1}^{r_{\mathbf{P}}} \oplus \mathbf{x}_{\mathbf{e}}\oplus(([\widetilde{y}]_{r_{\mathbf{P}}+1}^{k}\oplus \mathbf{e}_{j}^{\mathrm{ext}})\mathbf{P}_{\mathrm{r}}^{\top})   \ \ & \textup{if $\widetilde{\mathbf{c}}_{\mathbf{e}}^{(j)} \neq  \widetilde{\mathbf{c}}$}
        \end{cases}
        \ \ \ \text{and} \ \ \
        \mathbf{h}_2 = 
        \begin{cases}
            [\widetilde{e}]_{k+1}^{n-r_{\mathbf{P}}} \ \ & \textup{if $\widetilde{\mathbf{c}}_{\mathbf{e}}^{(j)} = \widetilde{\mathbf{c}}$} \\
            \mathbf{e}^{\mathrm{pri}} \mathbf{Q}_{\mathrm{r}}^{\top} \oplus \mathbf{e}_0   \ \ & \textup{if $\widetilde{\mathbf{c}}_{\mathbf{e}}^{(j)} \neq  \widetilde{\mathbf{c}}$}
        \end{cases}
    \end{equation}
    It can be seen that when $\widetilde{\mathbf{c}}_{\mathbf{e}}^{(j)} = \widetilde{\mathbf{c}}$, the difference pattern is simply given by $\widetilde{\mathbf{d}}_{\mathbf{e}}^{(j)} = \widetilde{\mathbf{e}}$. On the other hand, when $\widetilde{\mathbf{c}}_{\mathbf{e}}^{(j)} \neq \widetilde{\mathbf{c}}$, $\mathbf{h}_1$ will be a length-$r_{\mathbf{P}}$ random binary vector and $\mathbf{h}_2$ will be a length-$(n-k-r_{\mathbf{P}})$ random binary vector if $\mathcal{C}(n,k)$ is a random code (or having a binomial weight spectrum ). 
    
    Therefore, let $D_{h}$ represent a random variable representing the Hamming weight of $[\mathbf{h}_1 \ \ \mathbf{h}_2]$, i.e., $w(\mathbf{h}_1) + w(\mathbf{h}_2)$. Using the approach of \cite[Corollary 1]{yue2021revisit}, the $\mathrm{pmf}$ of $D_{h}$ can be approximated as 
    \begin{equation} \small \label{equ::Appx::Hdis}
    \begin{split}
        p_{D_{h}}(x) &\approx  \lambda\sum_{i=0}^{x} \! \binom{n-k}{i} p^i (1-p)^{n\!- \!k \!- \! i}  
             +  \left(1 - \lambda\right)\sum_{i=0}^{x} \! \binom{n-k}{i}  2^{k\!- \!n},   
        \end{split}
    \end{equation}
    for a integer $x$, $0 \leq x \leq n-k$, where $\lambda = \mathrm{P}(\mathbf{e}^{\mathrm{pri}})\cdot\mathrm{P}(\mathbf{e}_j^{\mathrm{ext}})$ and $p$ is given by (\ref{equ::Condition::avePe}). We note that the approximation in (\ref{equ::Appx::Hdis}) stands for that 1) the permutations $\Pi_{\mathbf{P}}$ and $\Pi_{\mathbf{Q}}$ are omitted and 2) not all block codes follow a strict binomial weight spectrum. For example, it has been shown in \cite{yue2021revisit} when the Polar codes are employed, the Hamming distance from incorrect codeword estimates to the received signal follows distributions discrepant from the binomial distribution.
    
    As the promising probability is defined as the probability that a TEP can result in a codeword estimate whose WHD to the received signal is less than the recorded minimum one, it can be denoted as $\mathrm{Pr}(\mathcal{D}_{\mathbf{e}}^{(j)} \!\!\leq\! \mathcal{D}_{\min})$. Utilizing the approximation method of \cite[Eq. (7)]{yue2021probability}, $\mathrm{Pr}(\mathcal{D}_{\mathbf{e}}^{(j)} \!\!\leq\! \mathcal{D}_{\min})$ can be approximated by
    \begin{equation} \small\label{equ::Appx::PproDef}
        \mathrm{Pr}(\mathcal{D}_{\mathbf{e}}^{(j)} \!\!\leq\! \mathcal{D}_{\min}) \approx \mathrm{Pr}\left(D_{h} \!\!\leq\! \frac{1}{\mathbb{E}_{\mathrm{P}}(\bm{\alpha})}(\mathcal{D}_{\min}-\dot{\mathcal{D}}_{\mathbf{e}}^{(j)})\right),
    \end{equation}
    where
        \begin{equation} \small  
          \mathbb{E}_{\mathrm{P}}[\bm{\alpha}] \triangleq \frac{1}{n-k}\left(\sum_{i=1}^{r_{\mathbf{P}}}\widetilde{\alpha}_i + \sum_{i=k+1}^{n-r_{\mathbf{P}}}\widetilde{\alpha}_i  \right),
        \end{equation}
    and $\dot{\mathcal{D}}_{\mathbf{e}}^{(j)})$ is the component of $\mathcal{D}_{\mathbf{e}}^{(j)}$ introduced by $\mathbf{e}^{\mathrm{pri}}$ and $\mathbf{e}_j^{\mathrm{ext}}$, i.e.,
    \begin{equation} \small  
      \dot{\mathcal{D}}_{\mathbf{e}}^{(j)}) = \sum_{\substack{r_{\mathbf{P}}<i \leq k \\ \widetilde{d}_{\mathbf{e},i}^{(j)}\neq 0}} \widetilde{\alpha}_{i} + \sum_{\substack{n-r_{\mathbf{P}}<i \leq n \\ \widetilde{d}_{\mathbf{e},i}^{(j)}\neq 0}} \widetilde{\alpha}_{i}.
    \end{equation} 
    
    The probability (\ref{equ::Appx::PproDef}) can be approximated computed by using the pmf (\ref{equ::Appx::Hdis}). Precisely, let $$\beta = \min(\max(0, \lfloor (\mathcal{D}_{\min}-\dot{\mathcal{D}}_{\mathbf{e}}^{(j)})/\mathbb{E}_{\mathrm{P}}(\bm{\alpha}) \rfloor),n-k),$$ and then (\ref{equ::Appx::PproDef}) is approximately calculated as
    \begin{equation} \small  \label{equ::Appx::PproFinal}
        \mathrm{Pr}(\mathcal{D}_{\mathbf{e}}^{(j)} \!\!\leq\! \mathcal{D}_{\min}) \approx \sum_{x=0}^{\beta}p_{D_{h}}(x).
    \end{equation}
    Let $\mathrm{P_d}(\mathbf{e}^{\mathrm{pri}},\mathbf{e}_j^{\mathrm{ext}})$ denote the right-hand side of (\ref{equ::Appx::PproFinal}), we obtain (\ref{equ::Condition::Ppro}).
    
       \vspace{-0.5em}
    \section{Derivation of (\ref{equ::Condition::Psuc})} \label{Appdix::Psuc}
   \vspace{-0.5em}
   
    The success probability is defined as the probability that a codeword estimate $\widetilde{\mathbf{c}}_{\mathbf{e}}^{(j)}$ is the correct estimate conditioning on $\widetilde{\mathbf{d}}_{\mathbf{e}}^{(j)}$, i.e., $\mathrm{Pr}(\widetilde{\mathbf{c}}_{\mathbf{e}}^{(j)} = \widetilde{\mathbf{c}}|\widetilde{\mathbf{d}}_{\mathbf{e}}^{(j)})$. As discussed in Appendix \ref{Appdix::Ppro}, $\widetilde{\mathbf{c}}_{\mathbf{e}}^{(j)}$ is the correct codeword estimate if and only if $\mathbf{e}^{\mathrm{pri}}$ and $\mathbf{e}_j^{\mathrm{ext}}$ eliminate the hard-decision errors over $[\widetilde{y}]_{r_{\mathbf{P}}+1}^{k}$ and $[\widetilde{y}]_{n - r_{\mathbf{P}}+1}^{n}$ respectively. Therefore, there exists 
    \begin{equation} \small \label{equ::Appx::PsucDef}
        \mathrm{Pr}(\widetilde{\mathbf{c}}_{\mathbf{e}}^{(j)} \!=\! \widetilde{\mathbf{c}}|\widetilde{\mathbf{d}}_{\mathbf{e}}^{(j)}) = \mathrm{Pr}(\mathbf{e}^{\mathrm{pri}} \!=\! [\widetilde{e}]_{n - r_{\mathbf{P}}+1}^{n},  \mathbf{e}_j^{\mathrm{ext}} \!=\! [\widetilde{e}]_{r_{\mathbf{P}}+1}^{k} | \widetilde{\mathbf{d}}_{\mathbf{e}}^{(j)}).         
    \end{equation}
    For the sake of simplicity, we denote the right-hand side of (\ref{equ::Appx::PsucDef}) as $\mathrm{P}(\mathbf{e}^{\mathrm{pri}},  \mathbf{e}_j^{\mathrm{ext}}| \widetilde{\mathbf{d}}_{\mathbf{e}}^{(j)})$.
    
    According to the Bayes' theorem, the conditional probability (\ref{equ::Appx::PsucDef}) can be further as
    \begin{equation} \small \label{equ::Appx::postP1}
    \begin{split}
        \mathrm{Pr}(\widetilde{\mathbf{c}}_{\mathbf{e}}^{(j)} = \widetilde{\mathbf{c}}|\widetilde{\mathbf{d}}_{\mathbf{e}}^{(j)}) = & \frac{ \mathrm{Pr}(\widetilde{\mathbf{d}}_{\mathbf{e}}^{(j)}|\widetilde{\mathbf{c}}_{\mathbf{e}}^{(j)} = \widetilde{\mathbf{c}}) \mathrm{Pr}(\widetilde{\mathbf{c}}_{\mathbf{e}}^{(j)} = \widetilde{\mathbf{c}})}{\mathrm{Pr}(\widetilde{\mathbf{d}}_{\mathbf{e}}^{(j)})} =  \left(1 + \frac{\mathrm{Pr}(\widetilde{\mathbf{c}}_{\mathbf{e}}^{(j)} \neq \widetilde{\mathbf{c}})}{\mathrm{Pr}(\widetilde{\mathbf{c}}_{\mathbf{e}}^{(j)} = \widetilde{\mathbf{c}})}\cdot \frac{\mathrm{Pr}(\widetilde{\mathbf{d}}_{\mathbf{e}}^{(j)}|\widetilde{\mathbf{c}}_{\mathbf{e}}^{(j)} \neq \widetilde{\mathbf{c}})}{\mathrm{Pr}(\widetilde{\mathbf{d}}_{\mathbf{e}}^{(j)}|\widetilde{\mathbf{c}}_{\mathbf{e}}^{(j)} = \widetilde{\mathbf{c}})} \right)^{-1}.
    \end{split}
    \end{equation}
    With a minor notation abuse, $\mathrm{Pr}(\widetilde{\mathbf{d}}_{\mathbf{e}}^{(j)})$ in (\ref{equ::Appx::postP1}) denotes the probability that the difference pattern between $\widetilde{\mathbf{c}}_{\mathbf{e}}^{(j)}$ and $\widetilde{\mathbf{y}}$ is computed as $\widetilde{\mathbf{d}}_{\mathbf{e}}^{(j)}$, by regarding $\widetilde{\mathbf{c}}_{\mathbf{e}}^{(j)}\oplus\widetilde{\mathbf{y}}$ as a random vector.
    
    To compute (\ref{equ::Appx::postP1}), we first obtain that $\mathrm{Pr}(\widetilde{\mathbf{c}}_{\mathbf{e}}^{(j)} = \widetilde{\mathbf{c}}) = \mathrm{P}(\mathbf{e}^{\mathrm{pri}},  \mathbf{e}_j^{\mathrm{ext}}) = \mathrm{P}(\mathbf{e}^{\mathrm{pri}})\mathrm{P}(\mathbf{e}_j^{\mathrm{ext}}) = \lambda$, and $\mathrm{Pr}(\widetilde{\mathbf{c}}_{\mathbf{e}}^{(j)} \neq \widetilde{\mathbf{c}}) = 1-\lambda$. Then, as discussed in Appendix \ref{Appdix::Ppro}, there exists $\widetilde{\mathbf{d}}_{\mathbf{e}}^{(j)} = \widetilde{\mathbf{e}}$ if $\widetilde{\mathbf{c}}_{\mathbf{e}}^{(j)} = \widetilde{\mathbf{c}})$. Thus, the conditional probability $\mathrm{Pr}(\widetilde{\mathbf{d}}_{\mathbf{e}}^{(j)}|\widetilde{\mathbf{c}}_{\mathbf{e}}^{(j)} = \widetilde{\mathbf{c}})$ is derived as
    \begin{equation} \small   \label{equ::Appx::postP2}
    \begin{split}
        \mathrm{Pr}(\widetilde{\mathbf{d}}_{\mathbf{e}}^{(j)}|\widetilde{\mathbf{c}}_{\mathbf{e}}^{(j)} \!=\! \widetilde{\mathbf{c}}) &= \prod_{\substack{1 <i \leq r_{\mathbf{P}} \\ \widetilde{d}_{\mathbf{e},i}^{(j)}\neq 0}}\mathrm{Pe}(\widetilde{i})\prod_{\substack{1<i \leq r_{\mathbf{P}} \\ \widetilde{d}_{\mathbf{e},i}^{(j)}\neq 0}} (1 - \mathrm{Pe}(\widetilde{i}))  \prod_{\substack{k <i \leq n-r_{\mathbf{P}} \\ \widetilde{d}_{\mathbf{e},i}^{(j)}\neq 0}}\mathrm{Pe}(\widetilde{i})\prod_{\substack{k <i \leq n-r_{\mathbf{P}}\\ \widetilde{d}_{\mathbf{e},i}^{(j)}\neq 0}} (1 - \mathrm{Pe}(\widetilde{i})) .
    \end{split}
    \end{equation}
    When $\widetilde{\mathbf{c}}_{\mathbf{e}}^{(j)} \neq \widetilde{\mathbf{c}}$, it is discussed in Appendix \ref{Appdix::Ppro} that $\mathbf{h}_1$ and $\mathbf{h}_2$ are binary random vectors. Therefore, we can obtain that $\mathrm{Pr}(\widetilde{\mathbf{d}}_{\mathbf{e}}^{(j)}|\widetilde{\mathbf{c}}_{\mathbf{e}}^{(j)} \!\neq\! \widetilde{\mathbf{c}}) \approx \frac{1}{2^{n-k}}$,
    where the approximation comes from assuming that $\mathcal{C}(n,k)$ is a random code. Finally, substituting (\ref{equ::Appx::postP2}), $\mathrm{Pr}(\widetilde{\mathbf{d}}_{\mathbf{e}}^{(j)}|\widetilde{\mathbf{c}}_{\mathbf{e}}^{(j)} \!\neq\! \widetilde{\mathbf{c}}) \approx \frac{1}{2^{n-k}}$ and $\mathrm{Pr}(\widetilde{\mathbf{c}}_{\mathbf{e}}^{(j)} = \widetilde{\mathbf{c}}) = \lambda$ into (\ref{equ::Appx::postP1}), we obtain (\ref{equ::Condition::Psuc}).

\ifCLASSOPTIONcaptionsoff
  \newpage
\fi



%

\begin{spacing}{1.5}
\bibliography{IEEEabrv,MyCollection_OSD,MyCollection_LP,MyCollection_Survey,MyCollection_Classic,MyCollection_ML,MyCollection_Math,MyCollection_GRAND}
\bibliographystyle{IEEEtran}  
\end{spacing}

%

\end{document}